\documentclass[%
 reprint,
superscriptaddress,
 amsmath,amssymb,
 pre,
]{revtex4-2}

\usepackage{graphicx}
\usepackage{dcolumn}
\usepackage{bm}
\usepackage{physics}
\usepackage{footnote}
\usepackage{tabularx}
\usepackage{graphicx}
\usepackage{hyperref}
\usepackage{cleveref}
\Crefformat{figure}{#2Fig.~#1#3}
\Crefmultiformat{figure}{Figs.~#2#1#3}{ and~#2#1#3}{, #2#1#3}{ and~#2#1#3}
\usepackage[FIGTOPCAP, normalsize]{subfigure}
\usepackage{tikz}
\usepackage{color}
\newcommand{\beq}{\begin{equation}\begin{aligned}}
\newcommand{\eeq}{\end{aligned}\end{equation}}
\newcommand{\V}{\abs{V}}
\newcommand{\E}{\abs{E}}

\begin{document}

\preprint{}

\title{Robustness and Stability of Spin Glass Ground States to Perturbed Interactions}

\author{Vaibhav Mohanty}\thanks{Email: mohanty@hms.harvard.edu}
\affiliation{Rudolf Peierls Centre for Theoretical Physics, University of Oxford, Oxford, OX1 3NP, UK}
\affiliation{MD-PhD Program and Program in Health Sciences and Technology, Harvard Medical School, Boston, MA 02125, USA\\and Massachusetts Institute of Technology, Cambridge, MA 02139, USA}
\author{Ard A. Louis}\thanks{Email: ard.louis@physics.ox.ac.uk}
\affiliation{Rudolf Peierls Centre for Theoretical Physics, University of Oxford, Oxford, OX1 3NP, UK}

\date{\today}

\begin{abstract}
Across many scientific and engineering disciplines, it is important to consider how much the output of a given system changes due to perturbations of the input. Here, we investigate the glassy phase of $\pm J$ spin glasses at zero temperature  by calculating the robustness of the ground states to flips in the sign of single interactions. For random graphs and the Sherrington-Kirkpatrick model, we find relatively large sets of bond configurations that generate the same ground state. These sets can themselves be analyzed as subgraphs of the interaction domain, and we compute many of their topological properties. In particular, we find that the robustness, equivalent to the average degree, of these subgraphs is much higher than one would expect from a random model. Most notably, it scales in the same logarithmic way with the size of the subgraph as has been found in genotype-phenotype maps for RNA secondary structure folding, protein quaternary structure, gene regulatory networks, as well as for models for genetic programming. The similarity between these disparate systems suggests that this scaling may have a more universal origin.    
\end{abstract}

\maketitle

\section{Introduction}

Systems in which the input can be represented as a sequence of characters appear across science and engineering almost universally and especially commonly in biology and computer science. In these fields, notions of system \textit{robustness} or \textit{sensitivity} can be defined to quantify the outputs' average resistance to small changes in the input sequences.  

In biology, the mapping from genotypes (which store the information) to phenotypes (which describe biological properties) can be abstracted as genotype-phenotype maps (GP maps). Examples include 4 letter RNA sequences and 20 letter protein sequences that can be mapped to their physical folded states, and gene-regulatory networks, which can, for example, be described by Boolean networks~\cite{kauffman_homeostasis_1969} where a set of weights represent the gene interaction strengths. The responses of such systems to changes in the input sequences have been extensively studied computationally and analytically~\cite{wagner_distributed_2005, wagner_robustness_2007, wagner_robustness_2008, aguirre_topological_2011, payne_constraint_2013, payne_robustness_2014, schaper_arrival_2014, greenbury_tractable_2014, greenbury_organization_2015, dingle_structure_2015, greenbury_genetic_2016, ahnert_structural_2017, weis_phenotypes_2018, nichol_model_2019, camargo_boolean_2020, hu_network_2020,manrubia_genotypes_2021,payne_causes_2019,schaper_epistasis_2012,dingle_phenotype_2022,wright_evolving_2022}.

For GP maps, an important concept is the set of genotypes (sequences) that map to a particular phenotype, often called the \textit{neutral set}. These present a number of commonalities across GP maps~\cite{greenbury_genetic_2016,ahnert_structural_2017,manrubia_genotypes_2021}: the neutral sets are typically highly connected so that they can be viewed as networks which can be traversed by single mutational steps, leading to enhanced evolvability~\cite{wagner_robustness_2008,payne_causes_2019}. Neutral sets are also called \textit{neutral networks}. In some cases, the neutral network can split into smaller component networks which are disconnected due to biophysical constraints~\cite{aguirre_topological_2011,schaper_epistasis_2012}. The neutral set size (NSS) can vary over many orders of magnitude, and is typically strongly biased, with a small fraction of the phenotypes taking up the majority of genotypes. Such phenotype bias can strongly affect evolutionary outcomes~\cite{schaper_arrival_2014,dingle_structure_2015,dingle_phenotype_2022, johnston_symmetry_2022}. 

A key property of the neutral set  for GP maps  is the  \textit{mutational robustness}, typically defined in the literature as the fraction $\rho_p$ of single-character mutations in a genotype that produce the same phenotype $p$, averaged over the neutral set of all genotypes that produce phenotype $p$. That is to say, for a biological input-output map $f(g)$ which takes in a genotype $g$ of $d$ characters chosen from an alphabet $K = \{K_0, \dots, K_{k-1}\}$, the robustness of phenotype $p$ is defined as
\beq
\rho_p(f)  = \frac{1}{\abs{\mathcal{G}_p} d(k-1)} \sum_{g\in \mathcal{G}_{p}} n_{p,g}
\label{eq:rob_def}
\eeq
where $\mathcal{G}_{p}$ is the  neutral set of all genotypes $g$ whose output is the phenotype $p$, and $n_{p,g}$ is the number of nearest-neighbors of $g$ mapping to $p$ defined as  the number of genotypes $g'$ satisfying $f(g') = f(g) = p$ that differ from $g$ by a Hamming distance of 1.   Thus, $\rho_p(f) \in [0,1]$ measures the mean probability that a mutation from $g \in \mathcal{G}_{p}$ to a neighboring genotype $g' \in \mathcal{G}_{p}$ results in the same phenotype $p$.

What all the biological models mentioned above hold in common is that they each have high levels of mean robustness to changes in their inputs, excepting pathological or adversarial examples. To quantify this statement, consider the naive uncorrelated expectation that $\rho_p \approx \abs{\mathcal{G}_p}/k^d$ because, in a large randomly-assigned GP map, the probability that a nearest-neighboring genotype yields phenotype $p$ is approximately equal to the probability that any genotype drawn at random from the entire input space yields phenotype $p$. This probability is $\abs{\mathcal{G}_p}/k^d$, so $n_{p,g} \approx d(k-1) \abs{\mathcal{G}_p}/k^d$, and $\rho_p \approx \abs{\mathcal{G}_p}/k^d$. And indeed this is true for completely uncorrelated and randomly-assigned GP maps~\cite{greenbury_genetic_2016}. However,  biology and computer science-inspired GP maps such as the  RNA secondary structure maps, protein folding maps, gene regulatory networks, linear genetic programs, and digital logic gate maps exhibit a remarkably similar observed scaling law for robustness $\rho_p \sim \mathcal{O}(\log \abs{\mathcal{G}_p})$ rather than $\rho_p \sim \mathcal{O}( \abs{\mathcal{G}_p})$~\cite{manrubia_genotypes_2021}. Such enhanced mutational robustness is necessary for the neutral sets to percolate~\cite{greenbury_genetic_2016} which, in turn,  is critical for the evolutionary process because it allows for neutral exploration of the neutral set, greatly enhancing the ability of a population to find novel phenotypic variation~\cite{wagner_robustness_2008}.  This enhanced robustness also facilitates the exploration of fitness landscapes~\cite{greenbury_structure_2022}. 

Robustness has a direct counterpart in computer science, the \textit{sensitivity}, which measures the likelihood that flipping a single input bit will alter the output bit of Boolean functions $f : \{0,1\}^d \rightarrow \{0,1\}$ that map binary sequences of length $d$ onto a single binary output.  In other words, low sensitivity corresponds to high robustness. Huang's~\cite{huang_induced_2019} famously short solution to the decades-old Sensitivity Conjecture~\cite{nisan_degree_1994} concerning induced subgraphs of the $n$-dimensional hypercube graph has recently brought a great deal of attention to the sensitivity analysis of Boolean functions.  While scaling laws for sensitivity are typically not measured in the manner done for biological robustness, there are many mathematical equivalencies between the two. In the literature (see e.g.,~\cite{nisan_degree_1994, huang_induced_2019,bernasconi_sensitivity_1996, chakraborty_sensitivity_2011}), quantities are defined such as local sensitivity, function sensitivity, and average sensitivity, which are similar to biological robustness.  In this paper, we do not explicitly use these definitions from sensitivity analysis, instead opting for the GP map inspired ones. But, we point out that there is a seemingly understudied connection between robustness in biological systems and sensitivity in Boolean functions.

Given the wide range of systems for which high robustness is observed, we ask here whether a similar phenomenon can be found in spin glasses, which have a rich history in statistical and condensed matter physics. They have been intensely studied since the 1970s~\cite{sherrington_solvable_1975, edwards_theory_1975},  and  have led to many important insights in physics and other related disciplines, including computer science~\cite{mezard_spin_1987,mezard_information_2009,nishimori_statistical_2001}. More recently, the spin glass Hamiltonian has been used as a phenomenological model for epistatic genotype-to-fitness landscapes in which different sites (\textit{e.g.} DNA, genes expressions, or amino acids) may couple to each other~\cite{guo_stochastic_2019, louie_fitness_2018, butler_identification_2016, barton_relative_2016, shekhar_spin_2013, hopf_mutation_2017,cocco_inverse_2018}. An important application has been to viruses~\cite{louie_fitness_2018, butler_identification_2016, barton_relative_2016, shekhar_spin_2013, hopf_mutation_2017}. By taking sequence data over time, inverse statistical physics methods can be employed to ``learn'' the interactions (couplings/bond configurations) between different sites. In the context of evolution, therefore, one can interpret the ground state of the spin glass energy landscape as the global fitness peak on an evolutionary fitness landscape. The interactions between spins in such a system can depend on a number of biological or environmental factors~\cite{shekhar_spin_2013, guo_stochastic_2019}. The interesting question in the context of fitness landscapes is, what is the robustness of the sequences of spins to mutations in the bond configurations (the epistatic couplings)?

 \begin{figure*}[t!]
\centering
\includegraphics[width=\textwidth]{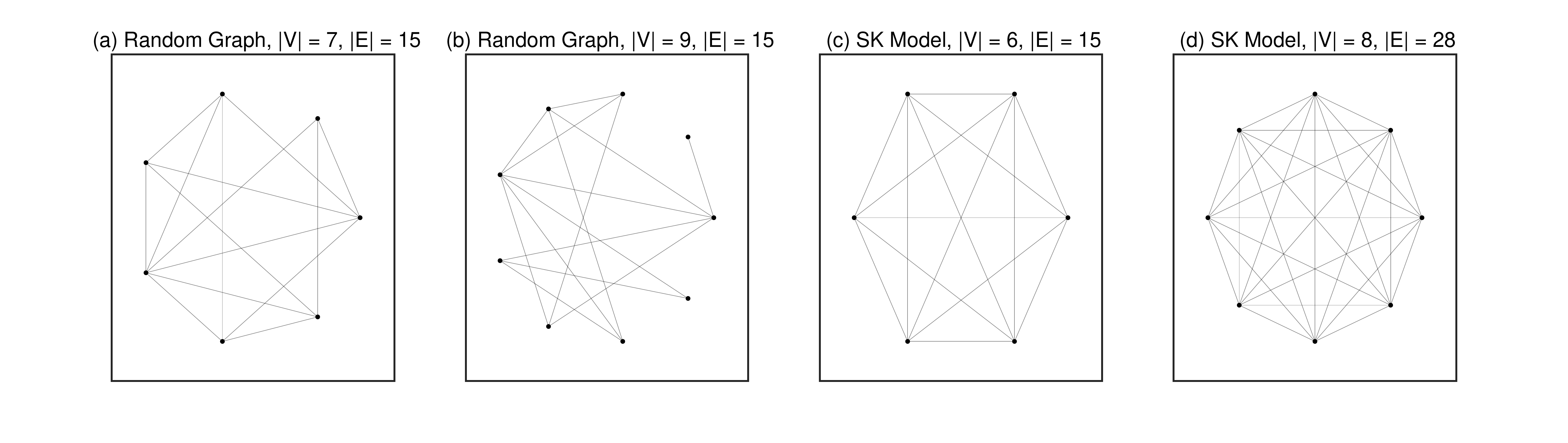}

\caption{$G(V,E)$ for (a and b) two representative examples of random graphs and (c and d) two representative examples of complete graphs (Sherrington-Kirkpatrick model). Each Ising spin $s_i$ is placed on a vertex, and each interaction $J_{ij}$ is placed on an edge. Each spin also experiences an external random field $h_i$.}
\label{graph_topologies}
\end{figure*}

In this paper, we consider this question by investigating the spin glass phase at $T = 0$ by calculating the robustness of $\pm J$ spin glass ground state configurations to a sign flip perturbation of a single bond. Our investigation is related to the concept of spin glass bond chaos or disorder chaos~\cite{bray_chaotic_1987}, which considers how the ground state of a spin glass changes when \textit{all} the couplings are perturbed by a small amount; notable investigations (see refs.~\cite{bray_chaotic_1987,aspelmeier_free-energy_2008,ney-nifle_chaos_1997, sasaki_temperature_2005, wang_bond_2016}) have been conducted on continuous $J$ spin glasses, whereas we consider the $\pm J$ spin glasses. Another similar concept is the recently-introduced  $\sigma$-criticality~\cite{newman_ground-state_2022}, which considers the effects on the ground state of only a single-bond perturbation, as we do here, but again this is a continuous $J$ spin glass. Our use of $\pm J$ spin glasses is an important distinction between our work and these previous studies because the discreteness of the bond configuration domain  of the $\pm J$ spin glass, modeled as a hypercube graph, allows us to investigate the topological features of the subgraphs formed by bond configurations mapping to the same ground state. The bond chaos and $\sigma$-criticality investigations are mainly interested in how the spin glass ground states change, and we consider these aspects here as well, but in this work we are mainly interested in the universality of the subgraph network features in the bond configuration domain.

In addition to spin glasses on random graphs, we examine special cases, namely the Sherrington-Kirkpatrick model and 1D Edwards-Anderson model. Various network topological properties are computed for interaction domain subgraphs comprised of bond configurations that all map to the same ground state spin configuration. We find that these subgraphs obey the same logarithmic scaling law between robustness and neutral set size as the analogous biological and computer science GP maps described above, suggesting that this high robustness may hold for a much wider set of physical systems. 

\section{Model and Methods}
\subsection{Spin Glass Model}
Consider an undirected, unweighted random graph $G(V,E)$ with vertex set $V$ and edge set $E$. We place Ising spins on each vertex, and each edge represents a nonzero interaction between spins. A spin configuration $s \in \{\pm 1\}^{\V}$ can be written as a sequence of $+1$ and $-1$ values, so it is essentially a binary sequence of length $\V$. A set of interactions (bond configuration) $J \in \{\pm 1\}^{\E}$ similarly is a sequence of $+1$ and $-1$ values of length $\E$. The spin glass Hamiltonian
\beq
\mathcal{H}_G(s;J) = -\sum_{\{i,j\}\in E} J_{ij} s_i s_j - \sum_{i\in V} h_i s_i \label{eq:Hamiltonian}
\eeq
contains couplings between all spins which are connected by an edge in $G$. The single-spin, external magnetic field interactions $h_i$ are independently and identically distributed uniformly on the interval $[-10^{-4}, +10^{-4}]$, noting that the external field $h_i \ll J_{ij} \in \{-1,+1\}$; we incorporate an external field in order to break the possible degeneracies of the spin glass ground state. In our numerical simulations described below, we find that this choice of distribution for $h_i$ is sufficient to ensure a unique ground state for all of our simulations.

The input-output map considered in our study is the spin glass ground state optimization function
\beq
\Gamma_G : \{\pm1\}^{\E} \rightarrow \{\pm1\}^{\V}
\eeq 
defined for the graph $G$. For a bond configuration $J$,  $\Gamma_G(J)$ outputs the ground state configuration $s$ that minimizes the Hamiltonian \cref{eq:Hamiltonian}.   The most common task in spin glass theory is to find $s$ given a particular bond configuration $J$. In this paper, we study an inverse problem, namely  the relationship between the set of all bond configurations $\{J\}$ that generate a particular output $s$. 

To efficiently represent this system, we note that the collection of all binary sequences of length $n$ can be represented by a $n$-dimensional undirected hypercube graph $Q_n(U,F)$. This is accomplished by mapping each binary sequence to a vertex in $Q_n(U,F)$ and placing edges between two vertices if the corresponding sequences have a Hamming distance of 1 between them. The hypercube graph has vertex set $U$ with $\abs{U} = 2^n$ and edge set $F$ with $\abs{F} = 2^{n-1} n$.

The domain of $\Gamma_G$ accordingly has a mapping to the $\E$-dimensional hypercube graph $Q_{\E}(U,F)$. In general, for graphs $G$ that can produce geometrical frustration in the spin glass, $\Gamma_G(J)$ follows no pattern and is difficult to calculate~\cite{barahona_computational_1982}, even more so because of the degeneracy-breaking external random field interactions $\{h_i\}$. But, because of frustration, two sets of bond configurations $J^{(i)}$ and $J^{(j)}$ corresponding to adjacent vertices in $Q_n(U,F)$ often have $\Gamma_G(J^{(i)}) = \Gamma_G(J^{(j)})$. The vertices corresponding to all $J$ such that $\Gamma_G(J) = s$ for some fixed $s$ induce a subgraph $H_s(U_s, F_s)$ of $Q_{\E}$. It follows that $\bigcup_s U_s = U$. In the GP map literature, a neutral network (or neutral set) for phenotype (here, the ground state) $s$ is the graph $H_s(U_s,F_s)$.

In this paper, we numerically compute topological properties of each neutral network of spin glass bond configurations which all map to the same ground state spin configuration. We consider multiple random graphs for $G$ as well as the fully-connected graph; the latter case is known as the Sherrington-Kirkpatrick (SK) model of a spin glass~\cite{sherrington_solvable_1975}. In our simulations, we also impose the condition that every node in $G$ has at least one neighbor. Additionally, in this paper we also consider the 1-dimensional Edwards-Anderson (EA) model~\cite{edwards_theory_1975}, for which the relationship between robustness $\abs{F_s}$ and vertex count $\abs{U_s}$ (equivalent to robustness) becomes analytically solvable. Our simulated spin glasses are relatively small because (1) we need to calculate the ground states exactly for each bond configuration, (2) every possible bond configuration is considered for each spin glass in order to accurately determine neutral network properties. In other words, the number of times \cref{eq:Hamiltonian} is computed in order to find all ground states for a random graph $G(V,E)$ scales as $O\left(2^{\abs{V}}\times 2^{\abs{E}}\right)$, which forces us to use relatively small systems. Our largest model ($\abs{V} = 8$, $\abs{E} = 28$) involves calculation of an exact ground state for every single bond configuration (over 268 million bond configurations); this is a similar size to systems for which exact ground state calculations like ours were carried out in the context of the spin glass literature (see \textit{e.g.}, ref.~\cite{boettcher_exact_2005}).

\subsection{Definitions of Topological Quantities}

The following parameters are computed for the neutral networks in bond configuration space:

\paragraph*{Robustness.} The neighbor count $n_{p,g}$ in \cref{eq:rob_def} is equivalent to the degree of a vertex $g \in U_s$ within a neutral network $H_s(U_s,F_s)$. The mean degree is related to the number of edges by
\beq
\sum_{v\in U_s} \deg(v) = 2\abs{F_s}.
\eeq
Here, we compute the robustness $\rho_s$, simply dividing the above quantity by the size of the subgraph $\abs{U_s}$ and by the length of the input sequence $\E$:
\beq
\rho_s = \phi_{ss} \equiv \frac{2\abs{F_s}}{\abs{U_s}\E} \in [0,1].
\label{eq:robustness}
\eeq
The notation $\phi_{ss}$ will become clear below when we also treat the transition probability $\phi_{rs}$  of a bond perturbation leading to a different ground state $r$. It is important to notice that the robustness is nothing other than the edge-to-vertex ratio  $\abs{F_s}/\abs{U_s}$ normalized to the range $[0,1]$ and also is equivalently the normalized mean degree within the neutral network $H_s$. In many ``real-world'' GP maps including RNA and protein folding, Boolean threshold networks, and genetic algorithms, it has been observed that $\rho_s \sim \log \abs{U_s}$ or equivalently $\abs{F_s} \sim \abs{U_s}\log \abs{U_s}$.   We will test this scaling for spin glass systems. 

\paragraph*{Transition probability.} We define the transition probability $\phi_{rs}$ to be the probability that, given a bond configuration which maps to ground state $s$, a single-bond perturbation changes the ground state from $s$ to $r$. Graph theoretically, we can think of this in the following way: consider two neutral networks of $Q_{\E}(U,F)$ called $H_s(U_s,F_s)$ and $H_r(U_r,F_r)$ so that $s$ and $r$ are two different spin configurations ($s\neq r$). Let $T_{rs} = T_{sr}$ be the set of edges connecting $H_s$ and $H_r$. The \textit{transition probability} of $s\rightarrow r$ due to a single-bond perturbation is
\beq
\phi_{rs} \equiv \frac{\abs{T_{rs}}}{\abs{U_s}\E} \in [0,1], \quad s\neq r.
\eeq
In many real-world GP maps, it has been observed that often $\phi_{rs} \sim \abs{U_r}$, or equivalently $\abs{T_{rs}} \sim \abs{U_r}\abs{U_s}$~\cite{schaper_arrival_2014,greenbury_genetic_2016}. We will also test this scaling for spin glass systems. 

We now see that, for the case where $s = r$, the transition probability $T_{ss}$, based on the definition above, simply counts the number of edges \textit{within} the induced subgraph $H_s(U_s,F_s)$, and therefore $T_{ss} = F_s$. In the definition of $\phi_{ss}$ in \cref{eq:robustness}, an additional factor of 2 comes from the double counting of edges; recall, the robustness is equivalent to the (normalized) mean degree, which means that the number of edges must be double-counted. From the definitions, it is easy to check that the normalization condition $\sum_{r} \phi_{rs} = 1$ holds, where the sum includes $r = s$.

\paragraph*{Rank-size plot.}  A rank-size plot is useful for visualizing how many orders of magnitude the neutral set sizes span. It may be used to deduce if there is a power law (\textit{i.e.}, generalized Zipf's Law) relationship between rank and neutral set size, as is seen for some GP maps.

\paragraph*{Degree distribution, clustering coefficients, assortativity, and betweenness centrality.} We also compute the degree distribution, clustering coefficients, assortativity, and betweenness centrality for the neutral networks studied here. Degree distributions give us information about the modality and skew of the degrees of the vertices in the neutral networks. Clustering coefficients, which turn out to be trivially zero for these systems, are informative about the number of neighboring nodes which share a neighbor. Assortativity measures correlation between the degrees of neighboring vertices. Betweenness centrality of a node measures the number of shortest paths in the network passing through that node; it is often compared with, and is expected to be positively correlated with, the degree of the node~\cite{oldham_consistency_2019}. Because our systems are small, generalizable trends regarding these quantities are more difficult to resolve, but calculating them assists our physical intuition, so we present them in the Supplementary Material along with their mathematical definitions.

\begin{figure*}[t!]

\includegraphics[width=\textwidth]{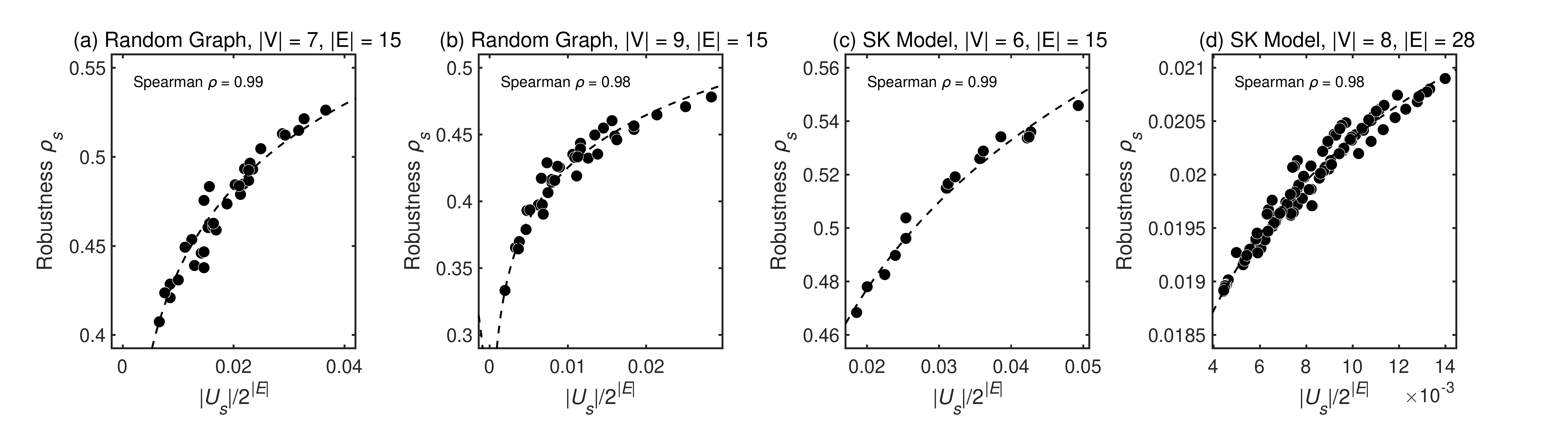}
\caption{Linear-linear plot of robustness $\rho_s$ versus normalized subgraph size $\abs{U_s}/2^{\E}$ (equivalent to NSS) for representative random graphs (a and b) and representative SK models (c and d). The dashed curve arises from the linear least squares fit calculated for \Cref{edgecount}, with the abscissa transformed to linear scale. A logarithmic relationship between robustness and NSS is established.
}
\label{linlin-edgecount}
\end{figure*}

\begin{figure*}[t!]

\includegraphics[width=\textwidth]{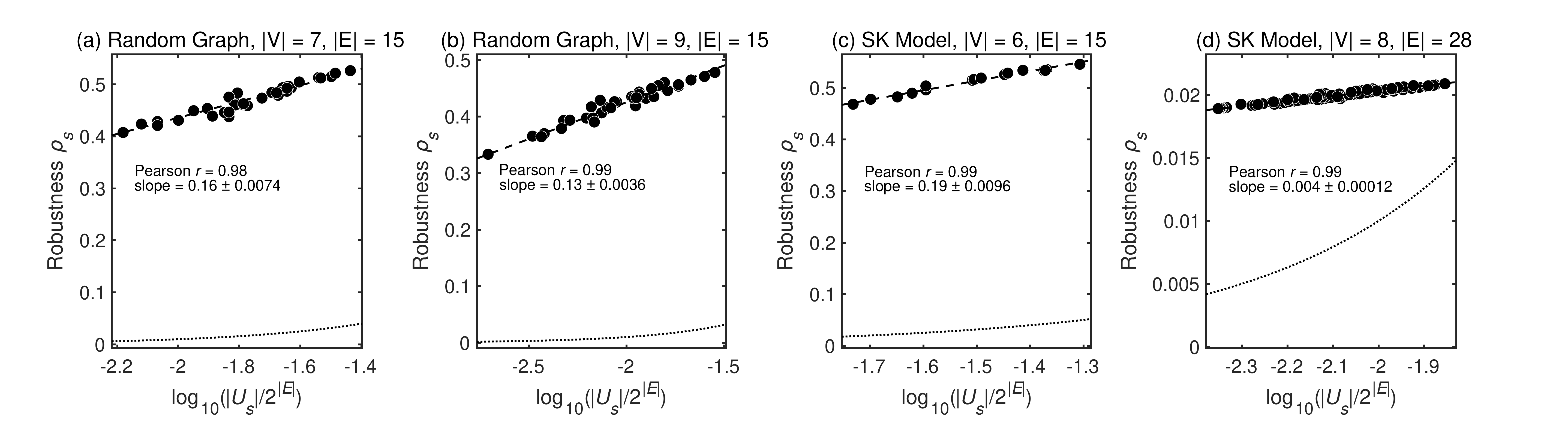}
\caption{Linear-log plot of robustness $\rho_s$ versus normalized subgraph size $\abs{U_s}/2^{\E}$ (equivalent to NSS) for representative random graphs (a and b) and representative SK models (c and d). The dashed line is the line of best of fit in the linear-log scale. A linear relationship between robustness and the log of NSS is established. Regression means and 95\% confidence intervals are given for the slope of the best fit lines. The dotted line shows the random null expectation for robustness. 
}
\label{edgecount}
\end{figure*}

\begin{figure*}[t]

\includegraphics[width=\textwidth]{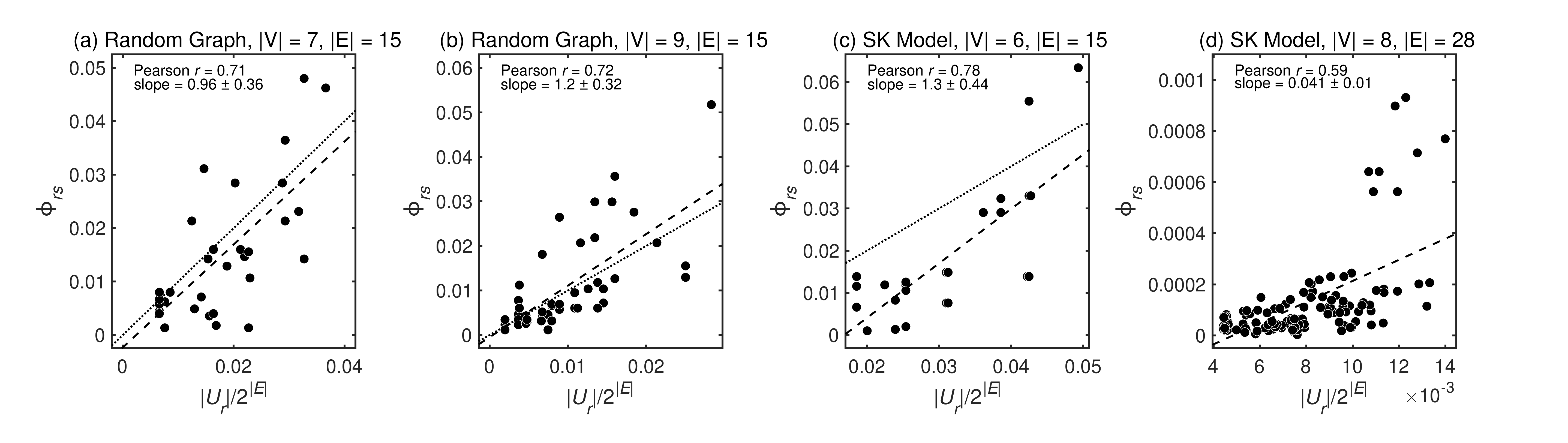}
\caption{Transition probability  $\phi_{rs}$ (for $r \neq s$ versus normalized subgraph size $\abs{U_s}/2^{\E}$ (equivalent to NSS) for representative random graphs (a and b) and representative SK models (c and d). The dashed line is the line of best of fit in the linear-log scale; the dotted line is the identity line (in which abscissa = ordinate). In panel (d), for the SK model, the line of best fit deviates most from the identity line (it is outside the viewing window), but the data are consistent with a linear fit, as for the other models. Regression means and 95\% confidence intervals are given for the slope of the best fit lines.}
\label{transition_probs}
\end{figure*}

\section{Results}
We now present results for the properties described in the previous section for the neutral networks $H_s(U_s,F_s)$ for spin glasses defined on random graphs as well as for the SK model. Two instances of random graphs and two SK models of differing sizes are used as the representative simulation examples for this main text; these graphs $G$ are shown in \Cref{graph_topologies}. To demonstrate the consistency of our results across many instances of the random graphs, several additional random graph instances and their network topological properties are found in the plots in the Supplementary Material. We also present the 1D EA model as a special case where we can calculate the exact relationship between robustness and neutral network size.

\subsection{Robustness}

For all topologies for $G$, we find that each induced subgraph  $H_s(U_s,F_s)$ has exactly one connected component, regardless of size.  Whether this will also hold for larger systems is unclear, but this lack of multiple component networks is different from, for example, the RNA GP map, where biophysical constraints (mainly that $\text{GC} \rightarrow \text{UA}$ and $\text{CG} \rightarrow \text{AU}$ are not possible by single point mutations) lead to fragmentation of the neutral networks~\cite{aguirre_topological_2011,schaper_epistasis_2012}. In \Cref{linlin-edgecount}, we plot the robustness of neutral networks versus the neutral set size. Spin glasses on random and complete graphs all show behavior consistent with the $\rho_s \sim \log \abs{U_s}$ relationship, which is also seen for the closely related scaling of robustness with neutral set size found for many GP maps. For these small models, the scale of the variation is too small to fully confirm the expected log-scaling, but the robustness is significantly enhanced when compared to an uncorrelated random null model (see \Cref{edgecount}).  This null model is calculated by  taking each spin glass ground state mapping $\Gamma_G(J)$ and randomizing the input-output pairings while keeping the subgraph size the same. A single-bond perturbation $J \mapsto J'$ will result in a spin configuration $s = \Gamma_G(J')$ being selected with probability $\abs{U_s}/2^{\E}$, regardless of $\Gamma_G(J)$. Thus, $\phi_{rs} \approx \abs{U_r}/2^{\E}$, even for $r = s$, in the null model.

\begin{figure*}[t]

\includegraphics[width=\textwidth]{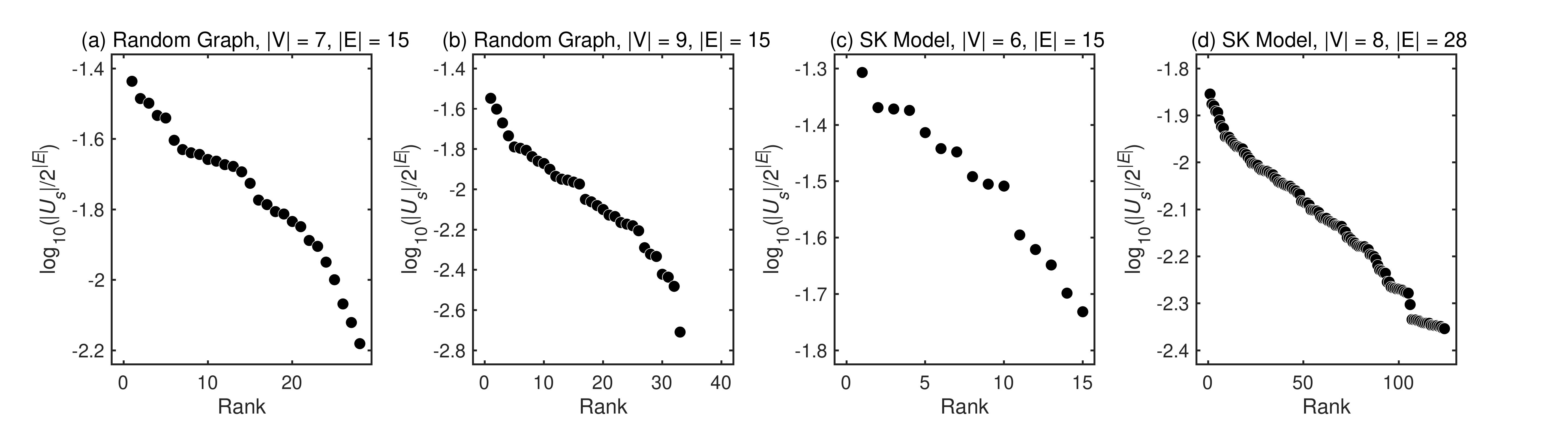}

\caption{Plot of normalized induced subgraph size $\log_{10}(\abs{U_s}/2^{\E})$ (equivalent to the NSS or frequency) versus the rank of the size for (a and b) random graphs and (c and d) SK models. All the models exhibit a similar rapid decay of the neutral set size.}
\label{logrank}
\end{figure*}

\subsection{Transition Probabilities}

Transition probabilities are plotted in \Cref{transition_probs} for the largest induced subgraph of each spin glass model. As found for biologically inspired GP maps~\cite{schaper_arrival_2014,greenbury_genetic_2016}, $\phi_{rs}$ is typically much smaller than the $\rho_s$; this is apparent from comparing the vertical axes in \Cref{transition_probs} and \Cref{edgecount}. As a null model, we again use
\beq
\phi_{rs} \approx \abs{U_r}/2^{\abs{E}}, \quad r \neq s.
\label{eq:transition_exp}
\eeq
Overall, for the random graphs, as can be seen in \Cref{transition_probs}, this null model curve matches the linear least squares fit, suggesting that vertices of subgraphs $H_r(U_r,F_r)$ (for $r \neq s$) with nonzero transition probability are approximately randomly distributed with frequency $\approx \abs{U_r}/2^{\E}$ in the neighborhoods of all vertices $v \in \abs{U_s}$. To interpret these findings about robustness and transition probabilities, consider a vertex $v \in H_s(U_s,F_s)$. In its set of nearest neighbors, this vertex is expected to see an overrepresentation (relative to the null model) of other vertices belonging to $H_s(U_s,F_s)$, and it sees a random assortment of other vertices $u \in H_r(U_r,F_r)$ for various $r \neq s$ with probability proportional to $\abs{U_r}/2^{\E}$. These findings are in accordance with other GP map studies.

For the SK models, there is less agreement between the null expectation and the observed lines of best fit. We speculate that the increased bond density in $G$ is responsible for this effect on transition probabilities, but our systems are currently too small to investigate this effect.

\subsection{Size-Rank Distributions} 

In the GP map literature, there has been a lot of interest in phenotype bias, the observation that the neutral set sizes can vary over many orders of magnitude, which can determine evolutionary outcomes~\cite{dingle_structure_2015,dingle_phenotype_2022} even when natural selection is also at play. In \Cref{logrank} we show that such phenotype bias also exists for spin glass systems. The rank plots show a consistent behavior independent of spin glass graph topology $G$.  Recent work on GP maps has suggested that there are two main classes of rank plots~\cite{greenbury_tractable_2014,manrubia_distribution_2017,weis_phenotypes_2018}. For the first class, the distribution of neutral set sizes obeys a Zipf-like power law, which arises from models in which input site ordering is strongly constrained (including Boolean neural networks~\cite{van_den_broeck_learning_1990,valle-perez_deep_2019}). The other class is  a log-normal distribution, which appears, for example, in RNA secondary structure GP maps~\cite{manrubia_distribution_2017,dingle_structure_2015}.  An open question is whether or not these spin glass systems also fall into one of these two classes. The current systems are still too small to conclusively answer this question, but the log-log rank versus size plots in Supplementary Material suggest a deviation from Zipf's law.

\subsection{Effects of Bond Perturbation on Ground State Spin Configuration}

Our single-bond perturbation numerical experiments are informative regarding the nature of the ground state in the $T = 0$ glassy phase. Our robustness calculations have already shown that a single-bond perturbation often leads to no change in the ground state, much more often than what the random null model predicts. Given that the robustness is equivalent to the normalized mean degree within a neutral network, we can understand robustness at the individual bond level by looking at the relationship between the degree of an individual vertex within the neutral network and various physical parameters such as ground state energy. We can ask, how is the degree within a neutral network related to the ground state energy? How many spin flips occur in the ground state energy when a single bond is perturbed? Is the unperturbed ground state a local minimum in the energy landscape generated by the new, perturbed bond configuration?

From the degree distributions within neutral networks with $\E = 15$ (shown in the Supplementary Material), we see that the range of degrees for the vertices in most neutral networks tends to span from 4 up to 15. In the Supplementary Material, we empirically confirm the expected result that the degree of a vertex is positively correlated with its betweenness centrality; that is, there tend to be a larger number of shortest paths passing through vertices with higher degree. Intuitively, one should then expect that vertices with higher degree have ground states which are more energetically stable to bond perturbation. In other words, the ground state energy is sufficiently low, and the minimum is deep such that a single-bond perturbation tends not to change the ground state. This trend is what we see in the numerical results presented in the Supplementary Material, where we also discuss the different scenarios when ground state change due to bond perturbation \textit{does} occur.  

\begin{figure*}[!t]
\includegraphics[width=\textwidth]{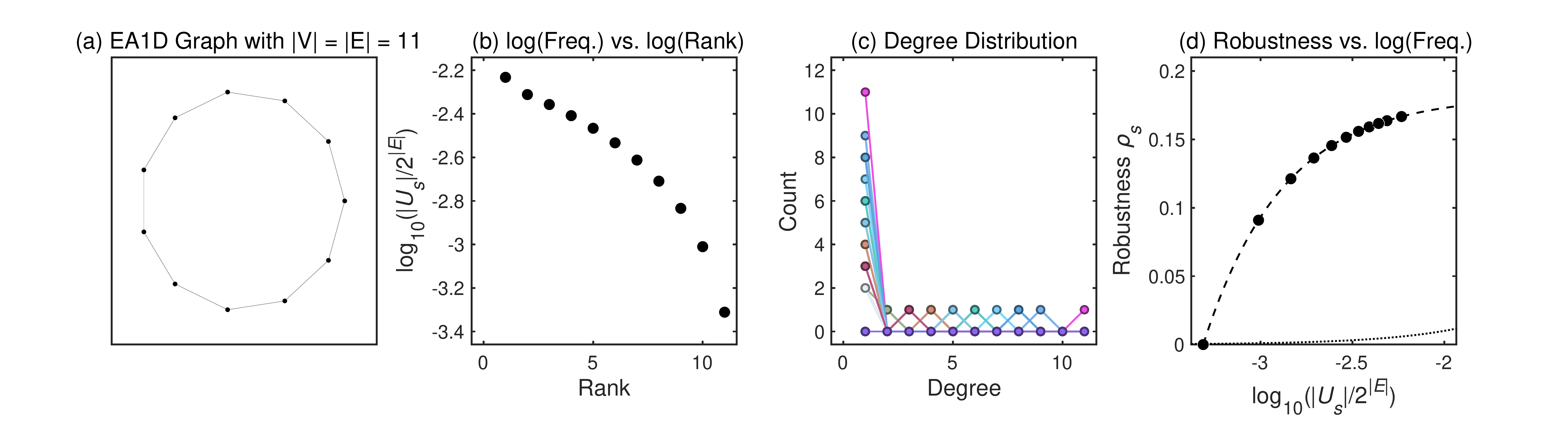}
\caption{Results for 1D Edwards-Anderson model: (a) graph representation of the 1D Edwards-Anderson model with $\abs{V} = 11$, (b) neutral set size versus rank plot on log-log scale, (c) degree distribution of neutral set vertices, and (d) robustness $\rho_s$ versus log of the normalized neutral set size $\abs{U_s}/2^{\abs{E}}$. The dashed line is the analytical result from \cref{eq:EA}.}
\label{eadata}
\end{figure*}

\subsection{Analytically Tractable Special Case: 1D Edwards-Anderson Model}
The Edwards-Anderson (EA) model is a special case which deserves individual treatment. The EA model, the original theory of spin glasses~\cite{edwards_theory_1975}, is simply a spin glass on a lattice with nearest neighbor interactions only.

For the 1D Edwards-Anderson (EA) model with periodic boundary conditions, the topology of which is shown in panel (a) of \Cref{eadata}, the behavior of $\rho_s$ is analytically tractable. Let us call a bond configuration with an even number of anti-ferromagnetic interactions ($J_{ij} = -1$) an \textit{even bond configuration} and a bond configuration with an odd number of anti-ferromagnetic interactions an \textit{odd bond configuration}.

Every even bond configuration has an exactly determinable ground state with no spin frustration, which can be found using the following algorithm:
\begin{enumerate}
\item[1.] Choose an arbitrary site $i$, and set the $i$-th spin to $s_i=+1$ or $s_i=-1$ arbitrarily.
\item[2.] Determine $s_{i+1}$ by setting $s_{i+1} = J_{i,i+1} s_i$, so $s_{i+1} = s_i$ if $J_{i,i+1} = +1$ or $s_{i+1} = -s_i$ if $J_{i,i+1} = -1$.
\item[3.] Continue to $s_{i+2}$ by similarly setting $s_{i+2} = J_{i+1,i+2} s_{i+1}$, and continue around the entire spin chain until we have set $s_{i-1}$. At this point one can check and confirm that $-J_{i-1,i}s_{i-1}s_i = -1$, so all bonds are satisfied, and each pairwise interaction \textit{decreases} the energy by the same amount ($= -1$).
\item[4.] If $\sum_{i=1}^{\V} h_i s_i < 0$, then all spins need to be flipped $s_i \rightarrow -s_i$. If $\sum_{i=1}^{\V} h_i s_i > 0$, then all the spins are kept as is. The ground state energy is exactly $E_{gs} = -\V - \sum_{i=1}^{\V} h_i s_i$, and we have now ensured $\sum_{i=1}^{\V} h_i s_i > 0$.
\end{enumerate}
Having an even number of anti-ferromagnetic interactions will ensure that the last spin $s_{i-1}$ does not experience frustration, and the bond between the last spin $s_{i-1}$ and the first spin $s_i$ contributes a negative amount to the energy: $-J_{i-1,i}s_{i-1}s_i = -1$. On the other hand, odd bond configurations will always have one bond at which there is frustration, and there will be a positive contribution to the energy $-J_{i-1,i}s_{i-1}s_i = +1$. The ground state for odd bond configurations will be identical to the ground state of one of the neighboring even bond configurations, but the energy will, of course, be higher. Exactly \textit{which} neighboring bond configuration that is depends on the random fields $h_i$.

Because an odd bond configuration has the same ground state as one of its neighboring even bond configurations, the neutral network $H_s(U_s,F_s)$ for every ground state must be a star graph $S_{\abs{U_s}-1}$ that has $\abs{U_s}$ nodes, with one central node corresponding to an even bond configuration, and $\abs{U_s} - 1$ peripheral nodes corresponding to odd bond configurations. The number of nodes of each star graph neutral network depends on the random fields $h_i$. Fig. \ref{eadata}(c) shows that the degree distribution matches our theory that all neutral networks are star graphs; for each neutral network, there is one node with $\abs{U_s} - 1$ neighbors, and $\abs{U_s} - 1$ nodes with one neighbor.

To calculate robustness, we note that the star graph $S_{\abs{U_s}-1}$ has $\abs{U_s}$ nodes and $\abs{F_s} = \abs{U_s} - 1$ edges. Plugging $\abs{F_s}$ into \cref{eq:robustness} immediately gives us that the robustness is
\beq
\rho_s = \frac{1}{\abs{U_s}\E} \sum_{v\in U_s} \deg(v) = \frac{2}{\E}\left(1 - \frac{1}{\abs{U_s}}\right). \label{eq:EA}
\eeq
The points in \Cref{eadata}(d) fall exactly along this curve. The maximum number of nodes in the star graph is $\abs{U_s} = \abs{E} + 1 = \V + 1$ because each even bond configuration has at most $\E = \V$ neighboring odd bond configurations. A size-rank plot is also shown for the 1D EA model in panel (b).

The 1D EA model serves as an example in which linear-log scaling is not seen for $\rho_s$. Nevertheless, the relationship between robustness and subgraph size found in the 1D EA model exhibits a $\rho_s$ that is significantly higher than would be expected from a random mapping $\Gamma_G$ of inputs to outputs, so this 1D model still exhibits enhanced robustness. For the 2D EA model we could not find an analytically tractable $\rho_s$.

\section{Discussion}

Our main result is that, by studying the mapping from spin glass bond configurations to the $T=0$ ground state, we observe several properties that are also observed for other input-output maps, such as the GP maps found in biology and computer science. These include (1) redundancy, in that many genotypes in the context of GP maps, or bond configurations in this case, map the the same output (a phenotype for the GP maps, or a ground state for the spin glasses), (2) phenotype frequency bias, in that the number of bond configurations (genotypes) mapping to the ground state (phenotype) vary significantly, and (3) enhanced mutational robustness, in that the probability of a single-bond perturbation changing the ground state is larger than expected from a random null model, and moreover that the robustness of a ground state to single bond perturbations often scales in the same way with the logarithm of the NSS (the number of bond configurations mapping to a ground state) as found for biological GP maps~\cite{greenbury_tractable_2014,greenbury_organization_2015,greenbury_genetic_2016,camargo_boolean_2020,hu_network_2020}. Another interesting result, also seen for GP maps~\cite{schaper_arrival_2014,greenbury_genetic_2016} is that, in contrast to the robustness, the transition probabilities, defined as the likelihood of a flip of the spin yielding a different ground state, do scale proportionally to the NSS, as one would expect from a random model.
The similarity to the GP map behavior suggests that there may be a more universal argument (based, for example, on algorithmic information theory~\cite{dingle_inputoutput_2018,dingle_generic_2020}) for these scaling properties.

Our spin glass models are relatively small because finding the ground state of a spin glass is typically computationally expensive and scales badly with system size. Depending on graph topology, finding the ground state of a spin glass can be NP hard~\cite{barahona_computational_1982}. Knowledge that the robustness of a ground state is large may potentially offer improvements to ground state-finding algorithms by providing a measure of stability of certain ground states as a function of parameter space. It would also be interesting to check some of our results on significantly larger graphs.  We find, for example, that all our subgraphs that map to the same ground state form only a single component; that is, they are connected by single-bond perturbations.  Will this percolation property hold for larger systems, or will these subgraphs start to fragment?   

Mapping epistatic interactions onto spin glass Hamiltonians to derive sequence-to-fitness maps has been especially important for the study of viral evolution~\cite{louie_fitness_2018, butler_identification_2016, barton_relative_2016, shekhar_spin_2013, hopf_mutation_2017}. These models typically have continuous $J$, so it will be interesting to see if the kind of scaling properties of robustness that we find here for the $\pm J$ spin glasses carry over to these more complex systems.  If, as we expect, a concept akin to high robustness persists, then this may have implications for the stability of fitness peaks to environmental changes, and may affect how easy it is to find continuously increasing paths to fitness maxima~\cite{greenbury_structure_2022}.

Another potentially interesting future direction of research is to explore these results about robustness in the complementary setting of the sensitivity of Boolean functions~\cite{nisan_degree_1994, huang_induced_2019,bernasconi_sensitivity_1996, chakraborty_sensitivity_2011}. It would be interesting to see whether similar high robustness/low sensitivity results can be found in this related context. 

\section{Acknowledgements}
The authors would like to thank Wilber Lim, Mehrana Nejad, Sloan Nietert, and Shuofeng Zhang for helpful discussions. VM has been supported by a Marshall Scholarship and award numbers T32GM007753 and T32GM144273 from the National Institute of General Medical Sciences. The content is solely the responsibility of the authors and does not necessarily represent the official views of the National Institute of General Medical Sciences, the National Institutes of Health, or the Marshall Aid Commemoration Commission.
\bibliography{bibs}

\end{document}


\preprint{}

\title{Supplemental Material for ``Robustness and Stability of Spin Glass Ground States to Perturbed Interactions''}

\author{Vaibhav Mohanty}\thanks{Email: mohanty@hms.harvard.edu.}
\affiliation{Rudolf Peierls Centre for Theoretical Physics, University of Oxford, Oxford, OX1 3NP, UK}
\affiliation{MD-PhD Program and Program in Health Sciences and Technology, Harvard Medical School, Boston, MA 02125, USA\\and Massachusetts Institute of Technology, Cambridge, MA 02139, USA}
\author{Ard A. Louis}\thanks{Email: ard.louis@physics.ox.ac.uk}
\affiliation{Rudolf Peierls Centre for Theoretical Physics, University of Oxford, Oxford, OX1 3NP, UK}

\date{\today}

\maketitle

\renewcommand{\theequation}{S\arabic{equation}}
\renewcommand{\thefigure}{S\arabic{figure}}

\section{Extended Figure for Transition Probabilities}

\begin{figure*}[b!]

\includegraphics[width=\textwidth]{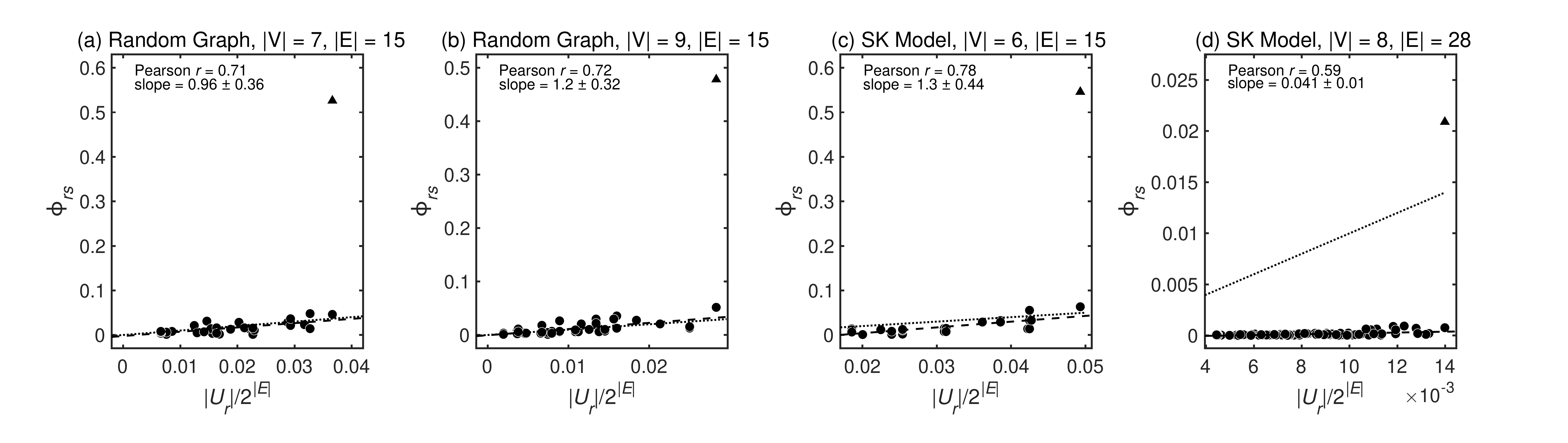}
\caption{Log-log plot of transition probability  $\phi_{rs}$ (for $r \neq s$ versus normalized subgraph size $\abs{U_s}/2^{\E}$ (equivalent to NSS) for representative random graphs (a and b) and representative SK models (c and d). The dotted line is the line of best of fit in the linear-log scale; the dashed line is the identity line (in which abscissa = ordinate). A linear relationship between robustness and the log of NSS is established. It is seen in the SK models that the line of best fit deviates more from the identity line, but linearity remains; in panel (d) the dashed line is outside the viewing window. Regression means and 95\% confidence intervals are given for the slope of the best fit lines.}
\label{supp_transition_probs}
\end{figure*}

\begin{figure*}[t]

\includegraphics[width=\textwidth]{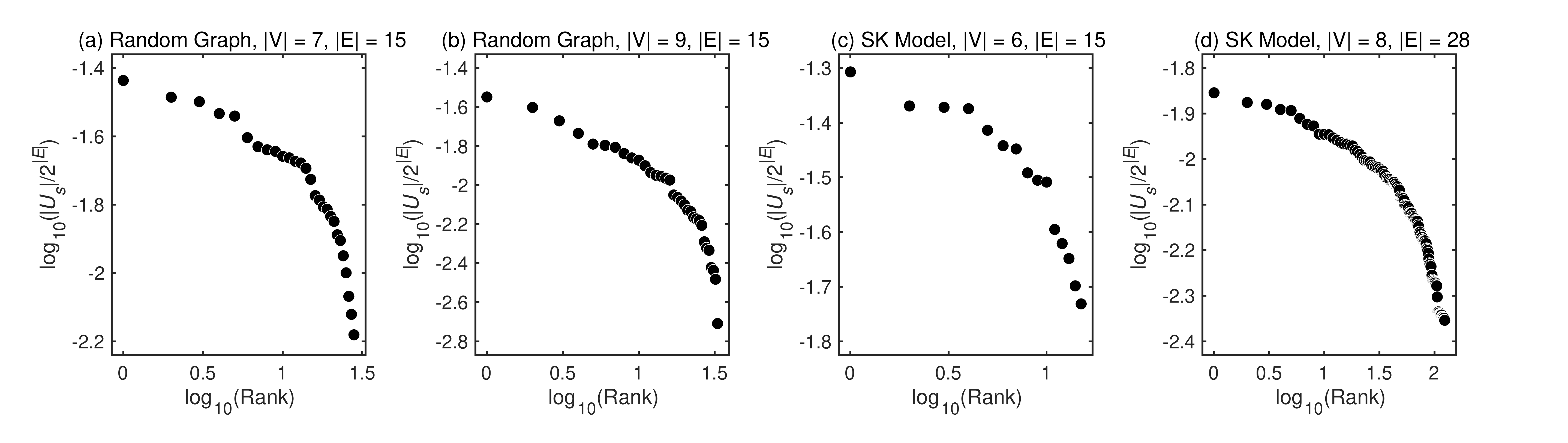}

\caption{Log-log of normalized induced subgraph size $\abs{U_s}/2^{\E}$ (equivalent to the NSS or frequency) versus the rank of the size for (a and b) random graphs and (c and d) SK models. Similar trends are shown between the plots. }
\label{loglogrank}
\end{figure*}

In the main text, we presented the transition probabilities $\phi_{rs}$ from the most frequently occurring ground state $s$ to the other ground states $r$ due to a single bond perturbation, with $r \neq s$. Here, we show in \Cref{supp_transition_probs} the same plot $\phi_{rs}$ with $\phi_{ss}$ indicated with a triangle. It is apparent that the ``robust'' transition---i.e. a lack of change in the ground state due to single bond perturbation---is much more likely than a transition to any other ground state.

\section{Log-log Size Versus Rank Plot}

A log-log plot of normalized neutral set size (i.e. frequency) versus the rank of that neutral set size is plotted in \Cref{loglogrank}. A deviation from a straight line suggests a deviation from Zipf's law.

\begin{figure*}[t]

\includegraphics[width=\textwidth]{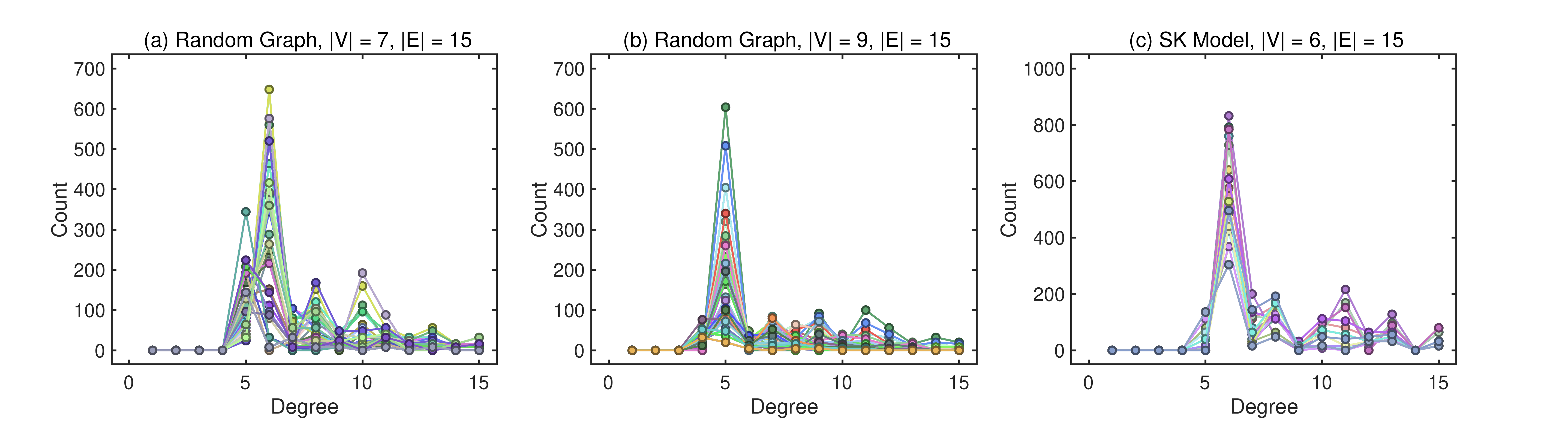}
\caption{Degree distributions of all induced subgraphs for (a and b) random graphs and (c) the $\abs{V} = 6$ SK model.}
\label{degdist}
\end{figure*}

\subsection{Subgraph Degree Distributions}

In \Cref{degdist}, we plot the degree distribution of each subgraph for the random graphs and the SK models. These degree distributions tend to have one large peak (and a few smaller peaks), with very few vertices attaining or coming close to attaining the maximum possible degree of $\E$. The mean of this degree distribution is of course proportional to the edge count found earlier.

\section{Clustering Coefficients} The local clustering coefficient defined at a vertex $w$ gives the ratio of all neighbors of $w$ connected to each other to  the ratio of all possible pairs of the neighbors of $w$, the latter of which is ${\deg(w) \choose 2}$. As such, $C(w)$ calculates the fraction of triangles involving $w$ out of all possible triangles involving $w$. Given an induced subgraph $H_s(U_s,F_s)$, the local clustering coefficient $C_s(w)$ for a vertex $w \in U_s$ is defined as
\beq
C_s(w) = \frac{2\abs{\{\{u,v\} \in F_s \,|\, u,v\in N_s(w) \wedge u \neq v\}}}{\deg(w)(\deg(w) - 1)},
\eeq
where $N_s(w) = \{v \in U_s \,|\, \{v,w\} \in F_s\}$ is the neighborhood of $w$, i.e. the set of all vertices connected to $w$.  An averaged clustering coefficient
\beq
\overline{C}_s = \frac{1}{\abs{U_s}}\sum_{w\in U_s} C_s(w)
\eeq
can also be defined for the entire induced subgraph.

In the systems we have considered in this paper, we note that the hypercube graph $Q_{\E}(U,F)$ has no triangles to begin with and thus has $C(w) = 0$ for all $w \in U$. It immediately follows that all local clustering coefficients are zero for all induced subgraphs. Our numerical simulations confirm the trivial result that clustering coefficients are always zero for induced subgraphs of hypercubes.

\begin{figure*}[t]
\includegraphics[width=\textwidth]{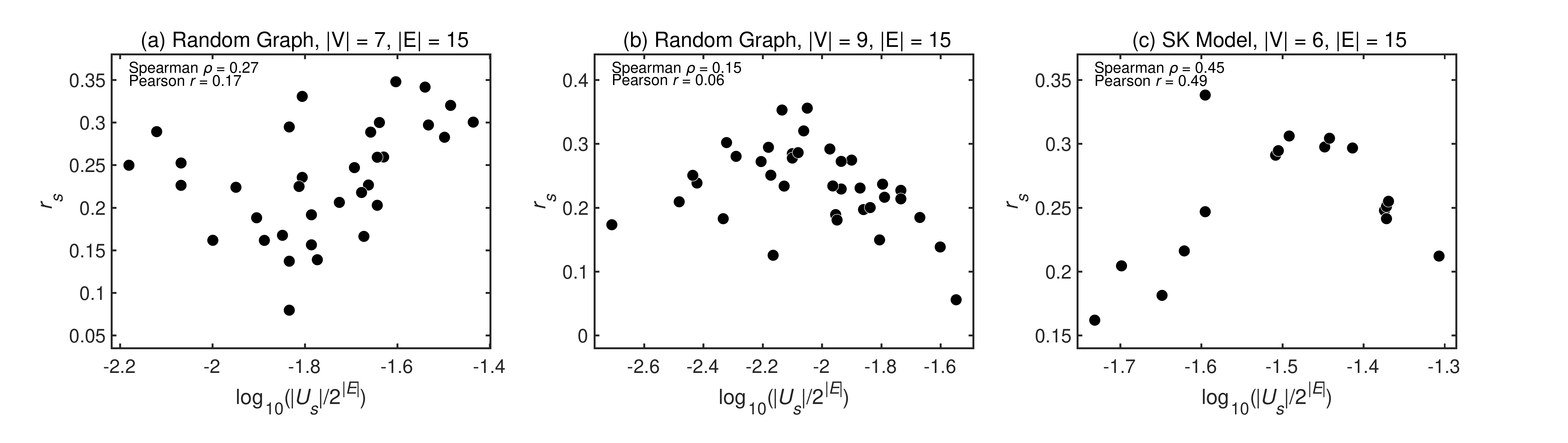}
\caption{Plot of assortativity $r_s$ of all induced subgraphs versus log of the normalized induced subgraph size $\abs{U_s}/2^{\E}$ (equivalent to NSS or frequency) for (a and b) random graphs and (c) the $\abs{V} = 6$ SK model. Little correlation is found, though perhaps correlation may be seen in larger systems than the ones we simulated here.}
\label{assortativity}
\end{figure*}

\begin{figure*}[t]

\includegraphics[width=\textwidth]{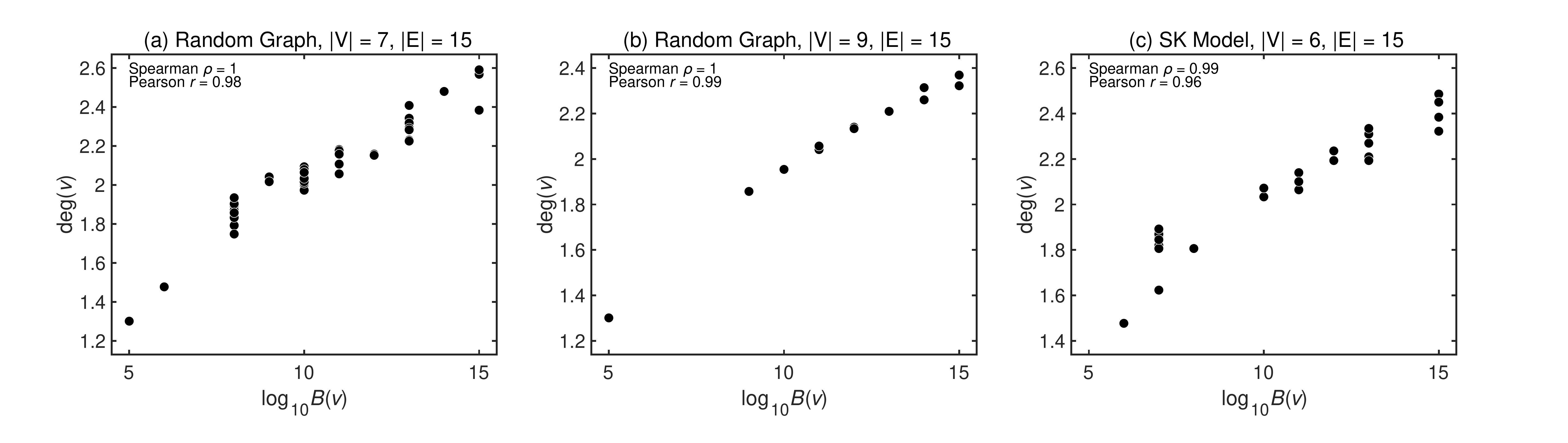}
\caption{Plot of vertex degree $v$ versus log of betweenness centrality $B(v)$ for all vertices $v$ in the largest induced subgraph (i.e. most frequently occurring ground state) for (a and b) random graphs and (c) the $\abs{V} = 6$ SK model.}
\label{betweenness}
\end{figure*}

\section{Assortativity}

A network's \textit{assortativity} is a measure of correlation between the degrees of two connected vertices. Typically, the Pearson correlation coefficient $r$ is used as a quantitative measure of assortativity. For a subgraph $H_s(U_s,F_s)$, this is calculated by finding \cite{aguirre_topological_2011}
\beq
r_s = \frac{\sum_{v\in U_s}\deg(v)^2\nndeg(v) - \frac{1}{2\abs{F_s}}\left(\sum_{v\in U_s}\deg(v)^2\right)^2}{\sum_{v\in U_s}\deg(v)^3 - \frac{1}{2\abs{F_s}}\left(\sum_{v\in U_s}\deg(v)^2\right)^2},
\label{eq:assortativity}
\eeq
where $\nndeg(v) = \frac{1}{\abs{N_s(v)}}\sum_{u\in N_s(v)} \deg(u)$ is the average degree of vertices in the neighborhood $N_s(v)$ of vertex $v \in U_s$. Networks with $r > 0$ are said to be assortative, and vertices with higher degree tend to be connected to vertices with higher degree. Accordingly, networks with $r < 0$ are said to be dissortative, and vertices with relatively high degree tend to be connected to vertices with relatively low degree.

In \Cref{assortativity} we show the assortativity, defined in \Cref{eq:assortativity}, of induced subgraphs plotted against the log of induced subgraph size.  The values are variable across the random graphs and negative for the SK model. The value of the assortativity itself is an indication of the correlation of the degree of a vertex and the degree of its neighbors. Positive (negative) $r_s$ of subgraph indicates that the degree of a vertex and the degree of its neighbors tend to be positively (negatively) correlated. In \cite{aguirre_topological_2011}, a weak positive correlation between assortativity and network size was found for RNA. It may be that our systems are too small to resolve such trends.

\section{Betweenness Centrality}

The \textit{betweenness centrality} $B_s(v)$ of a vertex $v\in U_s$ for the graph $H_s(U_s,F_s)$ is defined as
\beq
B_s(v) = \frac{1}{2} \sum_{u,w\in{U_s}} \frac{g(u,v,w)}{g(u,w)}, \quad u \neq v \neq w,
\eeq
where $g(u,w)$ is the number of shortest paths between $u$ and $w$ and $g(u,v,w)$ is the number of shortest paths between $u$ and $w$ that pass through $v$. $B_s(v)$ is often plotted against $\deg(v)$.

We also plot the betweenness centrality versus degree for the largest induced subgraph in \Cref{betweenness}. It is clear from the positive correlation found in all plots that vertices with higher degree tend to also be more central, i.e. there are a larger number of shortest paths traveling through that vertex. Such positive correlation has also been observed in induced subgraphs in RNA folding GP maps \cite{aguirre_topological_2011}.
\vfill
\pagebreak
\section{Effects of Bond Perturbation on Ground State Spin Configuration: Numerical Results}

\begin{figure*}[!t]
\includegraphics[width=\textwidth]{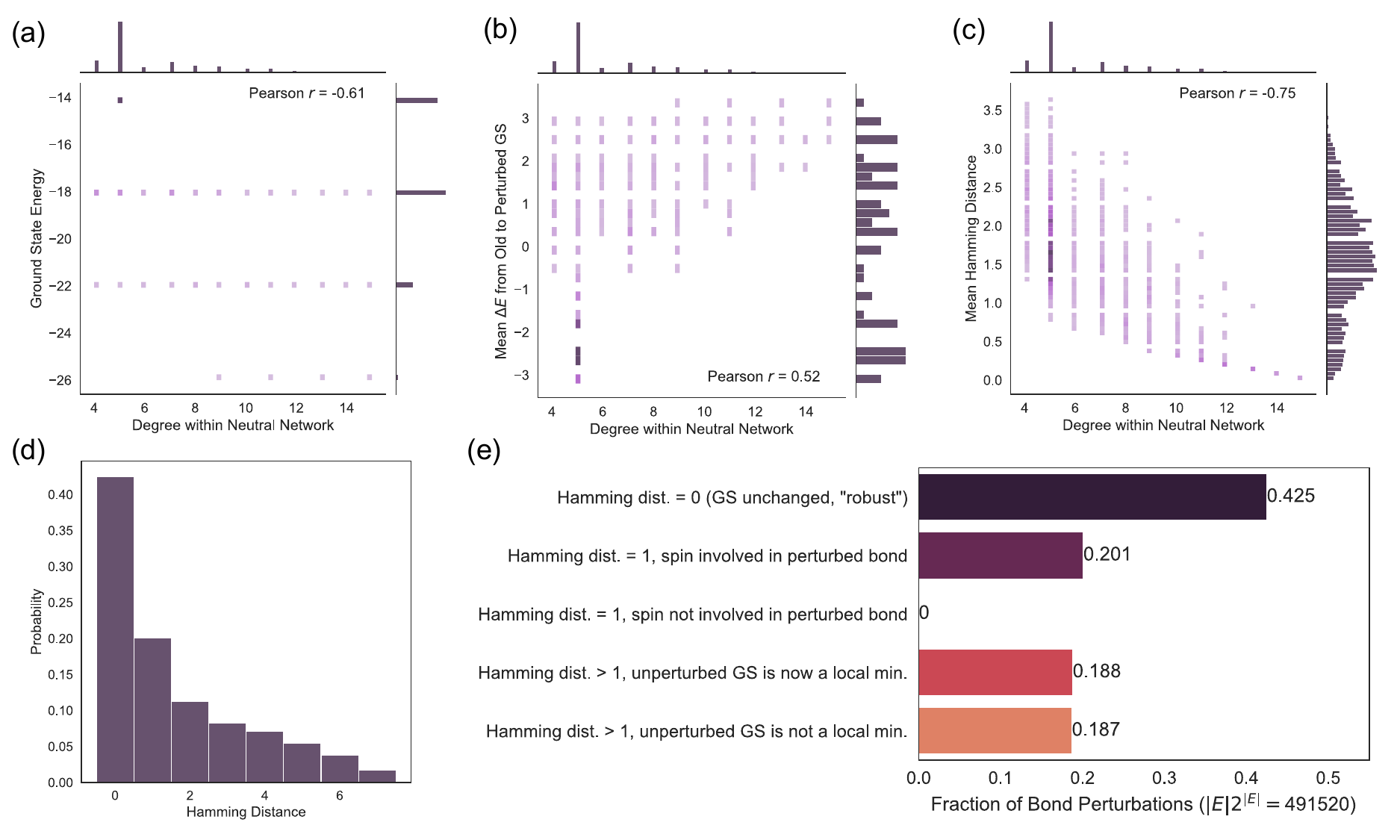}
\caption{Quantification of changes to ground states in the $T = 0$ glassy phase due to bond perturbations. All possible $\E 2^{\E} = 491,520$ bond configurations for the $\V = 9$, $\E = 15$ graph in main text Fig. 1(b) are exhaustively considered. (a) Ground state energy is negatively correlated with the degree of a vertex (representing a bond configuration) within its neutral network. This means bond configurations that have ground state spin configurations that are more resistant to single-bond perturbations tend to have lower ground state energies. (b) Change in ground state energy due to a single-bond perturbation, averaged over $\E$ possible perturbations, is positively correlated with degree within a neutral network. (c) Hamming distance between the ground state spin configuration for a bond configuration and the ground state spin configuration for the perturbed bond configuration is negatively correlated with degree within a neutral network. (d) The Hamming distance between the ground state spin configuration for a bond configuration and the ground state spin configuration for the perturbed bond configuration is monotonically decaying, with the plurality of mutations being robust (Hamming distance = 0). (e) When a single-bond perturbation changes the ground state spin configuration by a single spin, that spin is always involved in the perturbed bond. When a single-bond perturbation changes the ground state spin configuration by more than a single spin, the former ground state is often a local minimum in the energy landscape defined by the perturbed bond, but it does not have to be.}
\label{newfig9}
\end{figure*}

Our numerical results support this intuition. In \Cref{newfig9}(a), we show for the $\V = 9$, $\E = 15$ graph in main text Fig. 1(b) the that ground state energy is weakly negatively correlated with the degree of a vertex within a neutral network. For an individual bond configuration for this graph, there are $\E = 15$ possible single-bond perturbations. The higher ground state energies also correspond to shallower global minima, so many of these possible single-bond perturbations will destabilize the ground state and a new ground state emerges; the new ground state energy could be lower or higher than the original one. In this system, lower ground state energies tend to correspond to deeper global minima, and many of these possible single-bond perturbations will not change the ground state configuration; in these cases, the ground state energy will typically increase. As a result, we would expect to see that higher degree within a neutral network should be positively correlated with the change in the ground state energy due to bond perturbation; this is in fact what we see in \Cref{newfig9}(b).

In \Cref{newfig9}(c), we show how the Hamming distance between the new and old ground state spin configurations correlates negatively with the degree within the neutral network. When the degree is high, as we established, the ground state is more likely to be robust to bond perturbation; as a result, the Hamming distance is zero between the new and the old ground state. When the ground state is not robust to a bond perturbation, a few different possibilities emerge. One case is that the perturbation causes the ground state to become unstable locally such that changing only a single spin (involved in the perturbed bond) generates the new ground state. In \Cref{newfig9}(d) we see that this Hamming distance = 1 case is less common than a robust perturbation and more common than a ground state change with Hamming distance = $n$ for each $n > 1$. We see in \Cref{newfig9}(e) that there were zero instances in which a single spin change occurred in the ground state where that spin was \textit{not} connected to the perturbed edge.

\begin{figure*}[!h]
\includegraphics[width=\textwidth]{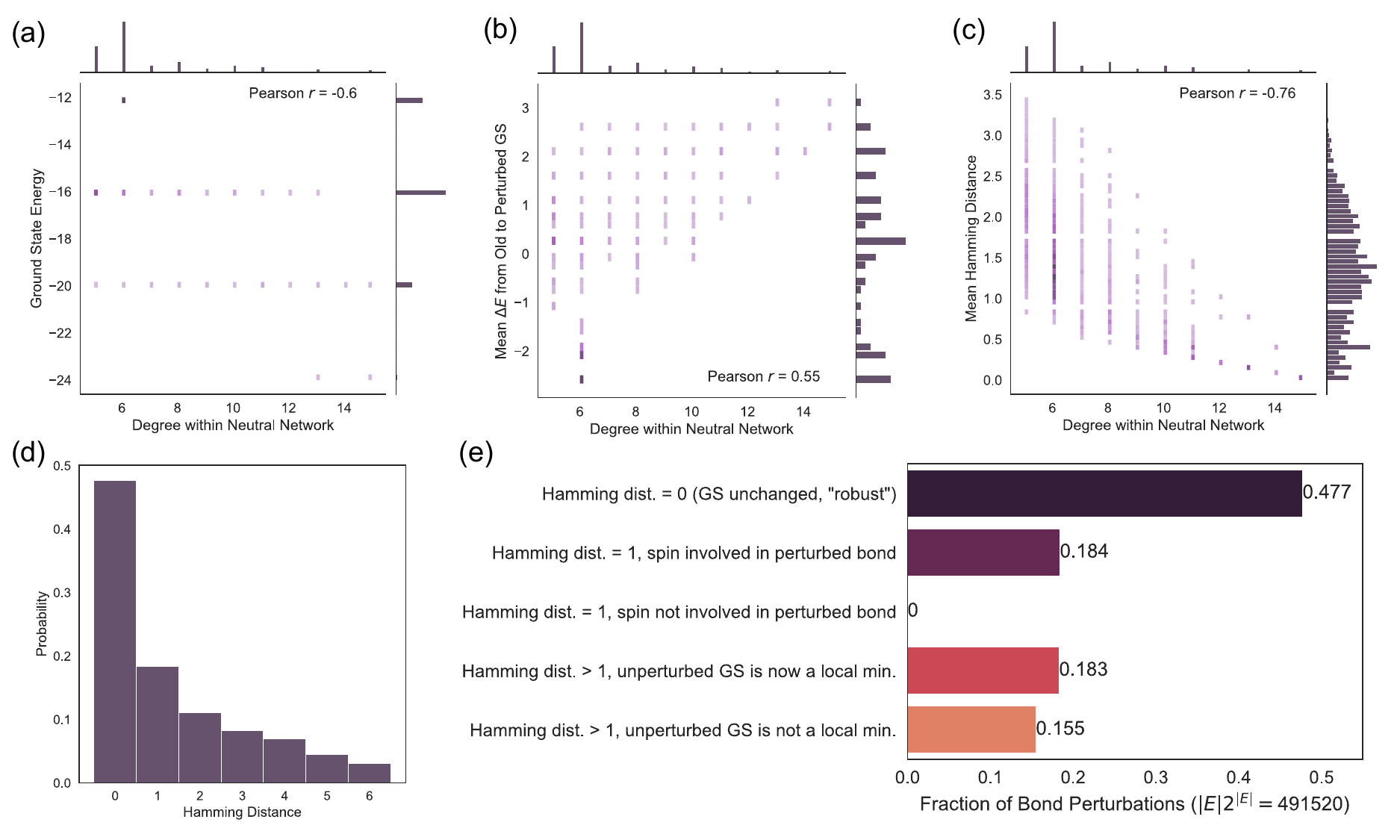}
\caption{Quantification of changes to ground states in the $T = 0$ glassy phase due to bond perturbations. All possible $\E 2^{\E} = 491,520$ bond configurations for the $\V = 7$, $\E = 15$ graph in main text Fig. 1(b) are exhaustively considered. (a) Ground state energy is negatively correlated with the degree of a vertex (representing a bond configuration) within its neutral network. This means bond configurations that have ground state spin configurations that are more resistant to single-bond perturbations tend to have lower ground state energies. (b) Change in ground state energy due to a single-bond perturbation, averaged over $\E$ possible perturbations, is positively correlated with degree within a neutral network. (c) Hamming distance between the ground state spin configuration for a bond configuration and the ground state spin configuration for the perturbed bond configuration is negatively correlated with degree within a neutral network. (d) The Hamming distance between the ground state spin configuration for a bond configuration and the ground state spin configuration for the perturbed bond configuration is monotonically decaying, with the plurality of mutations being robust (Hamming distance = 0). (e) When a single-bond perturbation changes the ground state spin configuration by a single spin, that spin is always involved in the perturbed bond. When a single-bond perturbation changes the ground state spin configuration by more than a single spin, the former ground state is often a local minimum in the new energy landscape defined by the perturbed bond configuration, but it does not have to be.}
\label{newfig7}
\end{figure*}

If the ground state spin configuration changes by more than 1 spin (Hamming distance $> 1$), then the original ground state has been destabilized, and flipping a spin adjoining the perturbed bond is not sufficient to obtain the new ground state. The new ground state can be found a greater Hamming distance away from the original ground state, though the probability that the new ground state requires $n$ spin flips decreases with $n$, as observed in \Cref{newfig9}(d). The former ground state spin configuration is often a local minimum in the new energy landscape defined by the perturbed bond configuration, though it does not necessarily have to be, as shown in \Cref{newfig9}(e). For our example case using the graph in main text Fig. 1(b), the probability of the old ground state being a local minimum in the perturbed energy landscape is nearly the same as the probability of it not being a local minimum; however, as our other simulations show, the relationship between these probabilities is non-trivial. In the Supplementary Material, we show the same calculations for the $\V = 7$, $\E = 15$ spin glass on the graph shown in main text Fig. 1(a); there, the probability of the old ground state being a local minimum in the new energy landscape is 0.183 while the probability of not being a local minimum in new energy landscape is 0.155; the former is 118\% of the latter. Overall, between the cases where a single spin flips and the cases where multiple spins flip in the ground state, there does not seem to be one scenario that drastically dominates over the other.

\begin{widetext}
\section{Network Topological Properties for Additional Random Graph Instances}
The following pages contain additional random graph $G$ topologies, frequency-rank plots of the ground state configurations, robustness versus ground state frequency plots in linear-log and linear-linear scale, corrected weighted transition probability versus frequency plot for the most frequently observed ground state (as the starting state), degree distributions, assortativity plots, and betweenness centrality versus degree for the most frequently observed ground state.

\vfill

\pagebreak
\includegraphics[height = \textheight]{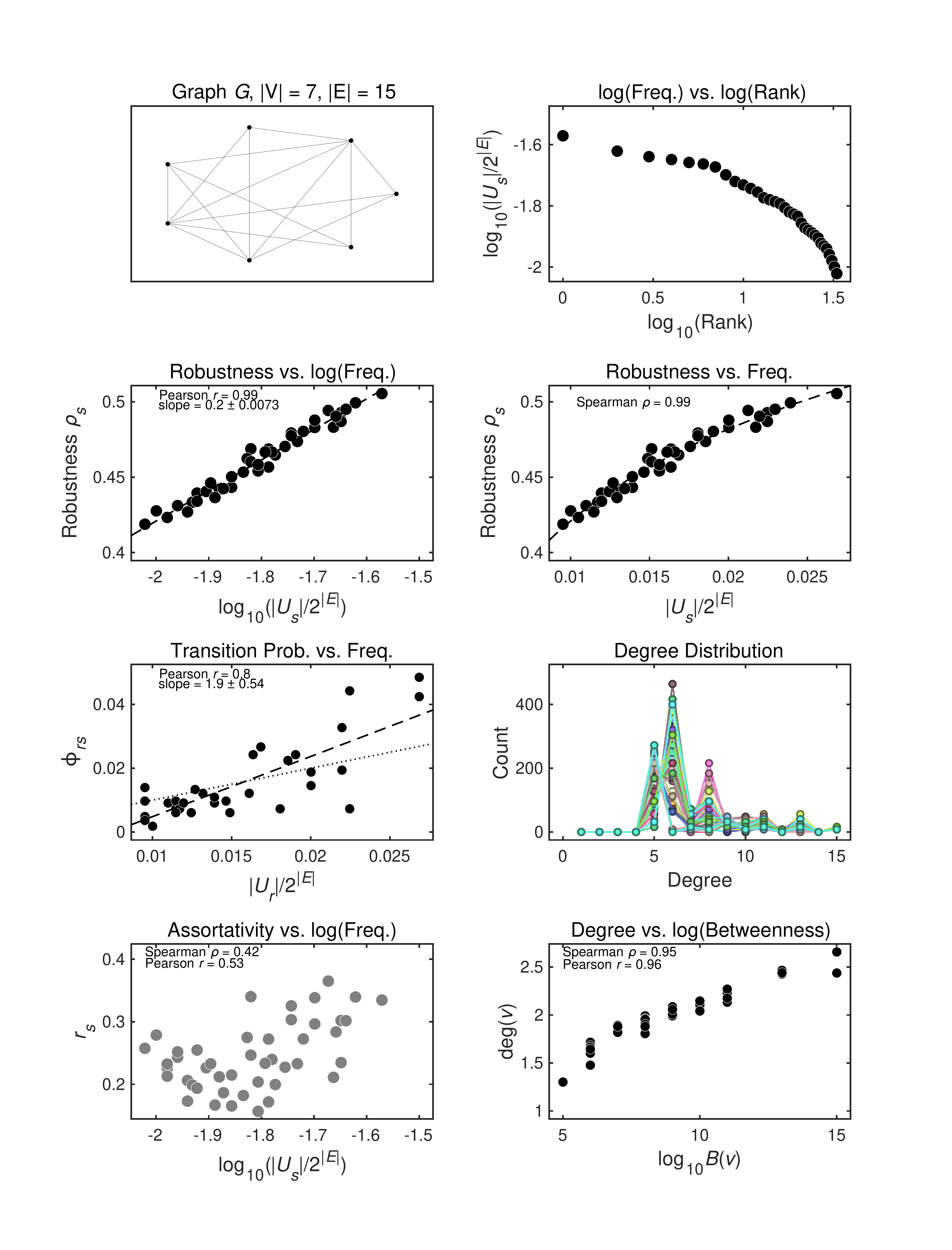}
\pagebreak
\includegraphics[height = \textheight]{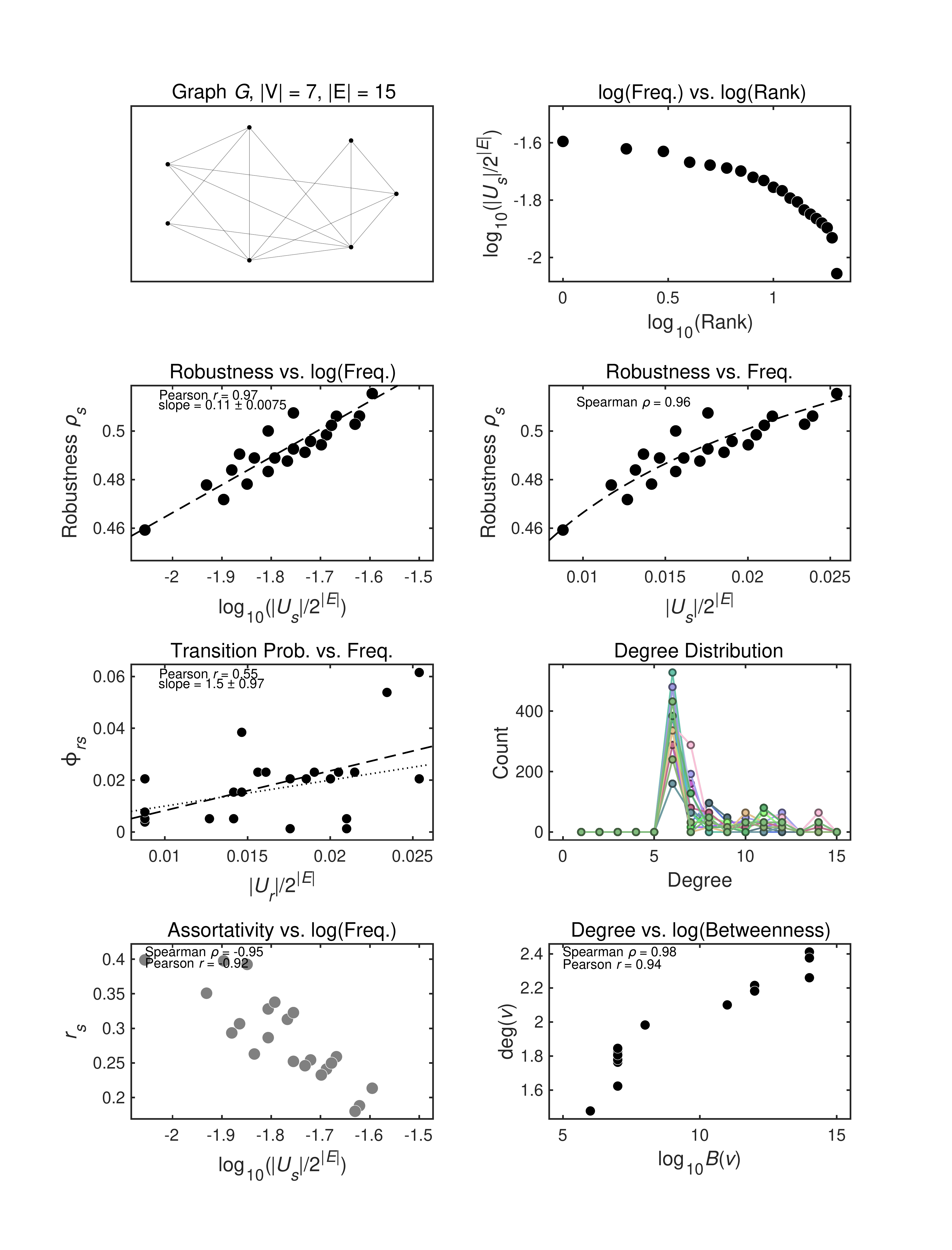}
\pagebreak
\includegraphics[height = \textheight]{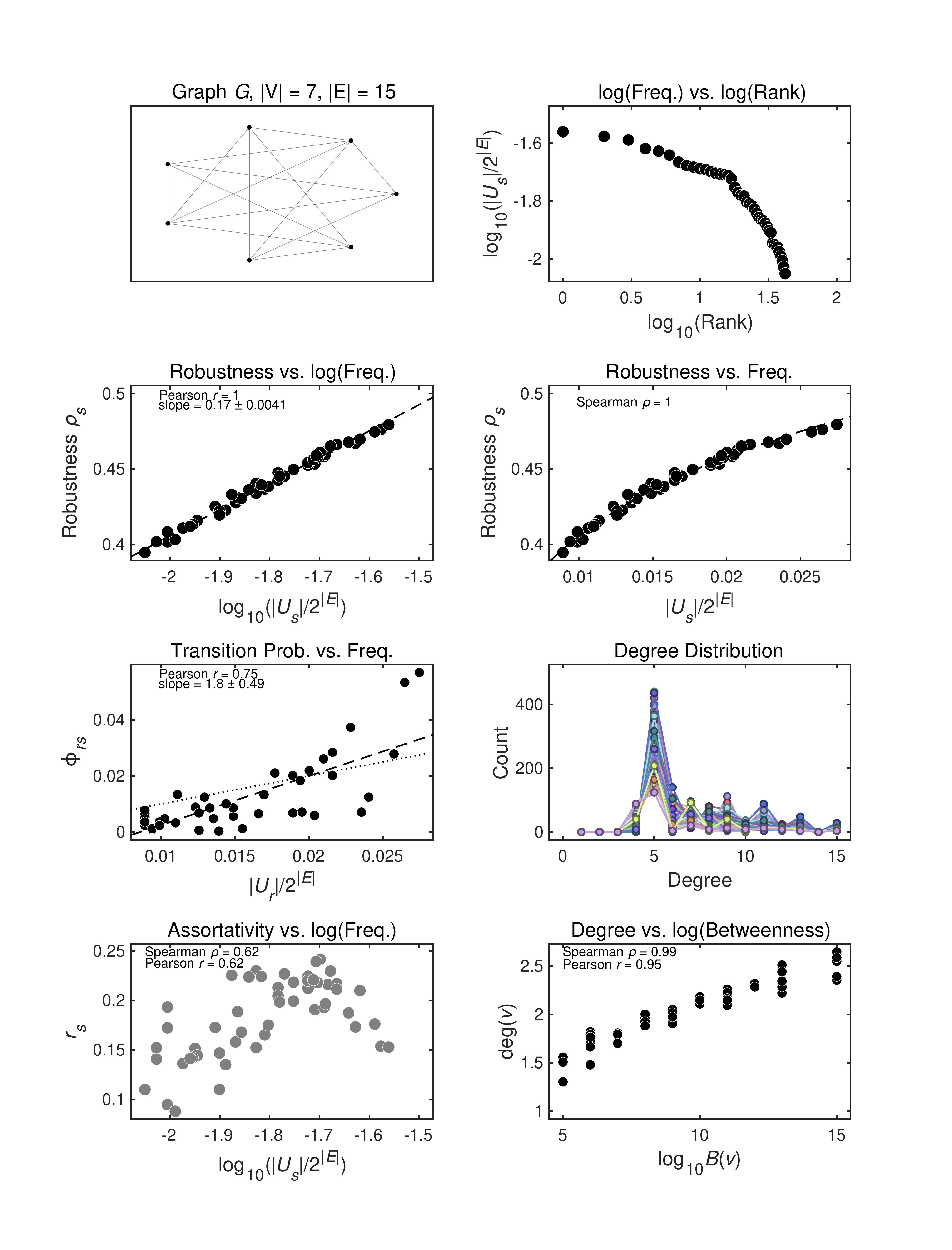}
\pagebreak
\includegraphics[height = \textheight]{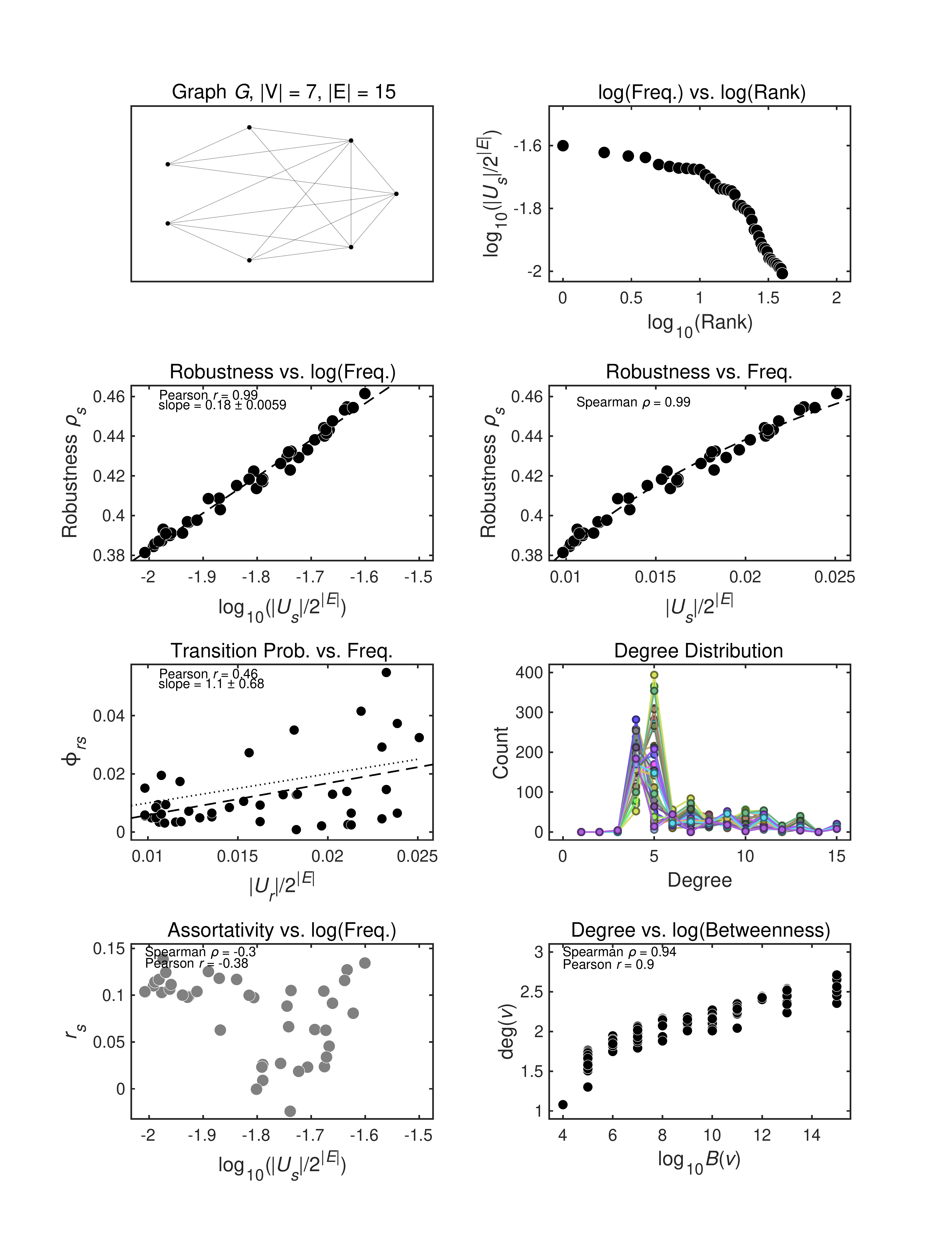}
\pagebreak
\includegraphics[height = \textheight]{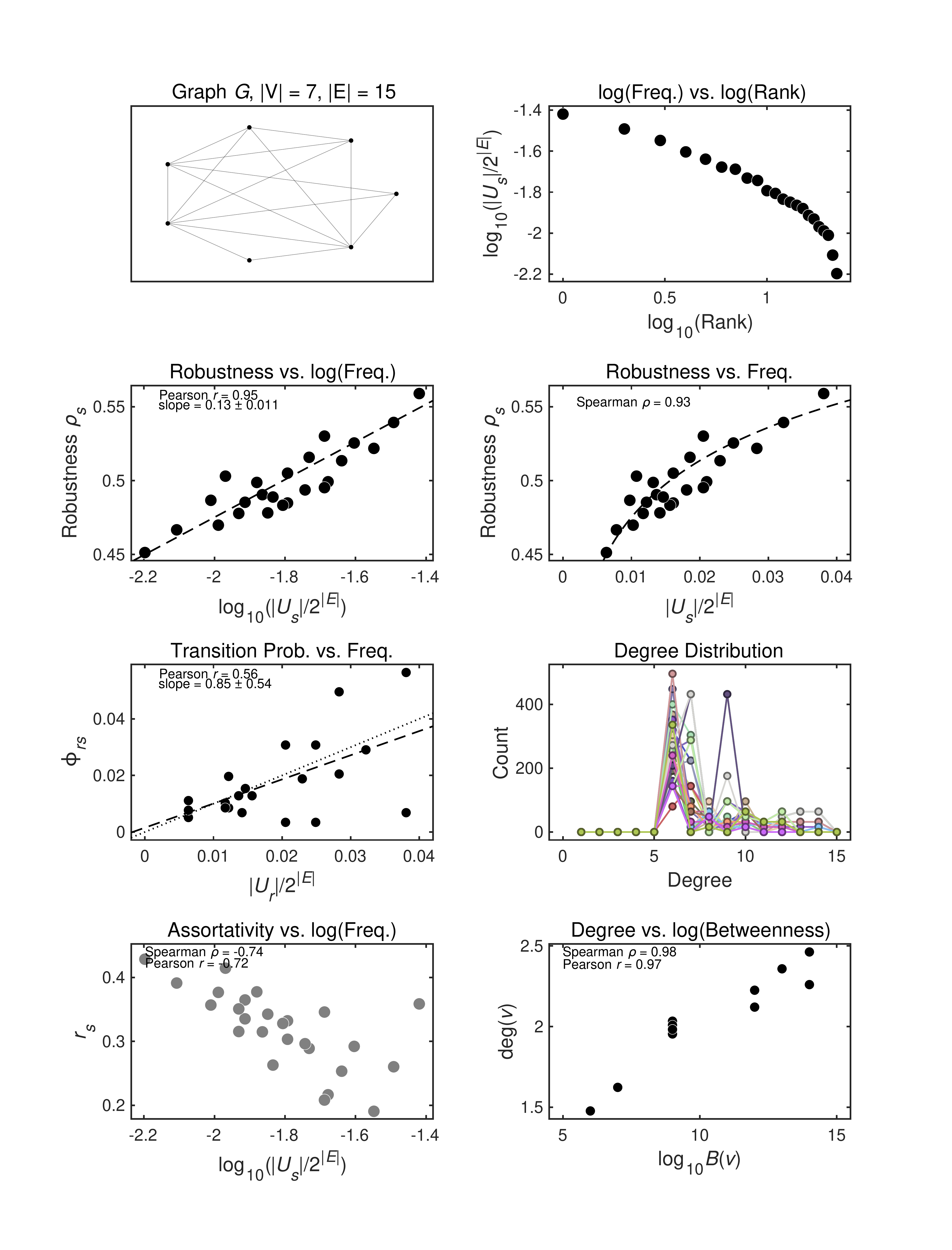}
\pagebreak
\includegraphics[height = \textheight]{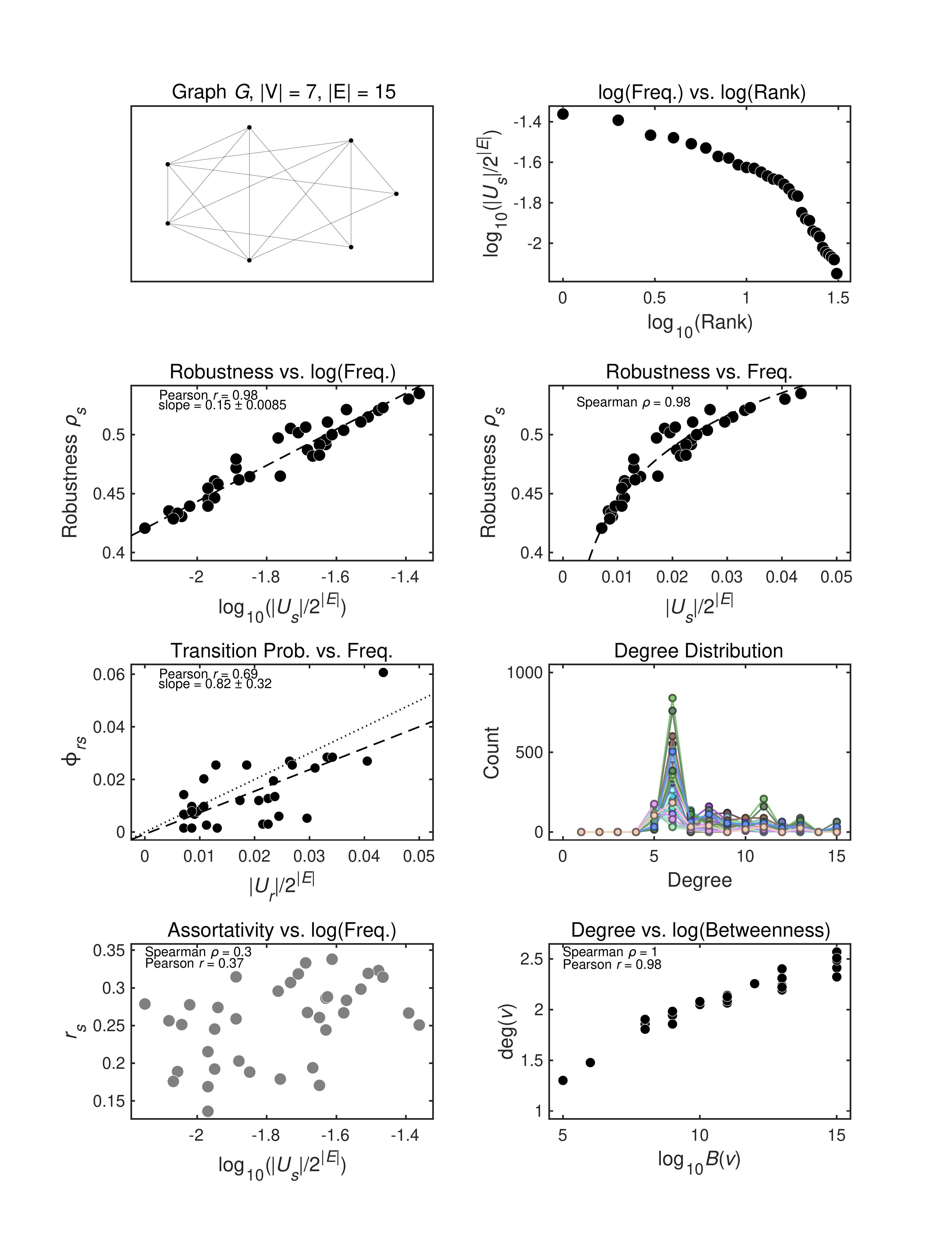}
\pagebreak
\includegraphics[height = \textheight]{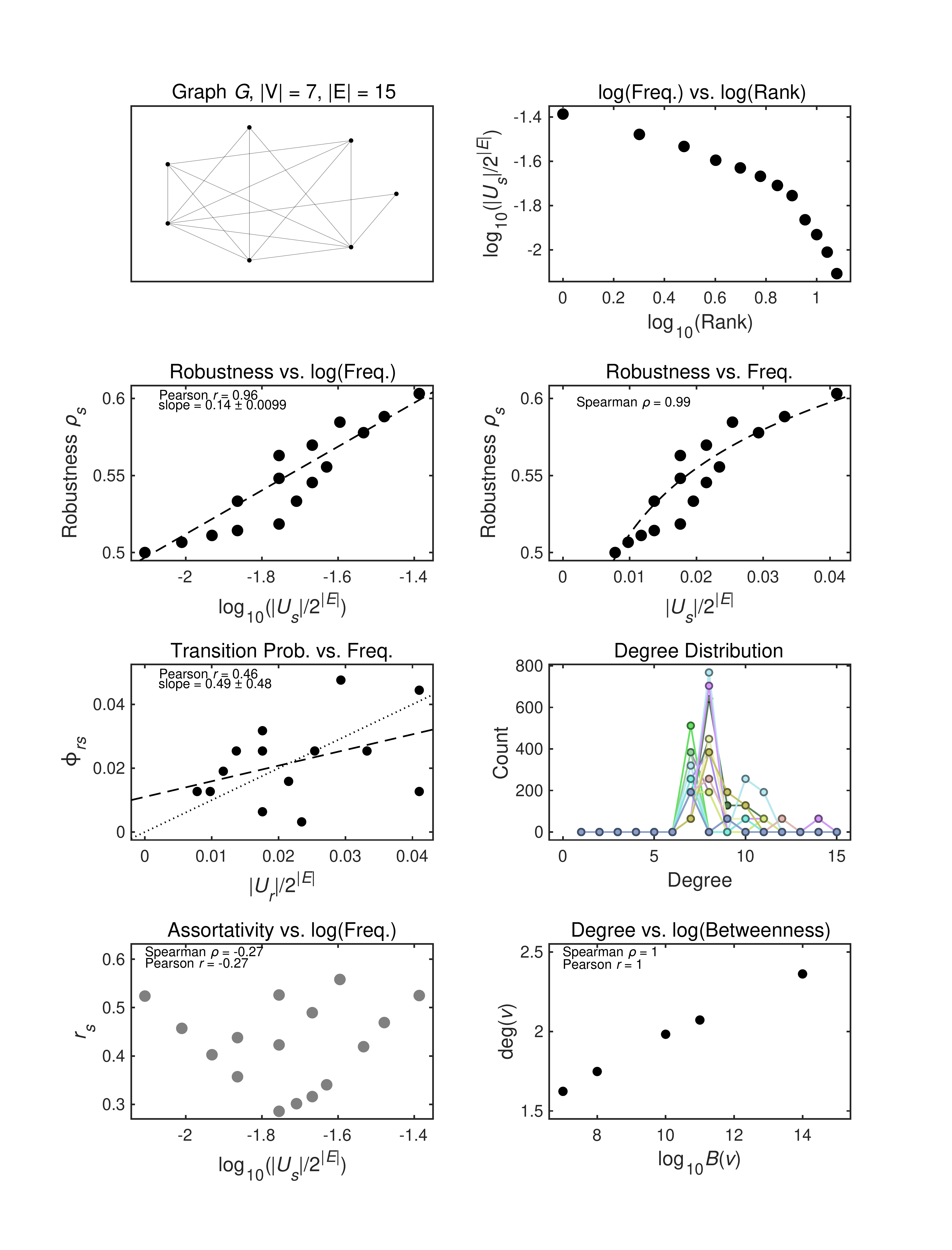}
\pagebreak
\includegraphics[height = \textheight]{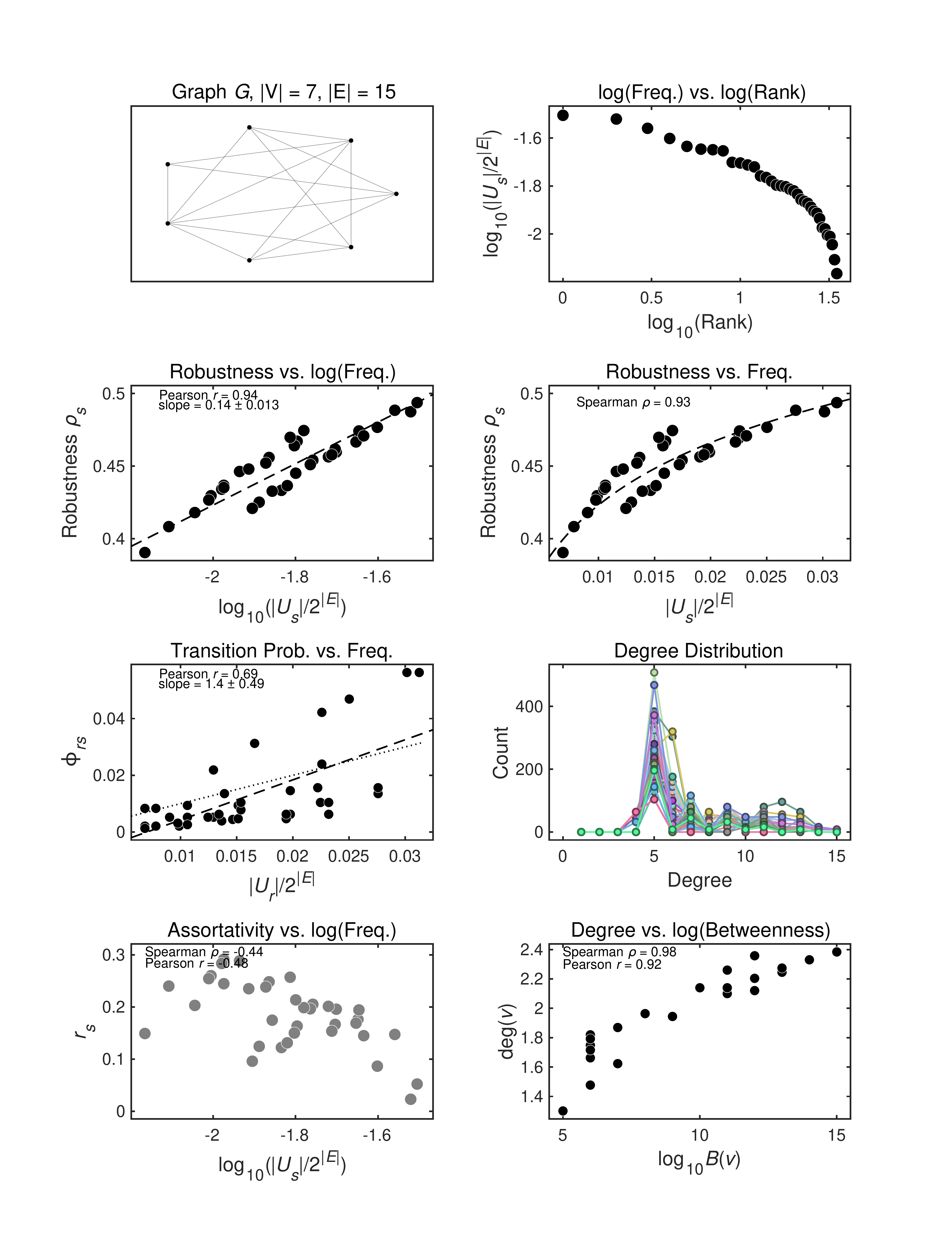}
\pagebreak
\includegraphics[height = \textheight]{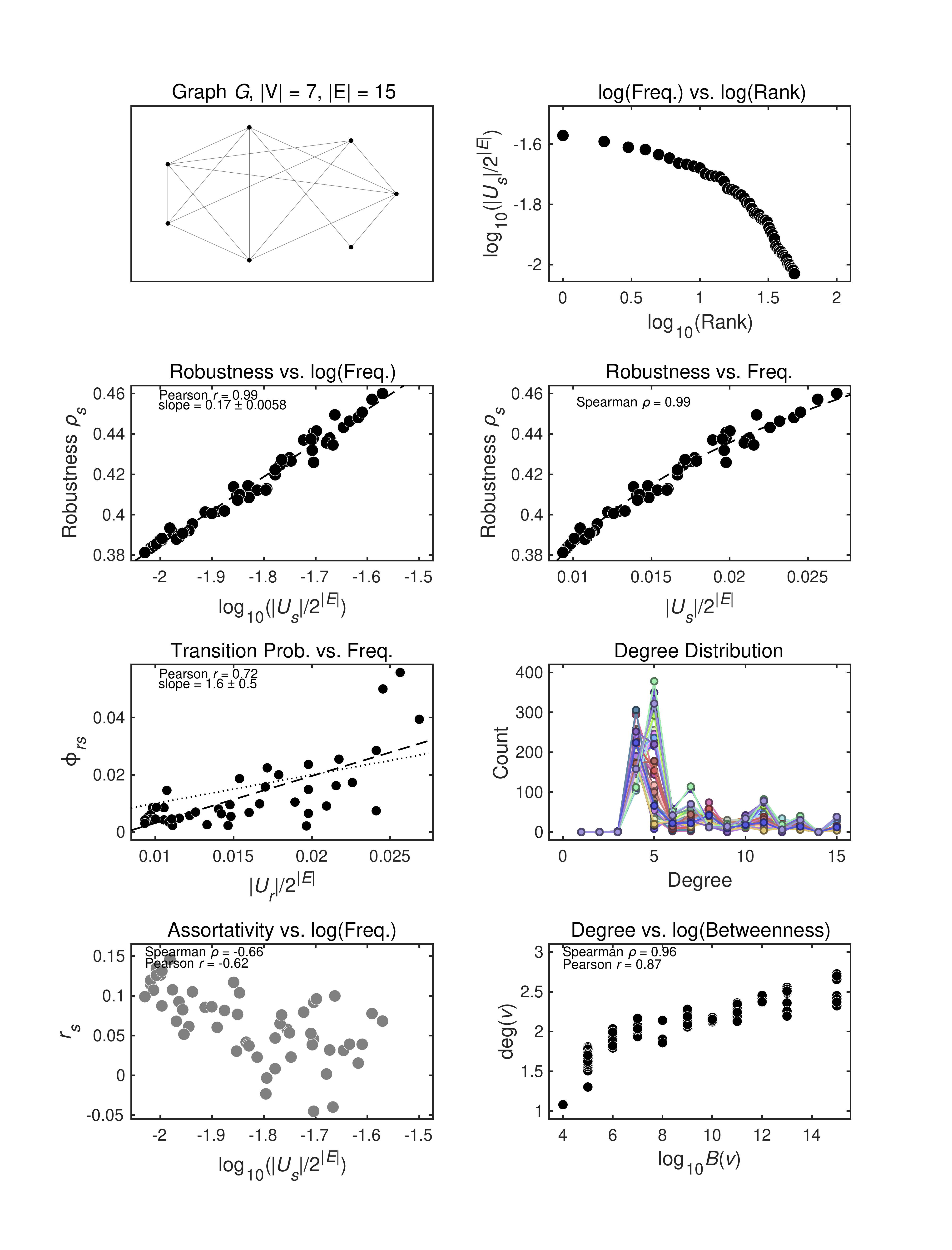}
\pagebreak
\includegraphics[height = \textheight]{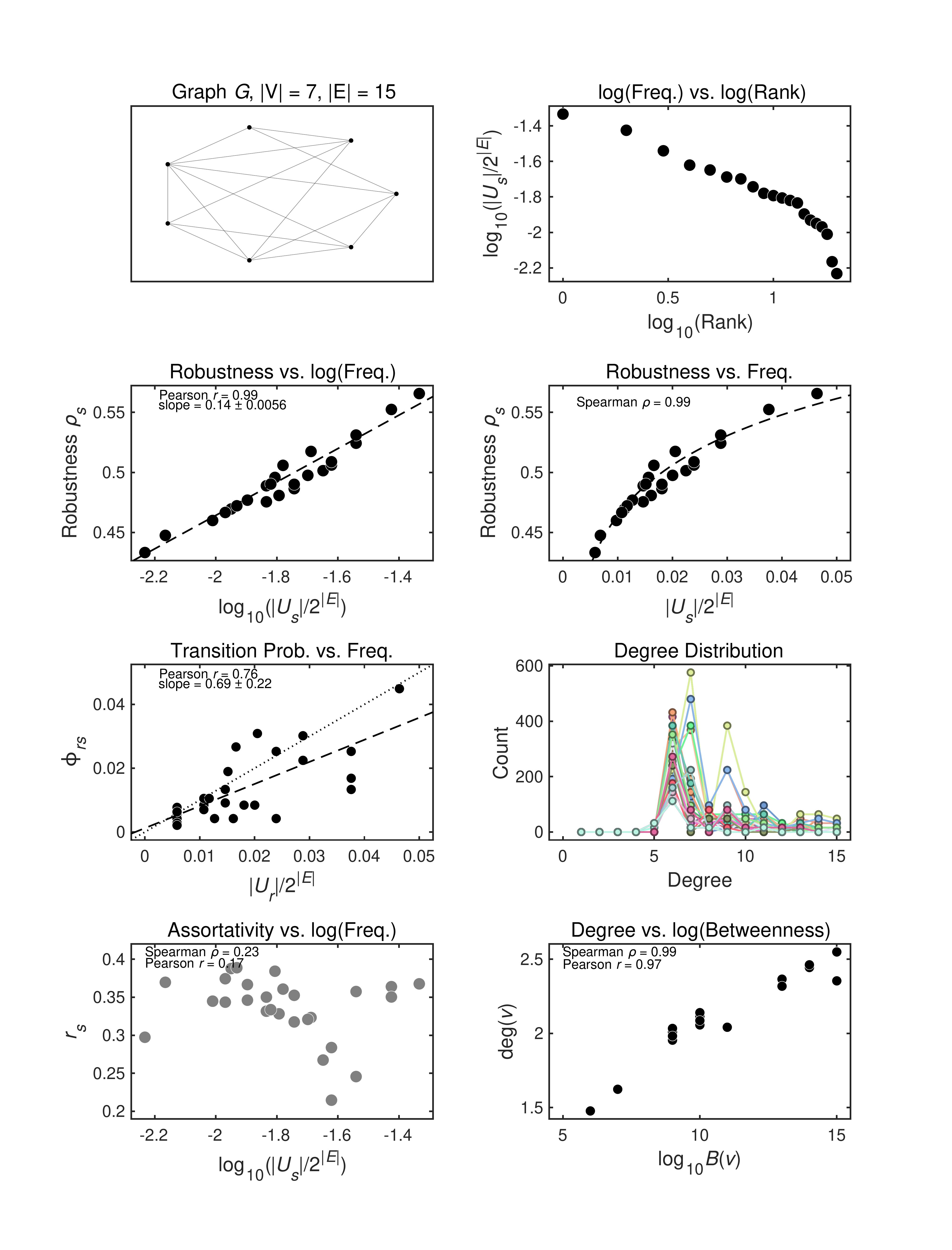}
\pagebreak
\includegraphics[height = \textheight]{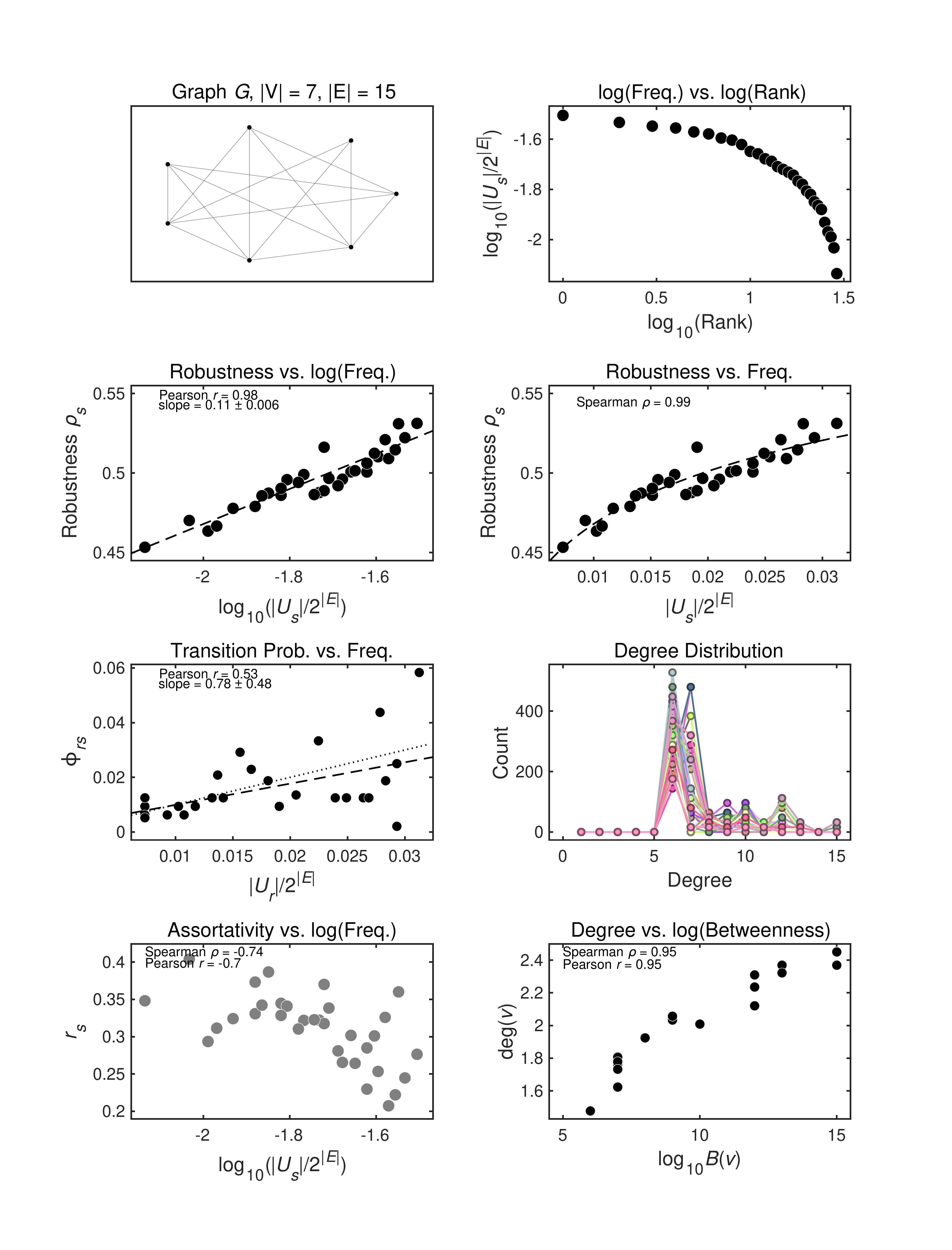}
\pagebreak
\includegraphics[height = \textheight]{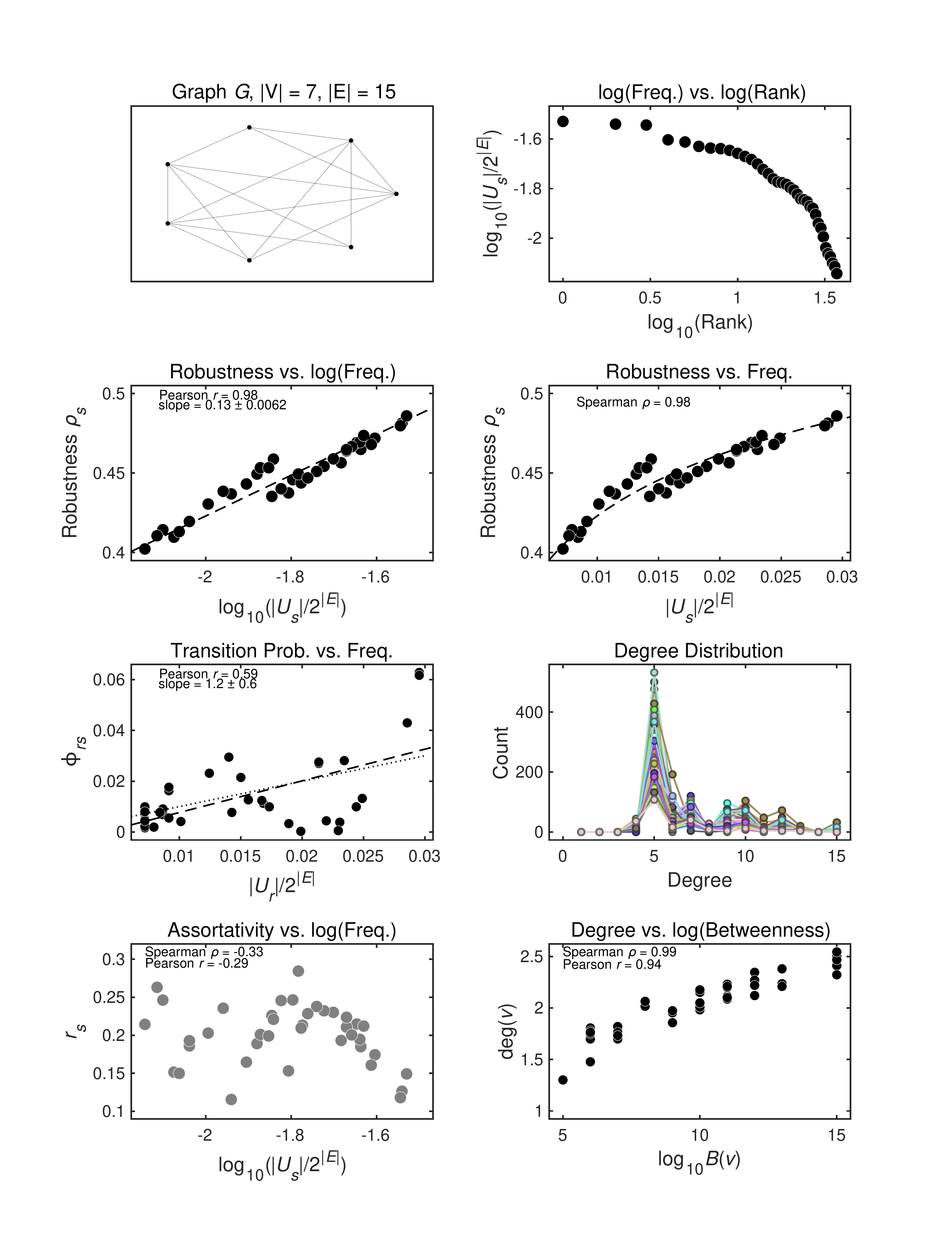}
\pagebreak
\includegraphics[height = \textheight]{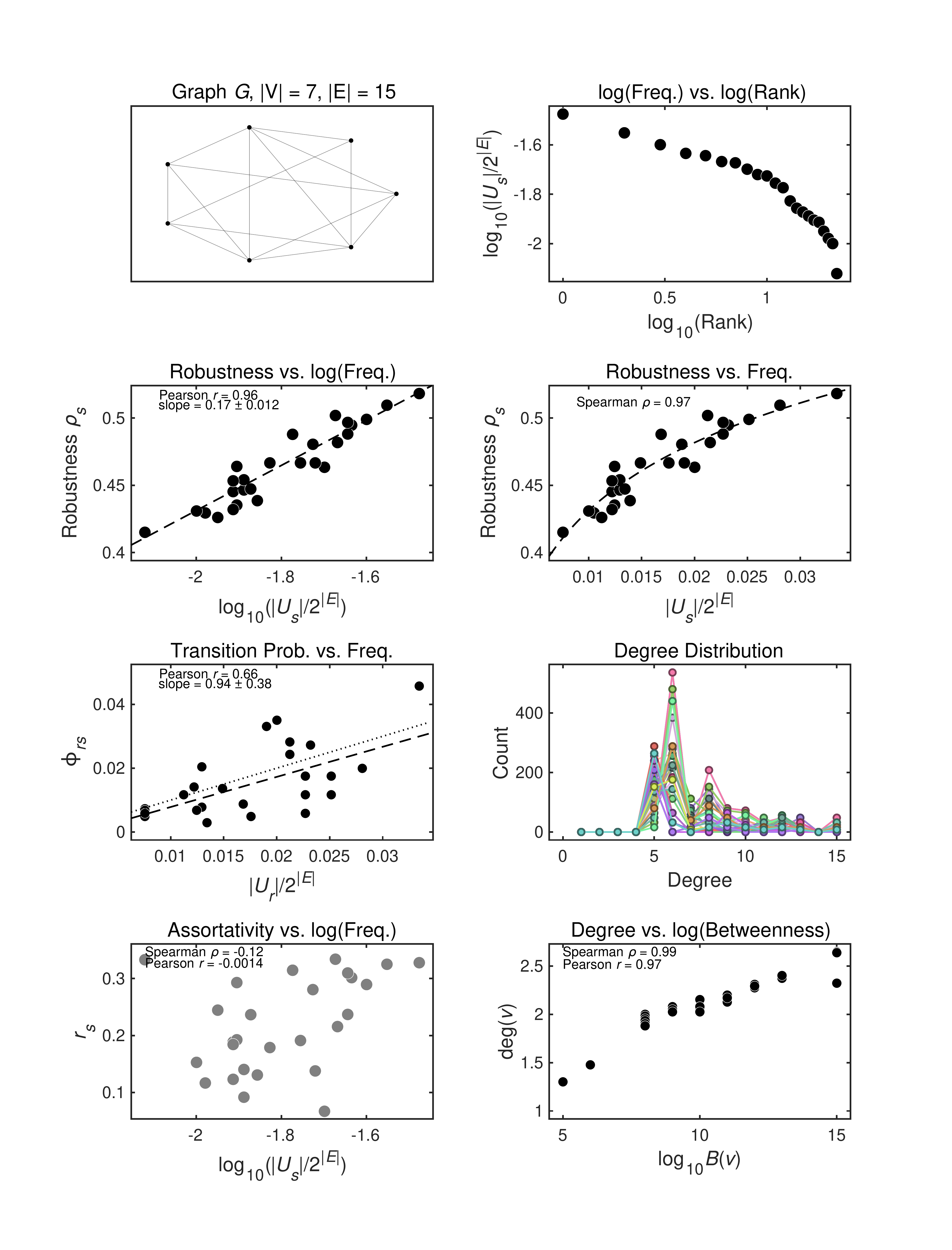}
\pagebreak
\includegraphics[height = \textheight]{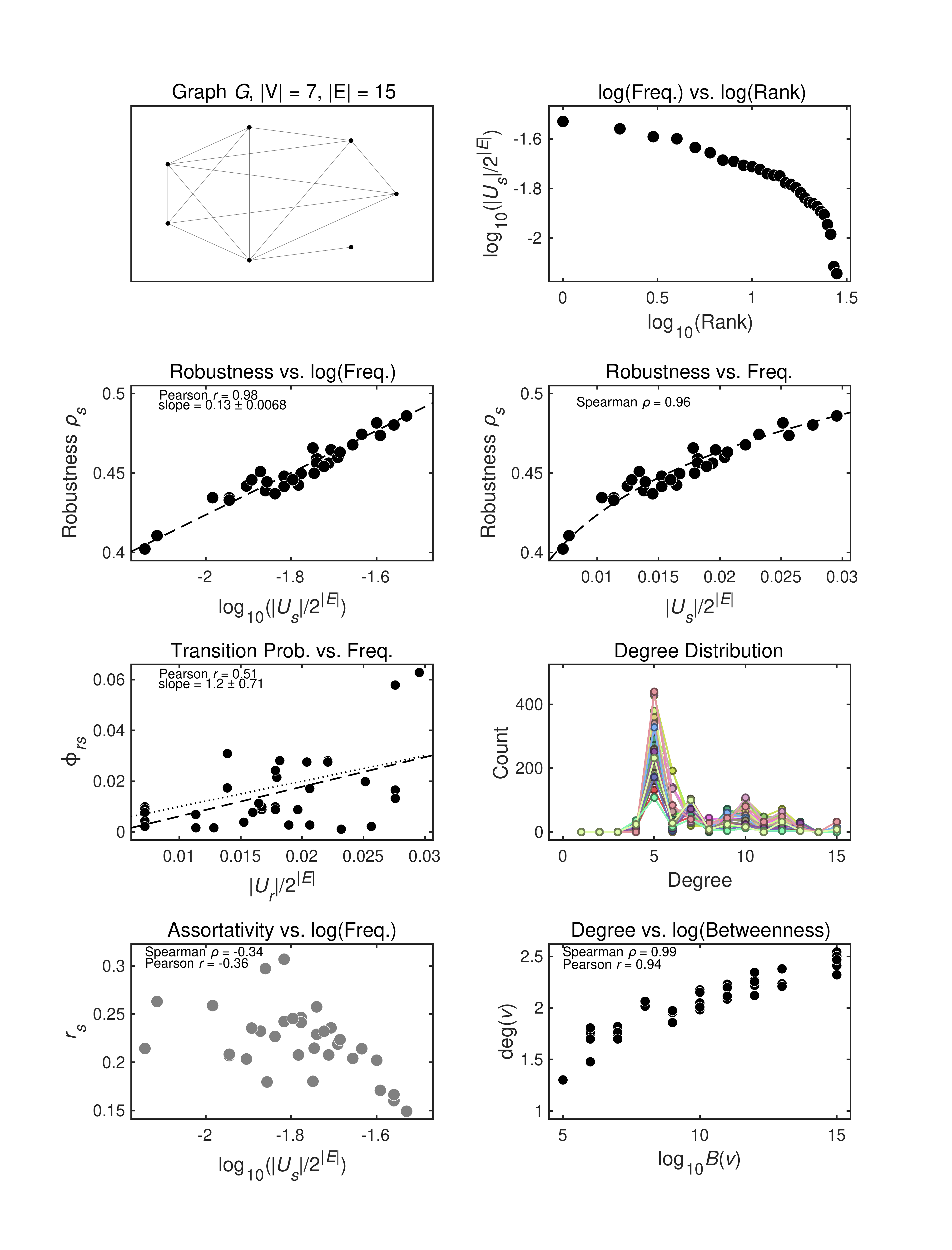}
\pagebreak
\includegraphics[height = \textheight]{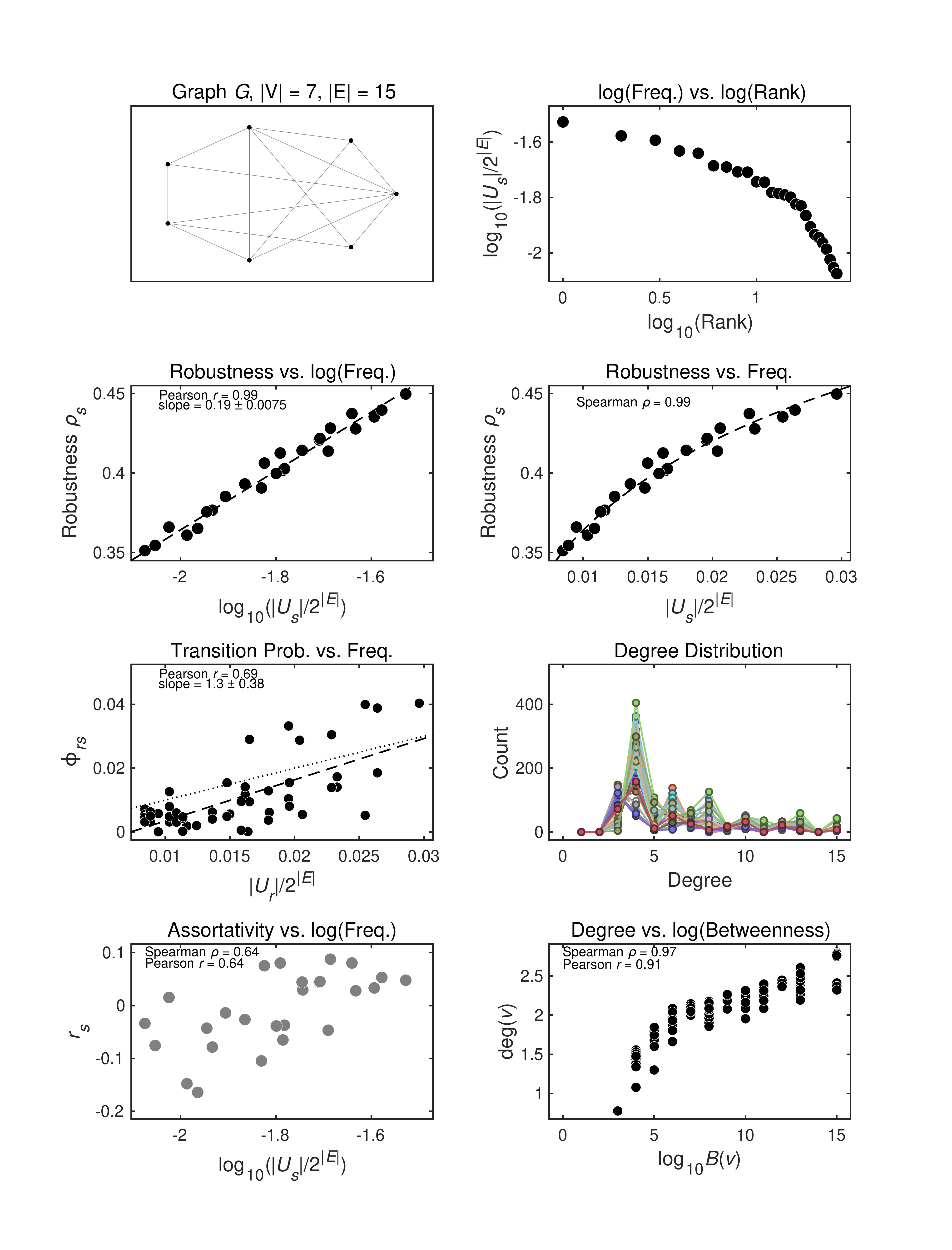}
\pagebreak
\includegraphics[height = \textheight]{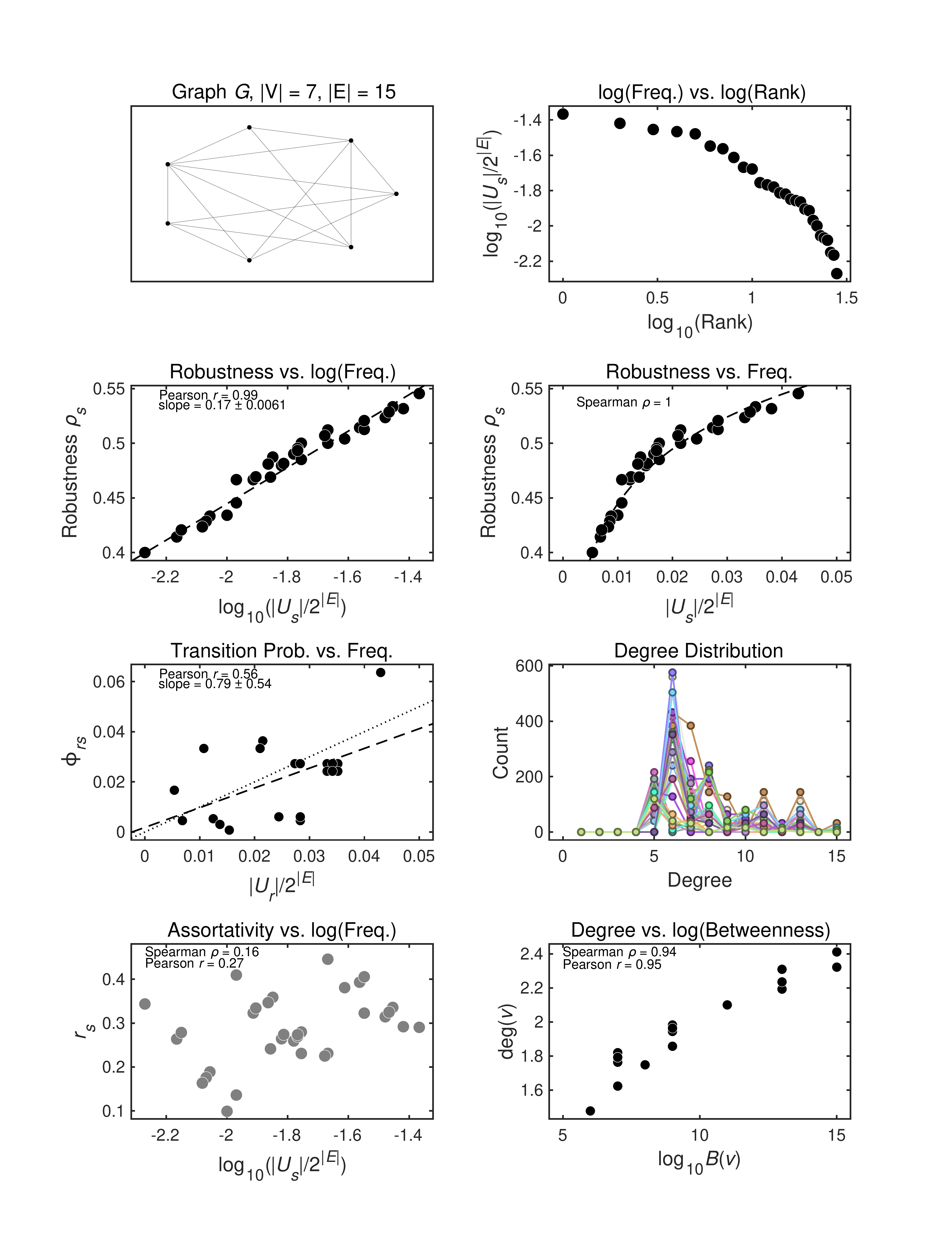}
\pagebreak
\includegraphics[height = \textheight]{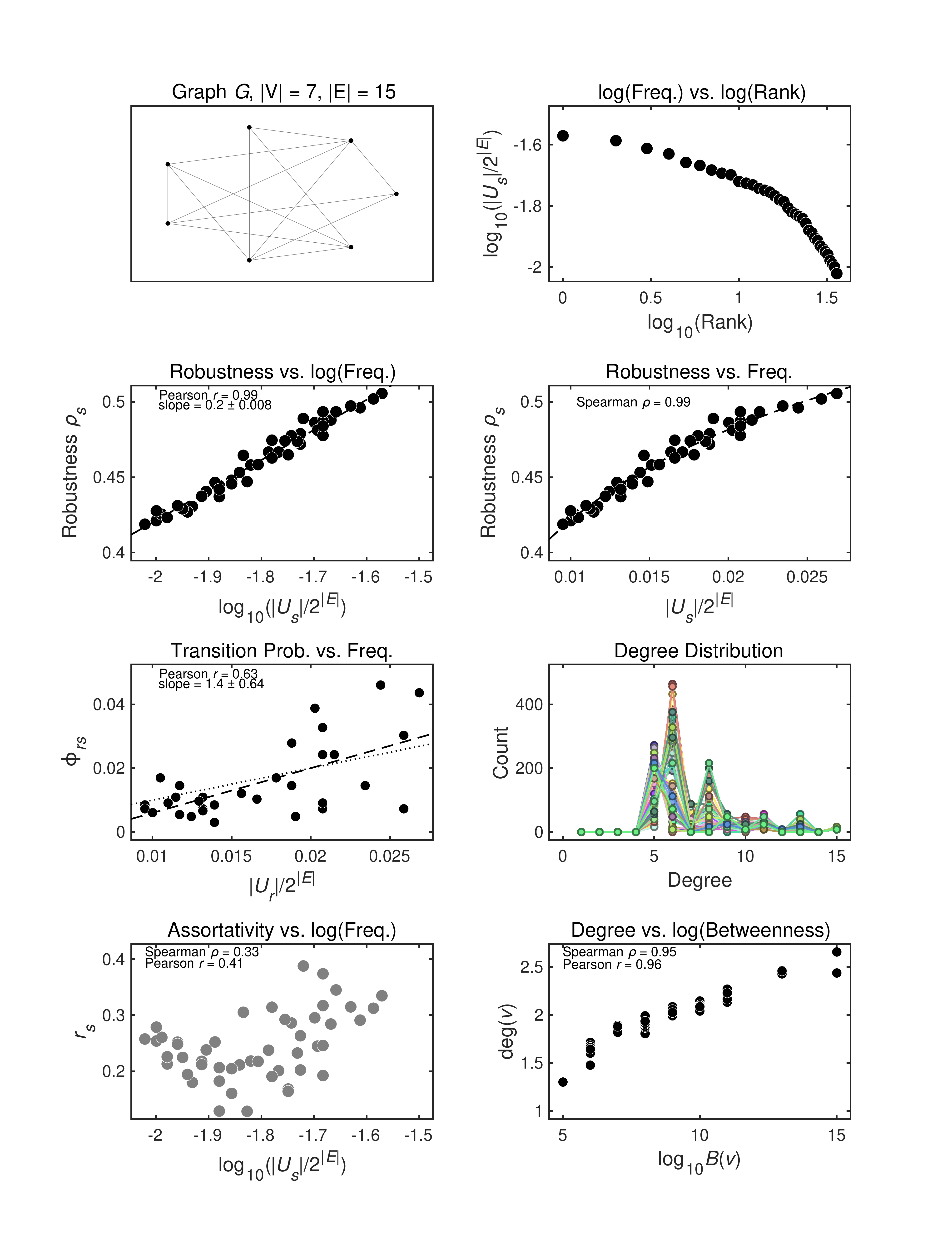}
\pagebreak
\includegraphics[height = \textheight]{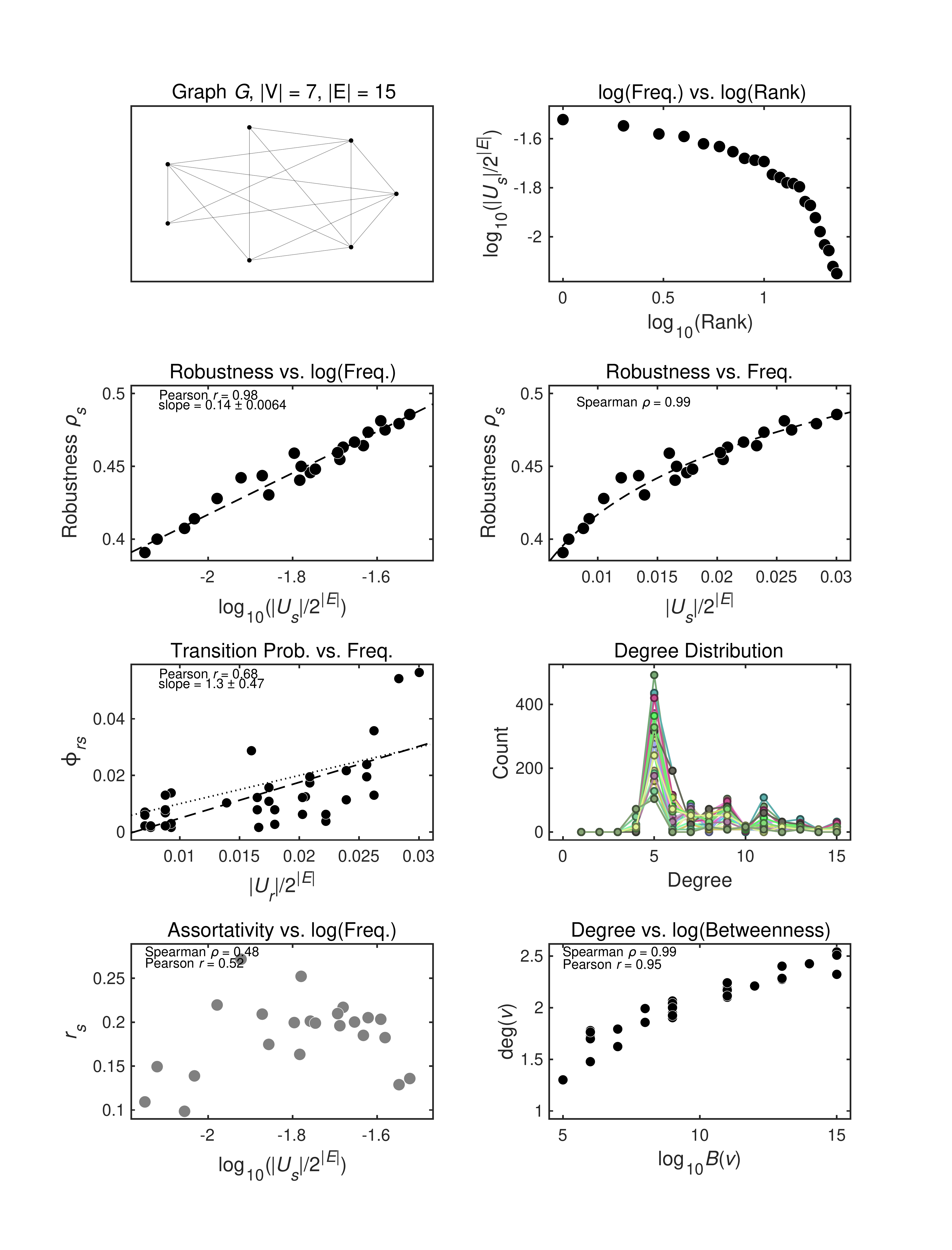}
\pagebreak
\includegraphics[height = \textheight]{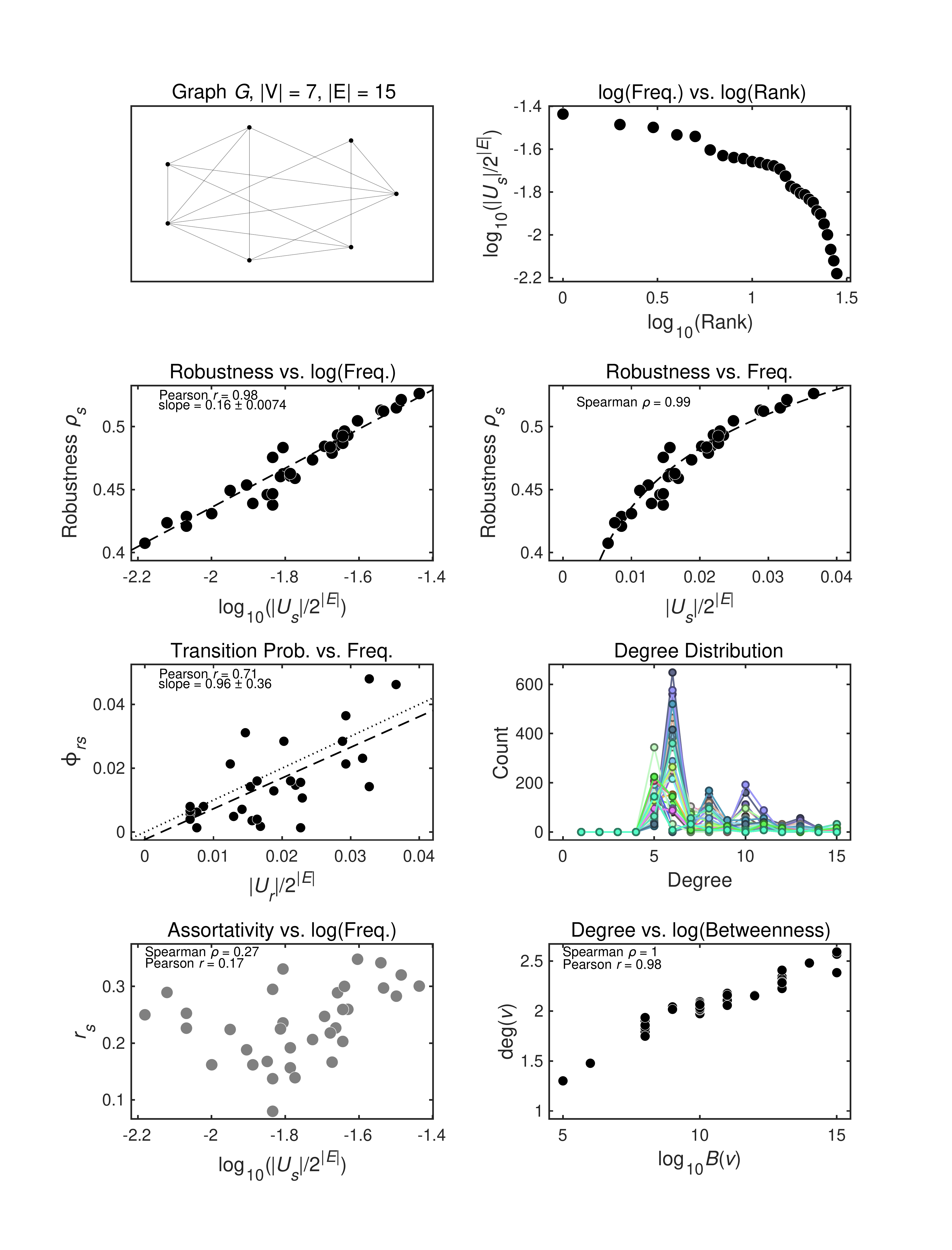}
\pagebreak
\includegraphics[height = \textheight]{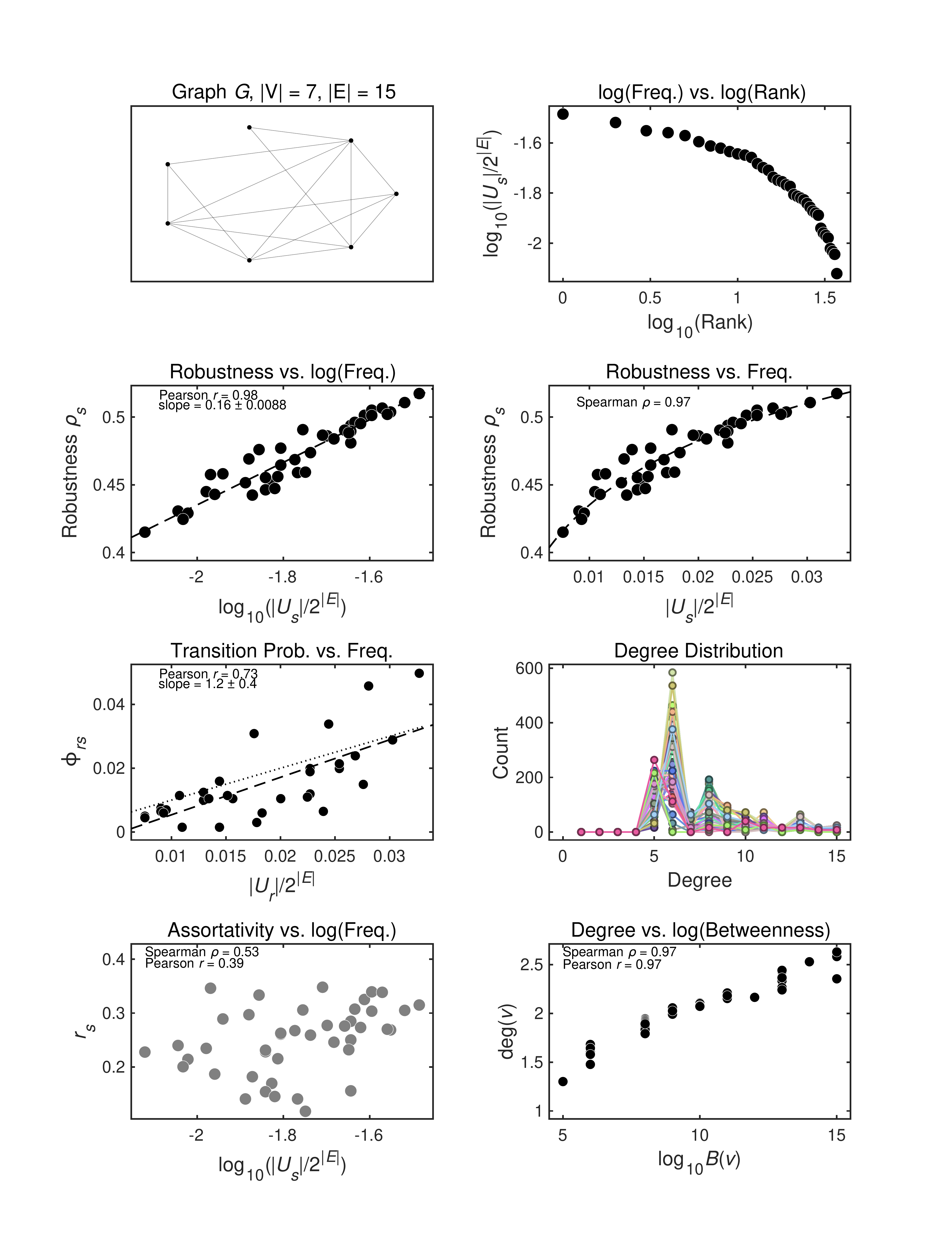}
\pagebreak
\includegraphics[height = \textheight]{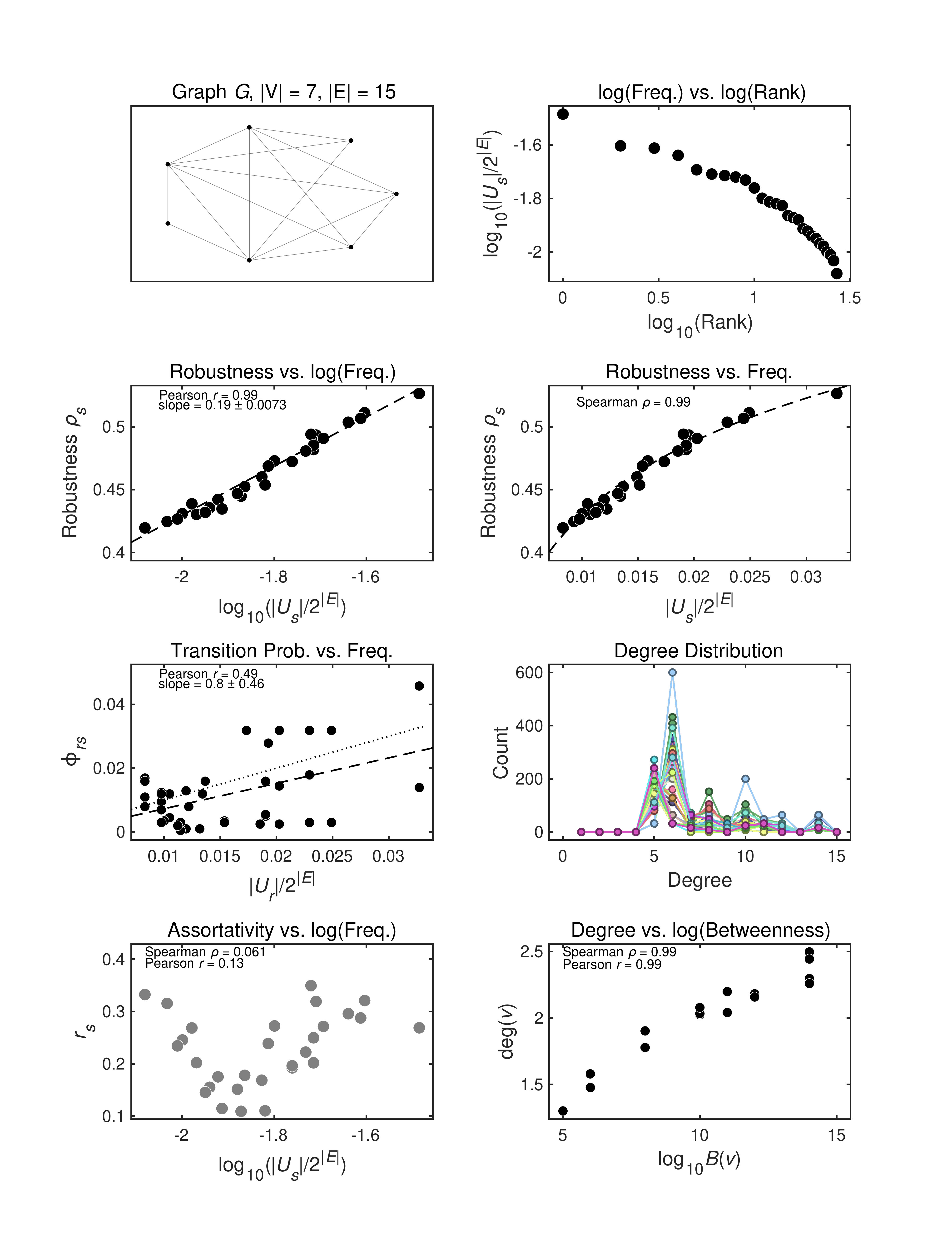}
\pagebreak
\includegraphics[height = \textheight]{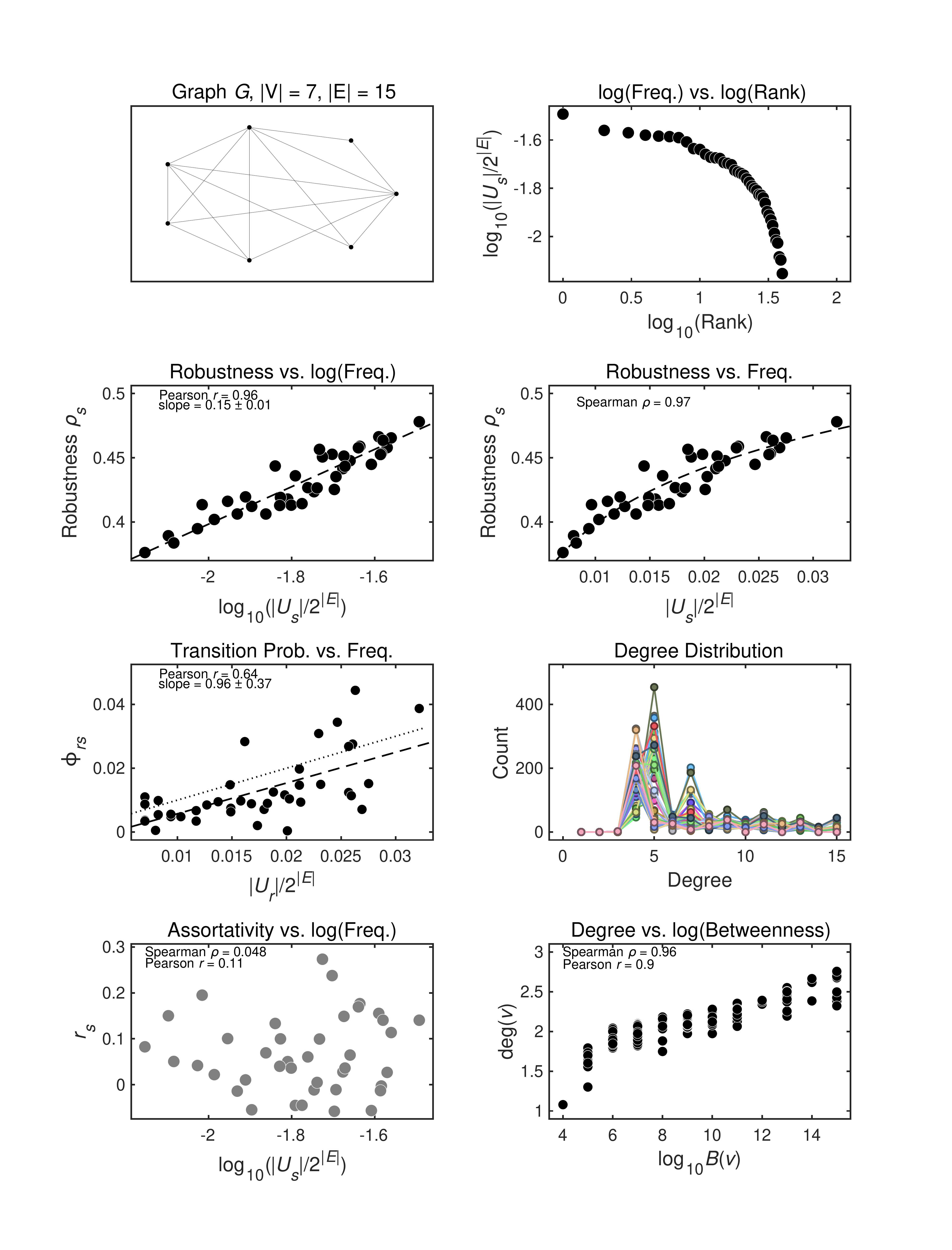}
\pagebreak
\includegraphics[height = \textheight]{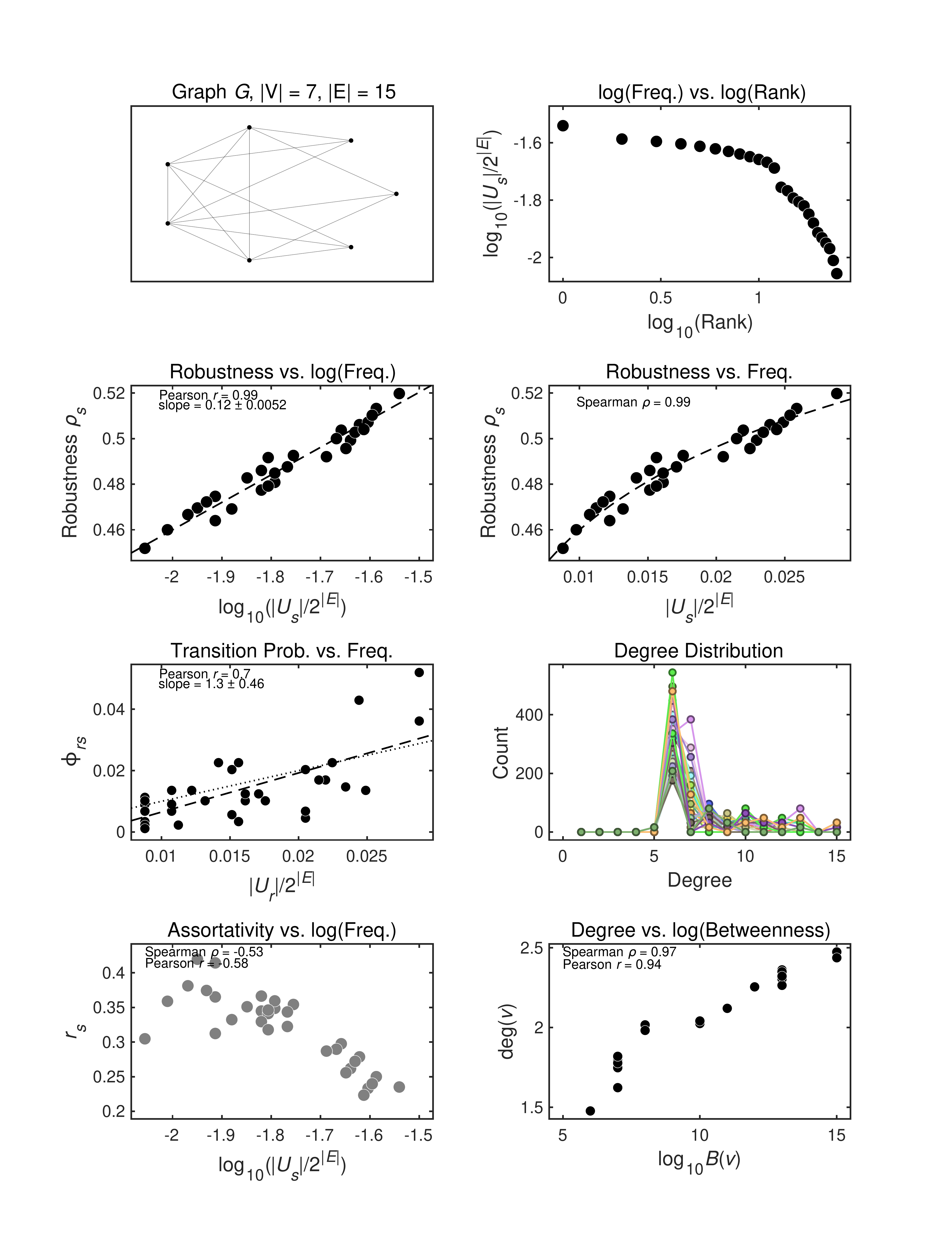}
\pagebreak
\includegraphics[height = \textheight]{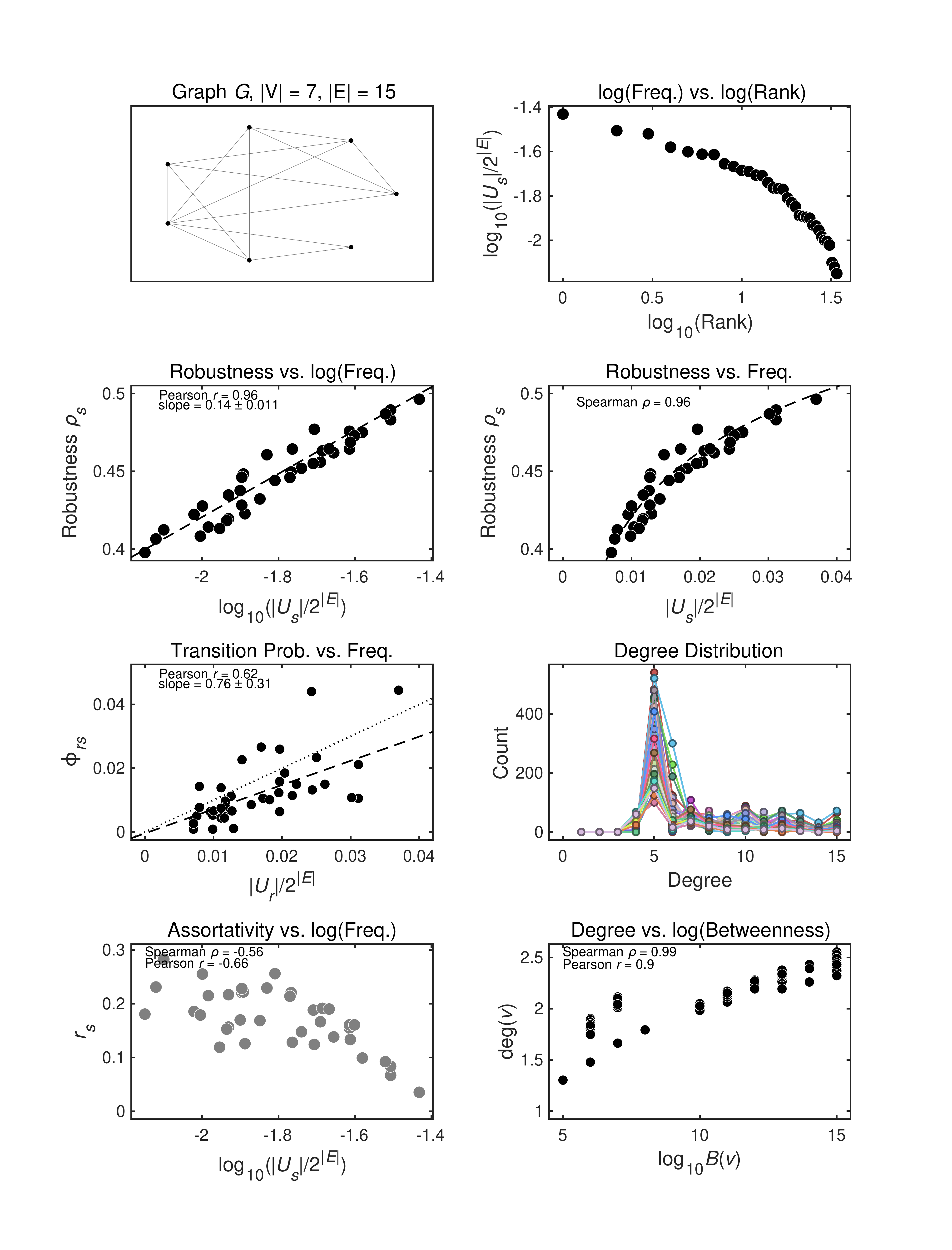}
\pagebreak
\includegraphics[height = \textheight]{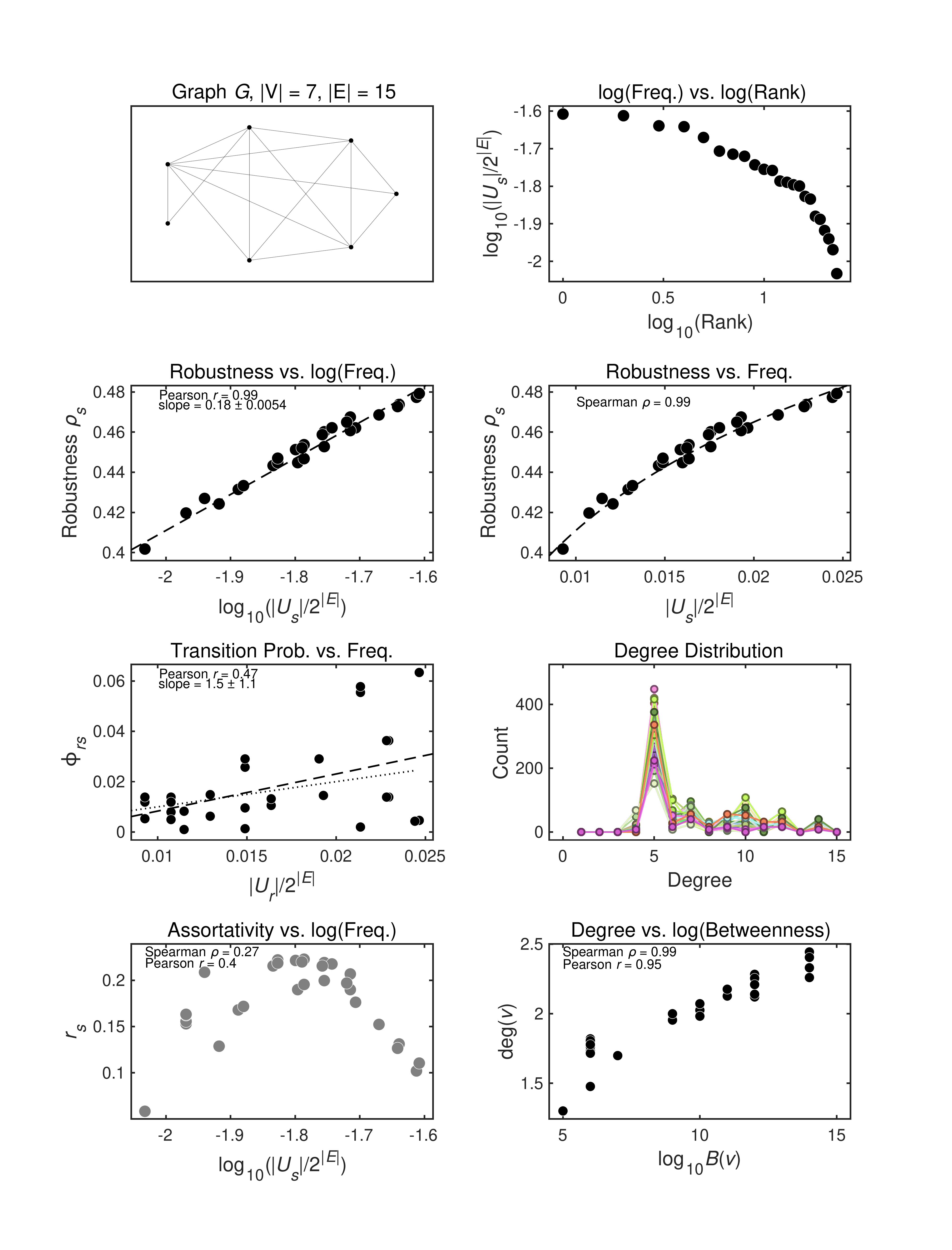}
\pagebreak
\includegraphics[height = \textheight]{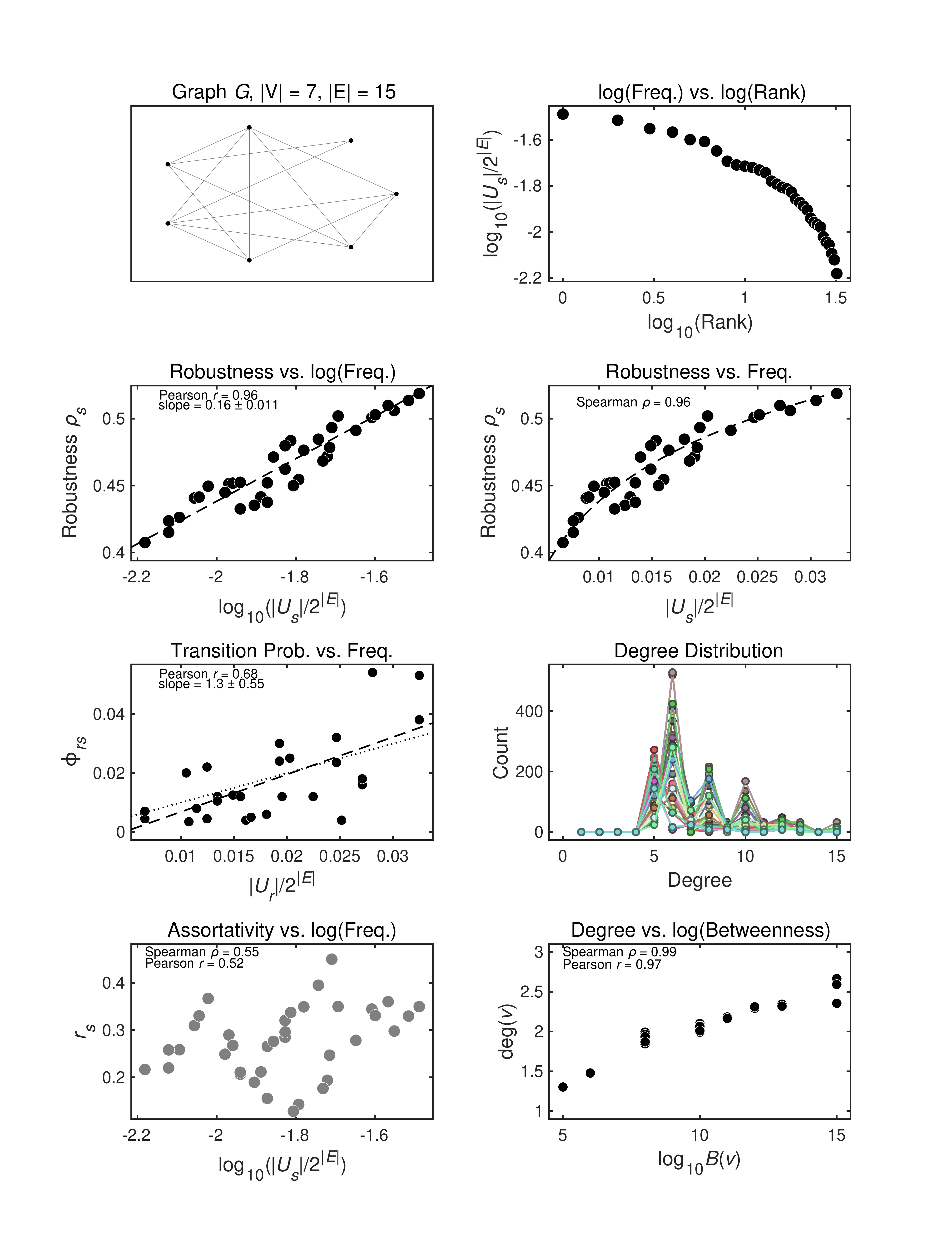}
\pagebreak
\includegraphics[height = \textheight]{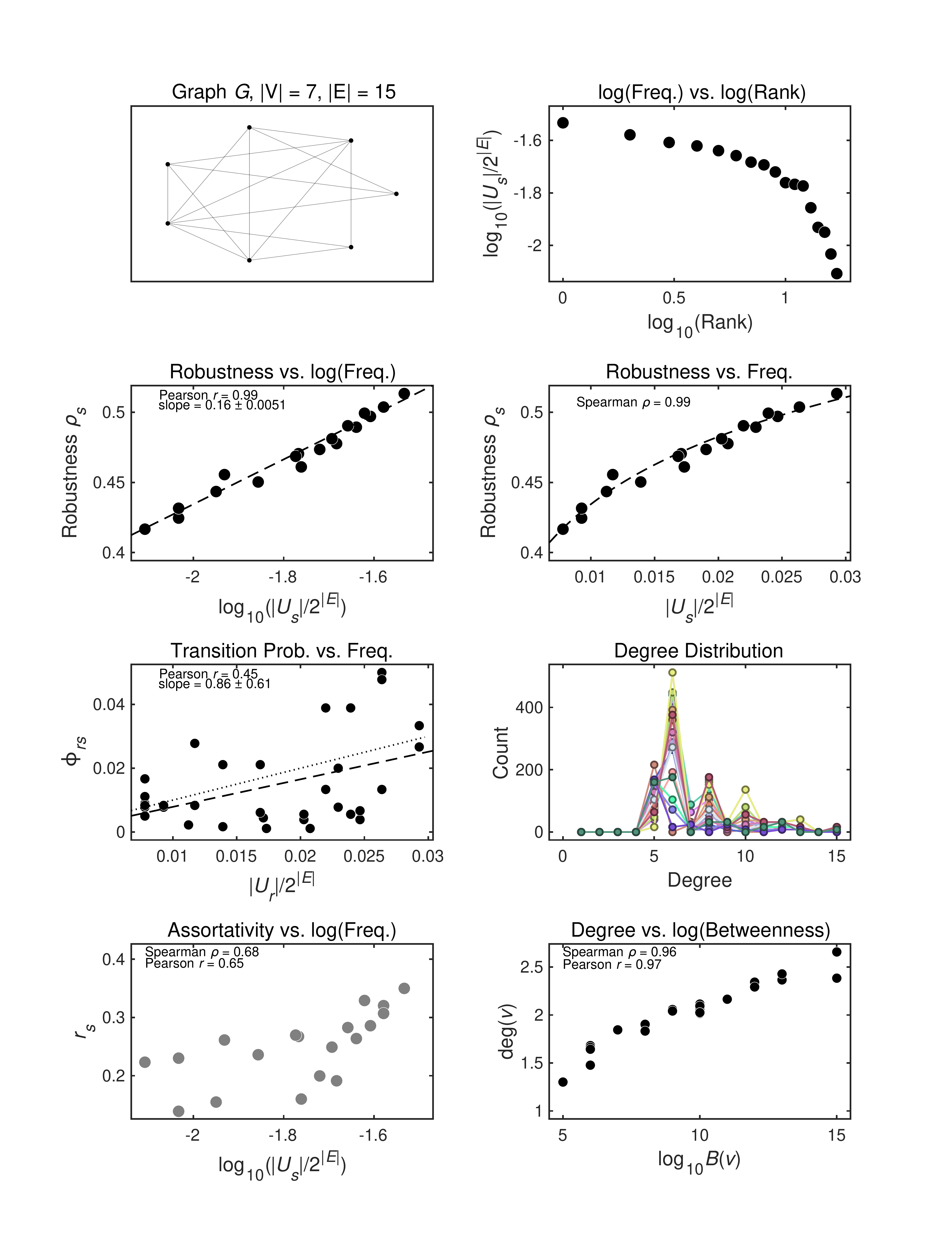}
\pagebreak
\includegraphics[height = \textheight]{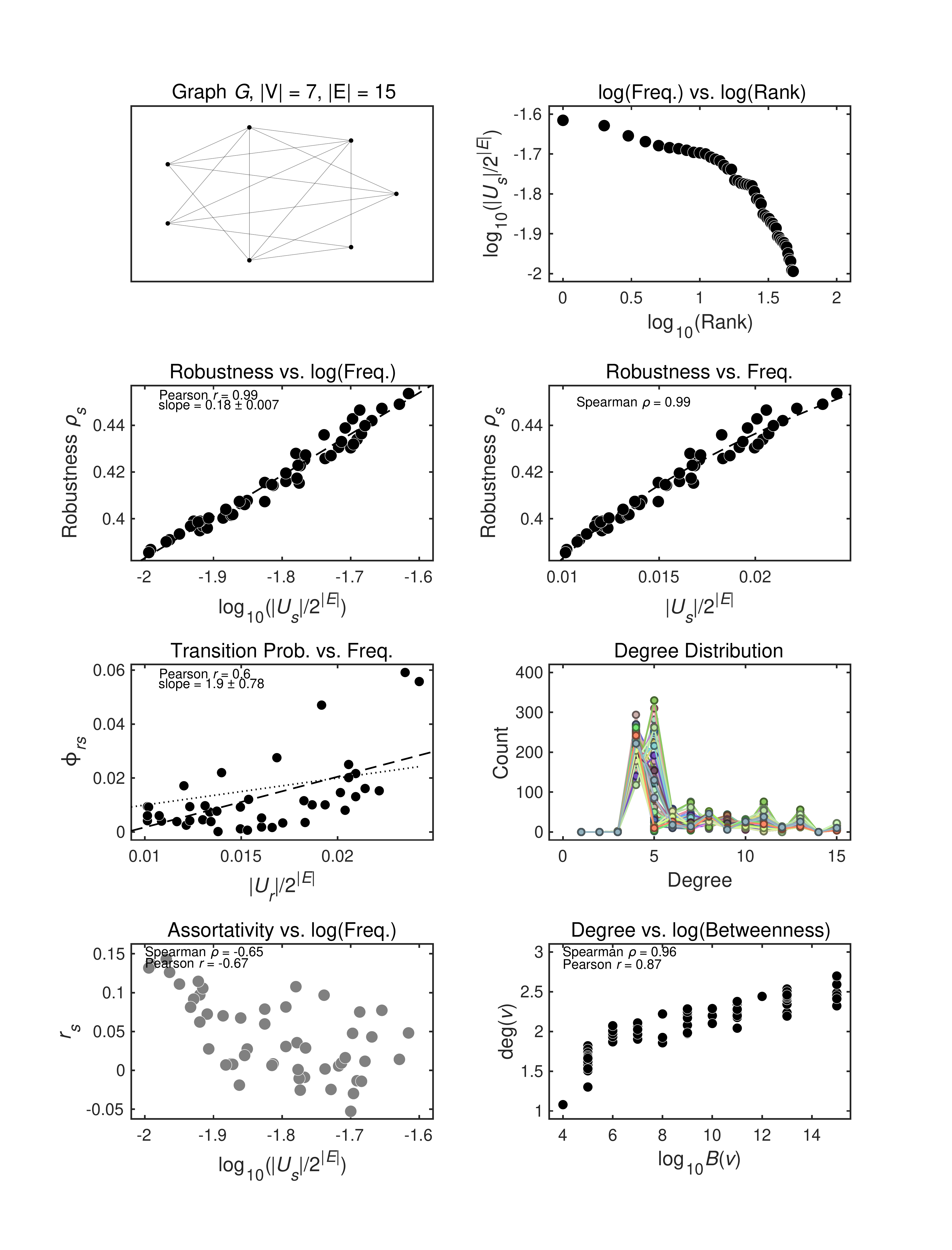}
\pagebreak
\includegraphics[height = \textheight]{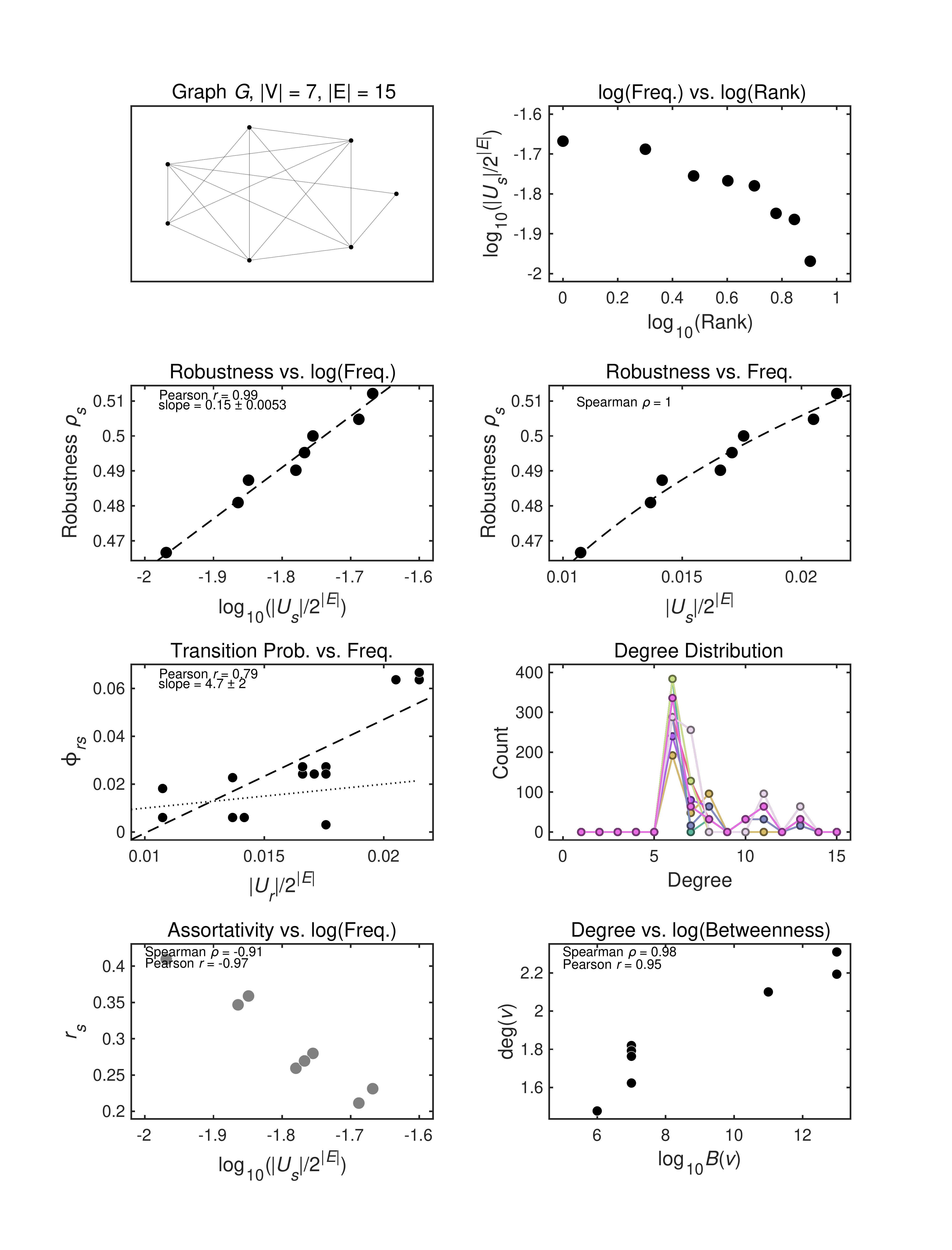}
\pagebreak
\includegraphics[height = \textheight]{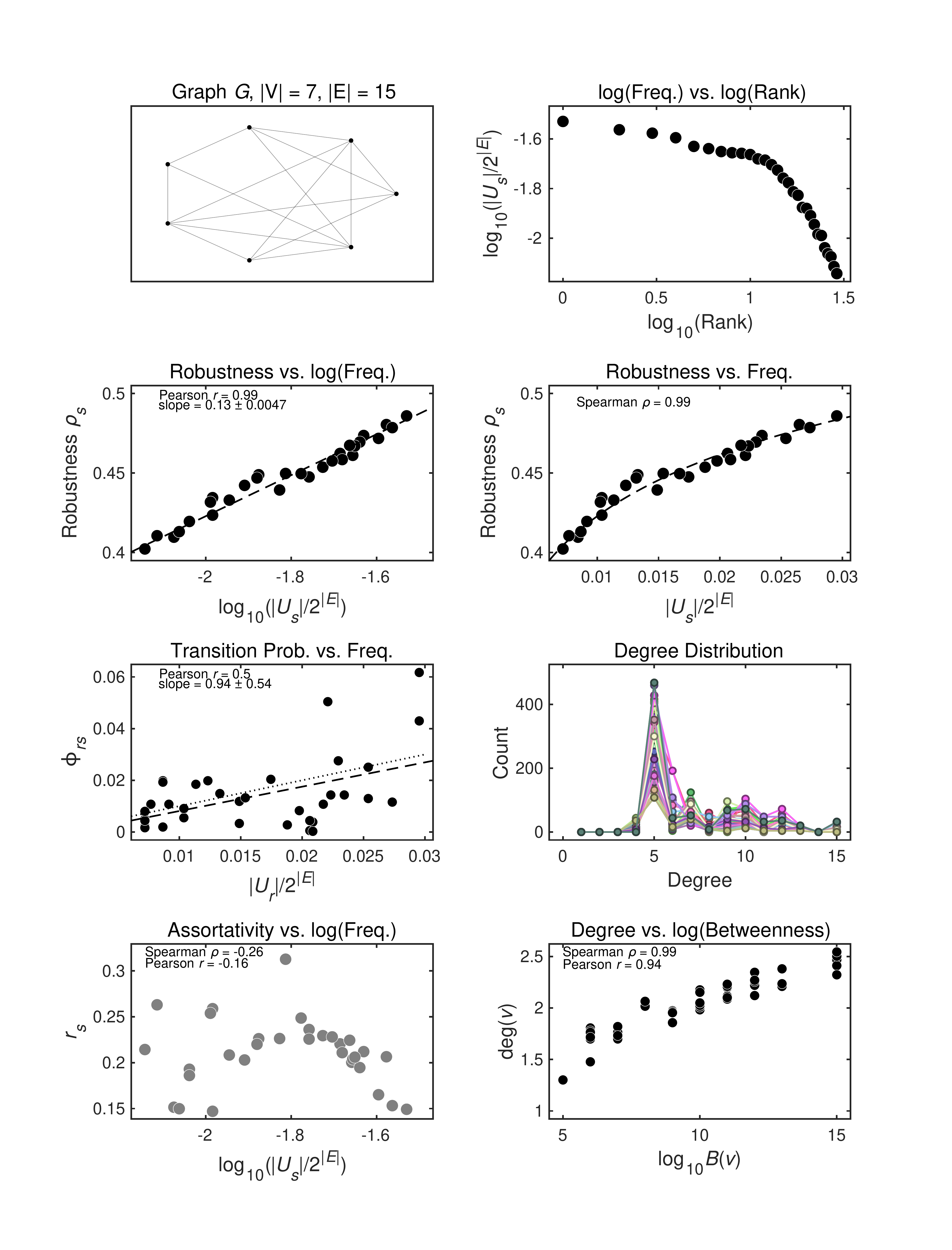}
\pagebreak
\includegraphics[height = \textheight]{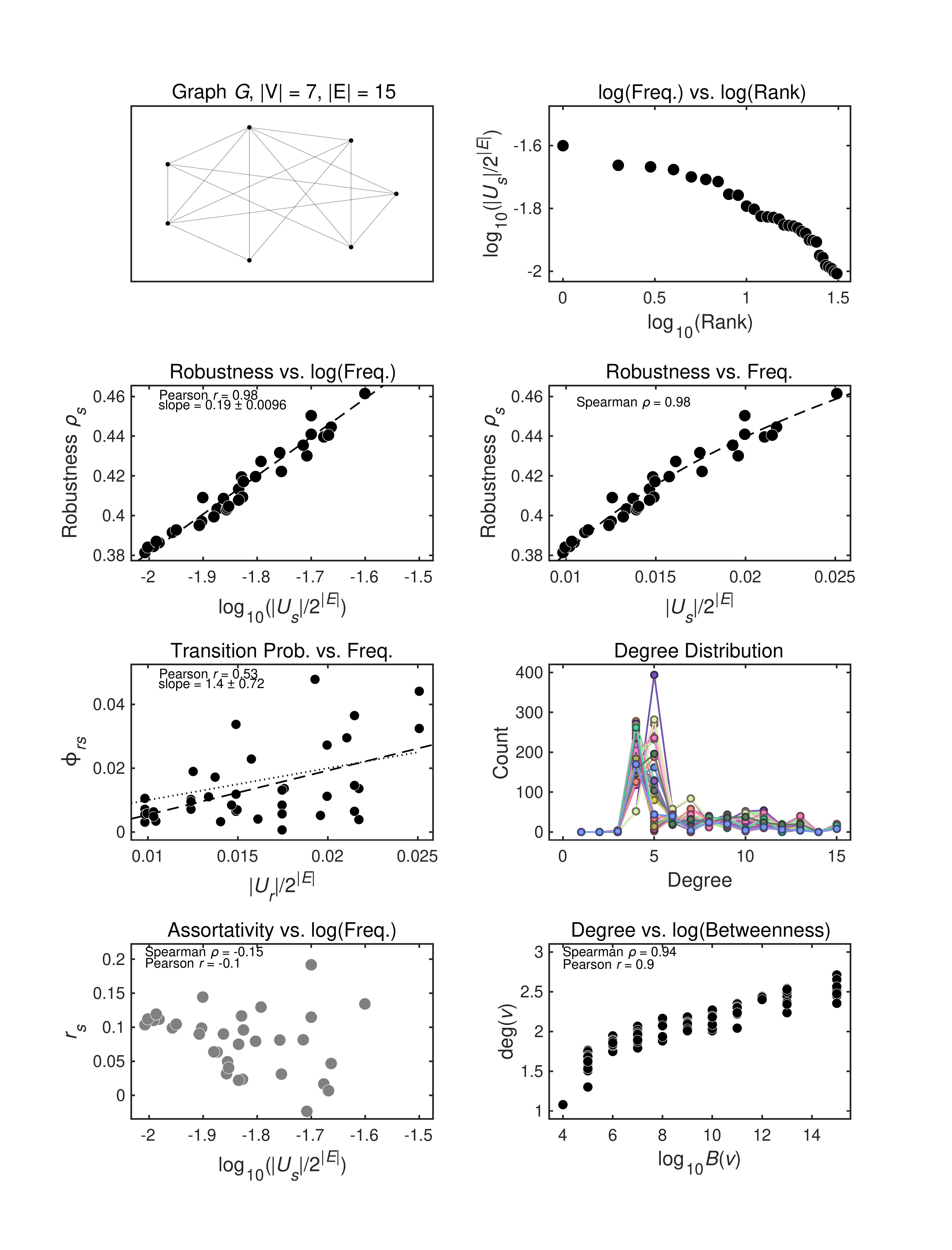}
\pagebreak
\includegraphics[height = \textheight]{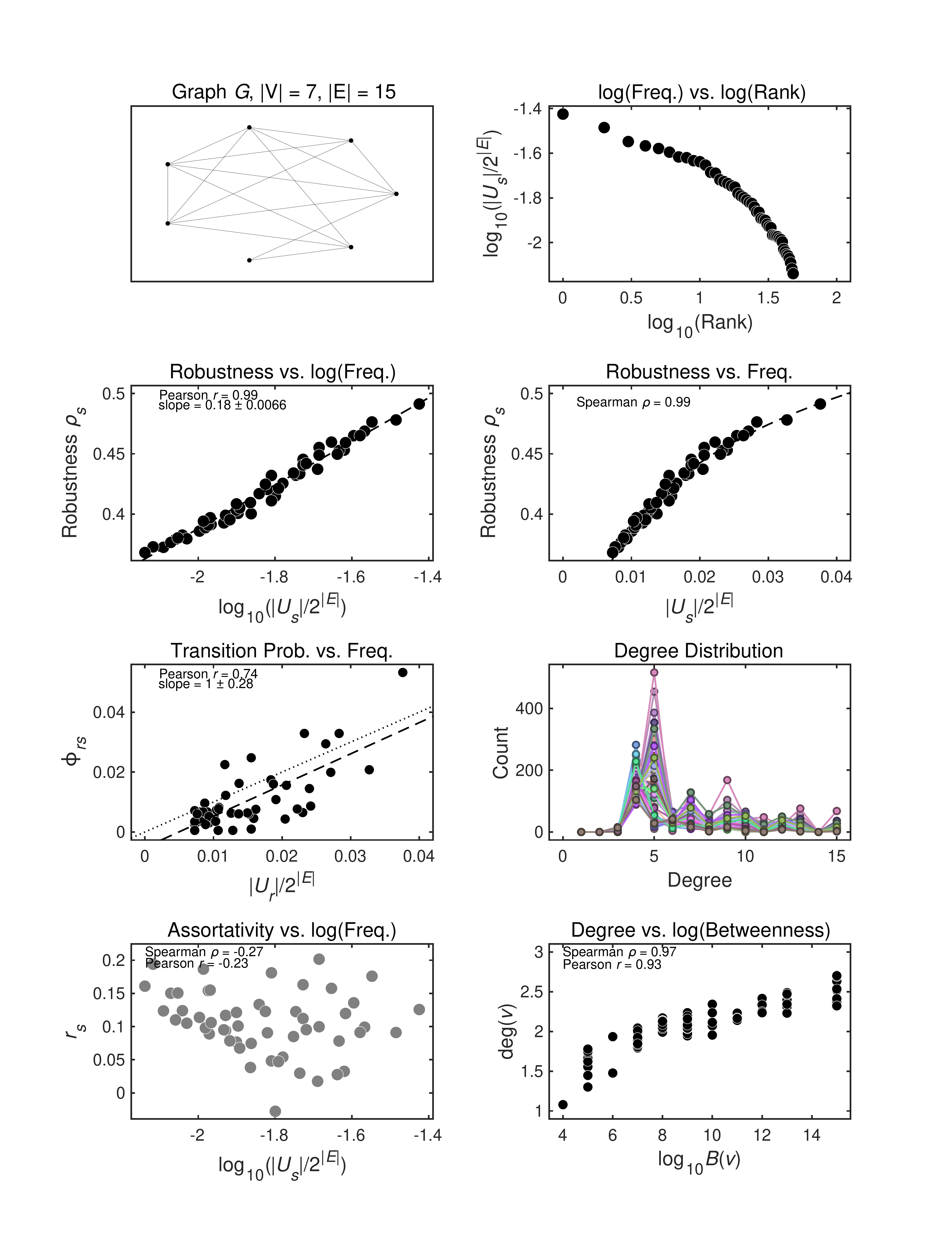}
\pagebreak
\includegraphics[height = \textheight]{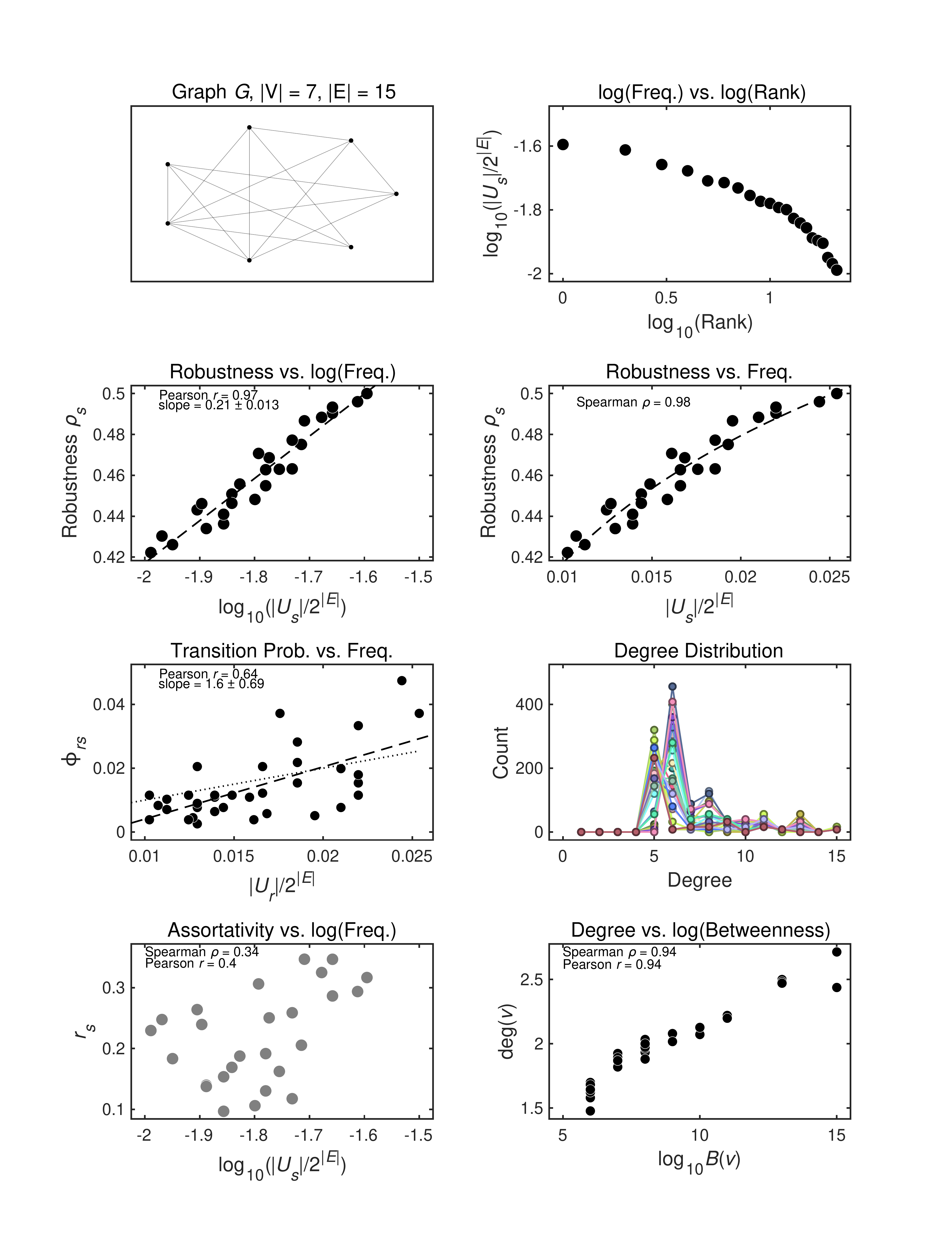}
\pagebreak
\includegraphics[height = \textheight]{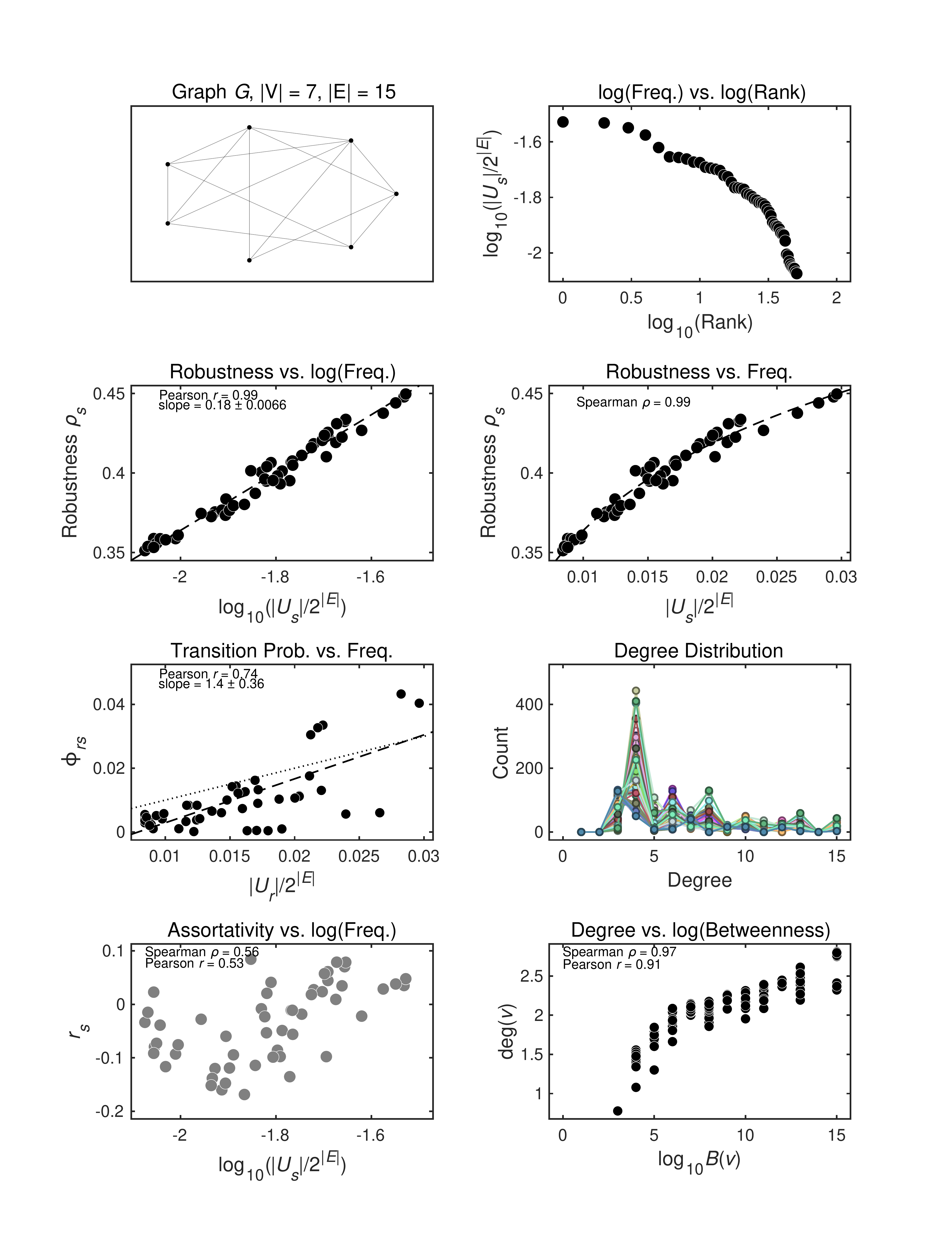}
\pagebreak
\includegraphics[height = \textheight]{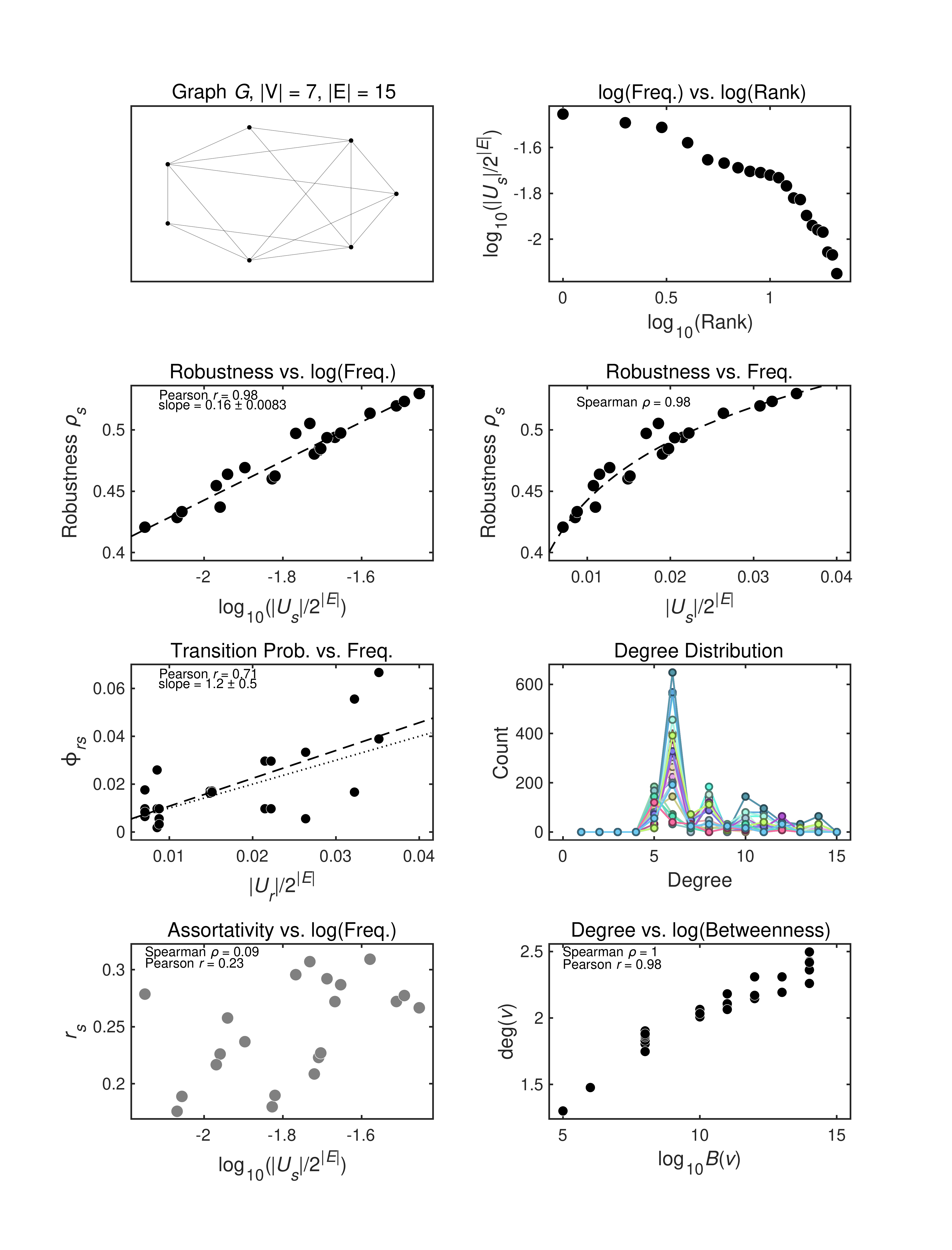}
\pagebreak
\includegraphics[height = \textheight]{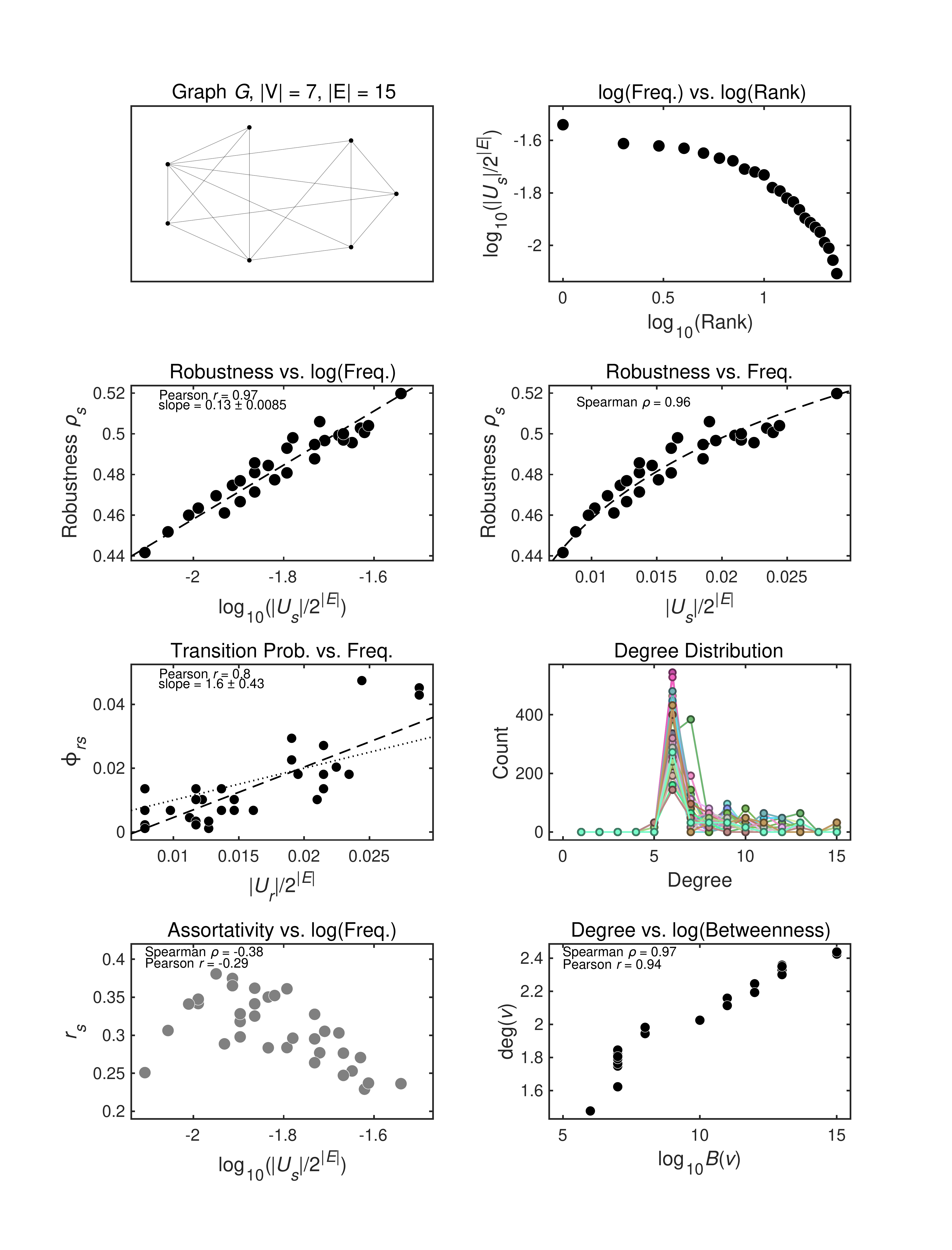}
\pagebreak
\includegraphics[height = \textheight]{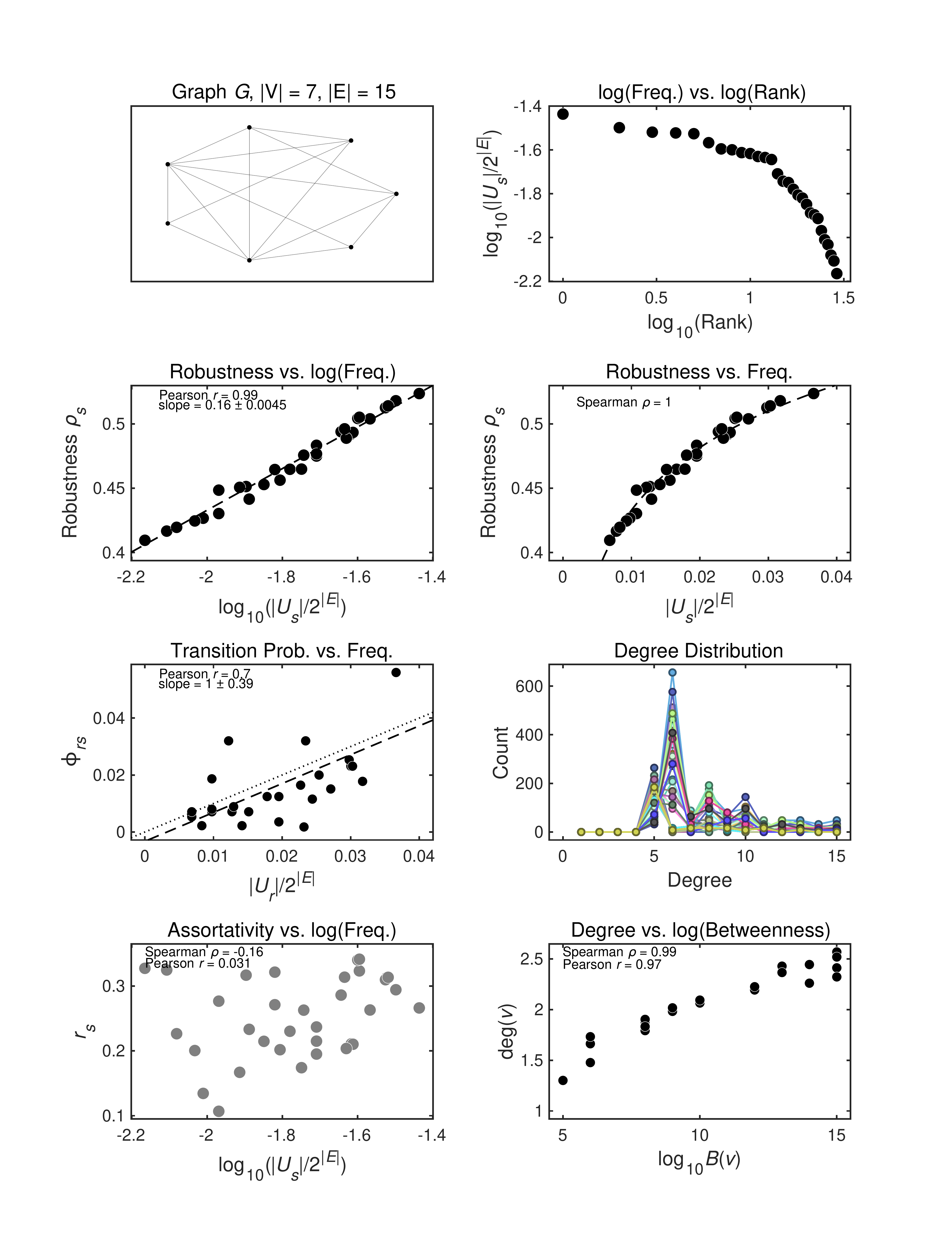}
\pagebreak
\includegraphics[height = \textheight]{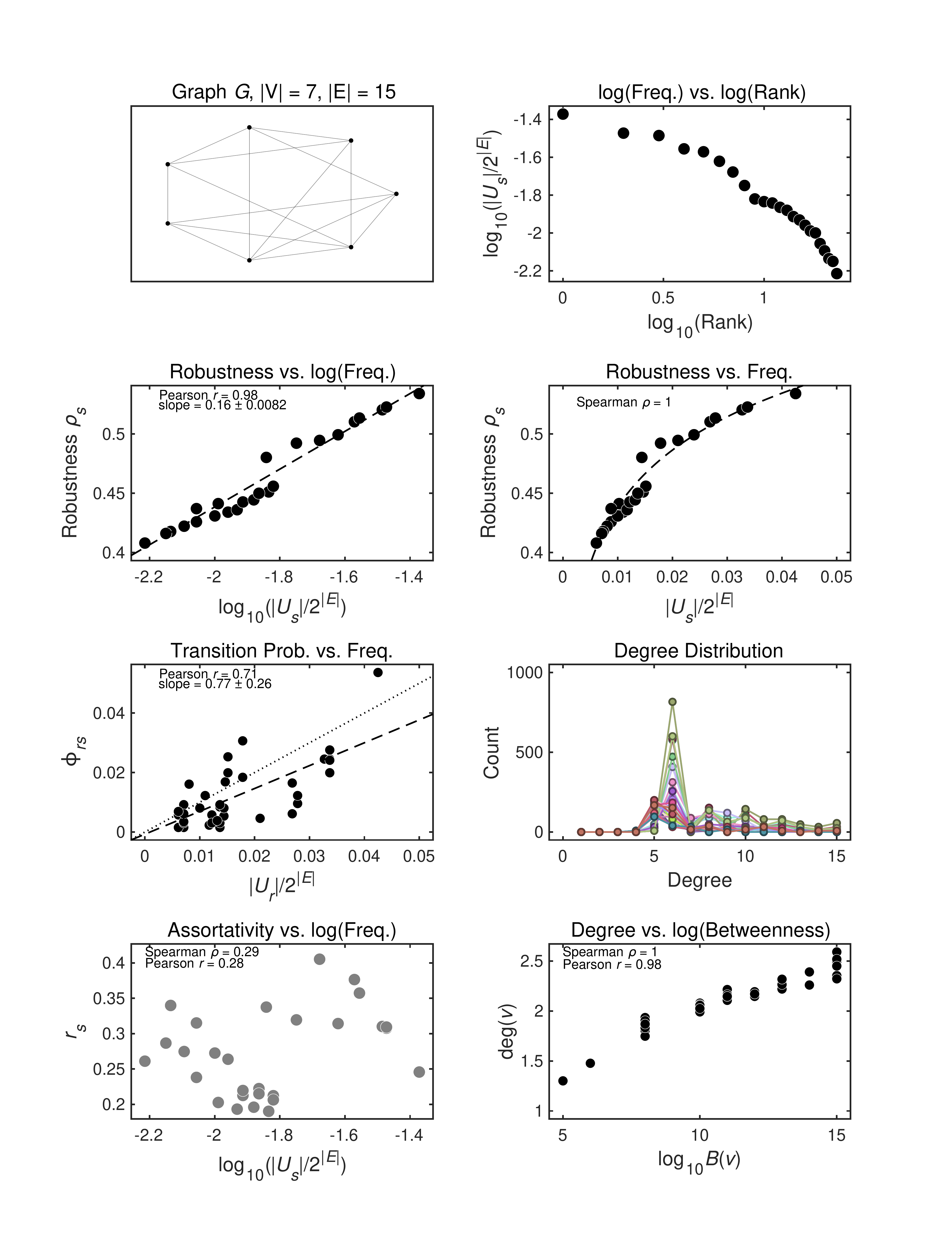}
\pagebreak
\includegraphics[height = \textheight]{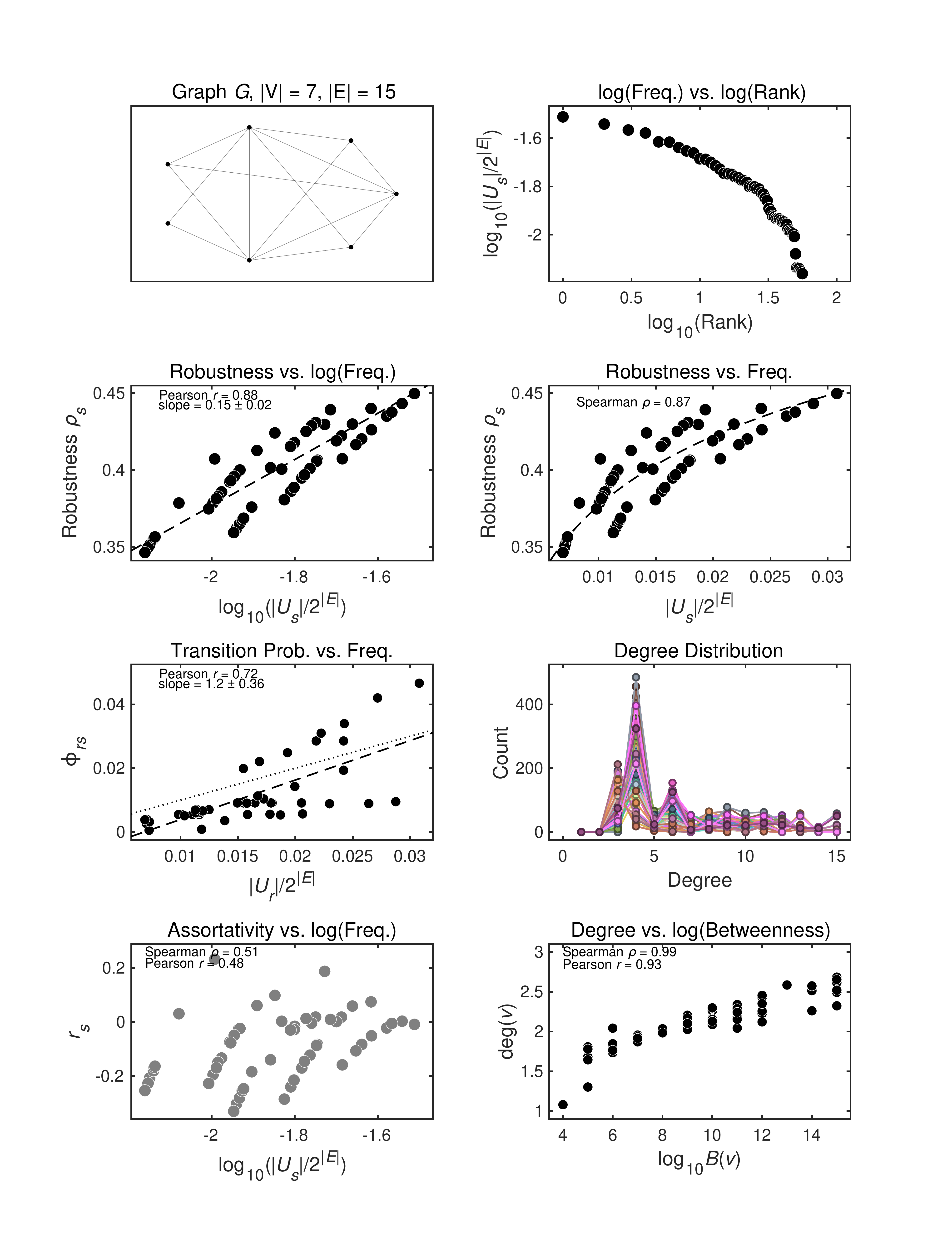}
\pagebreak
\includegraphics[height = \textheight]{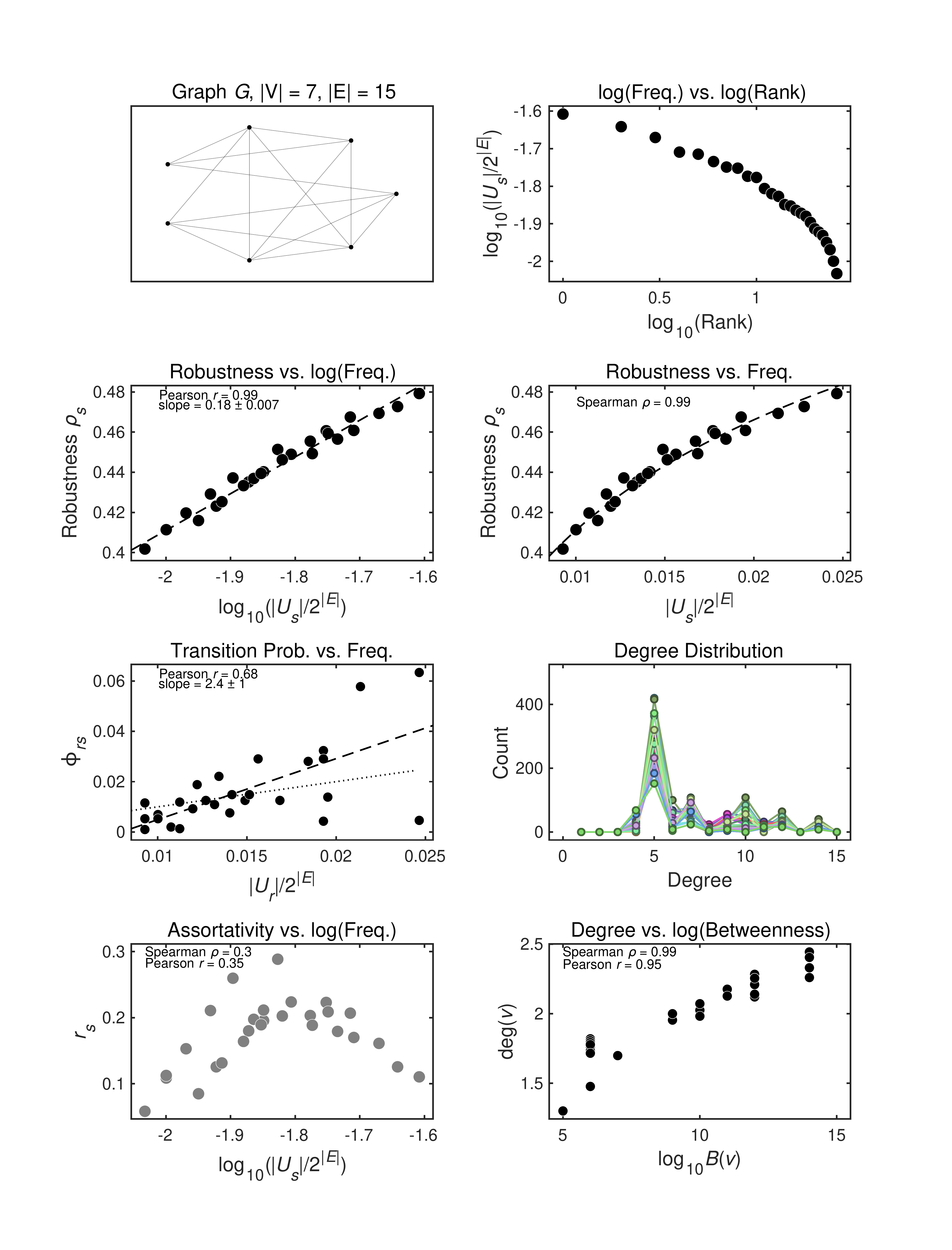}
\pagebreak
\includegraphics[height = \textheight]{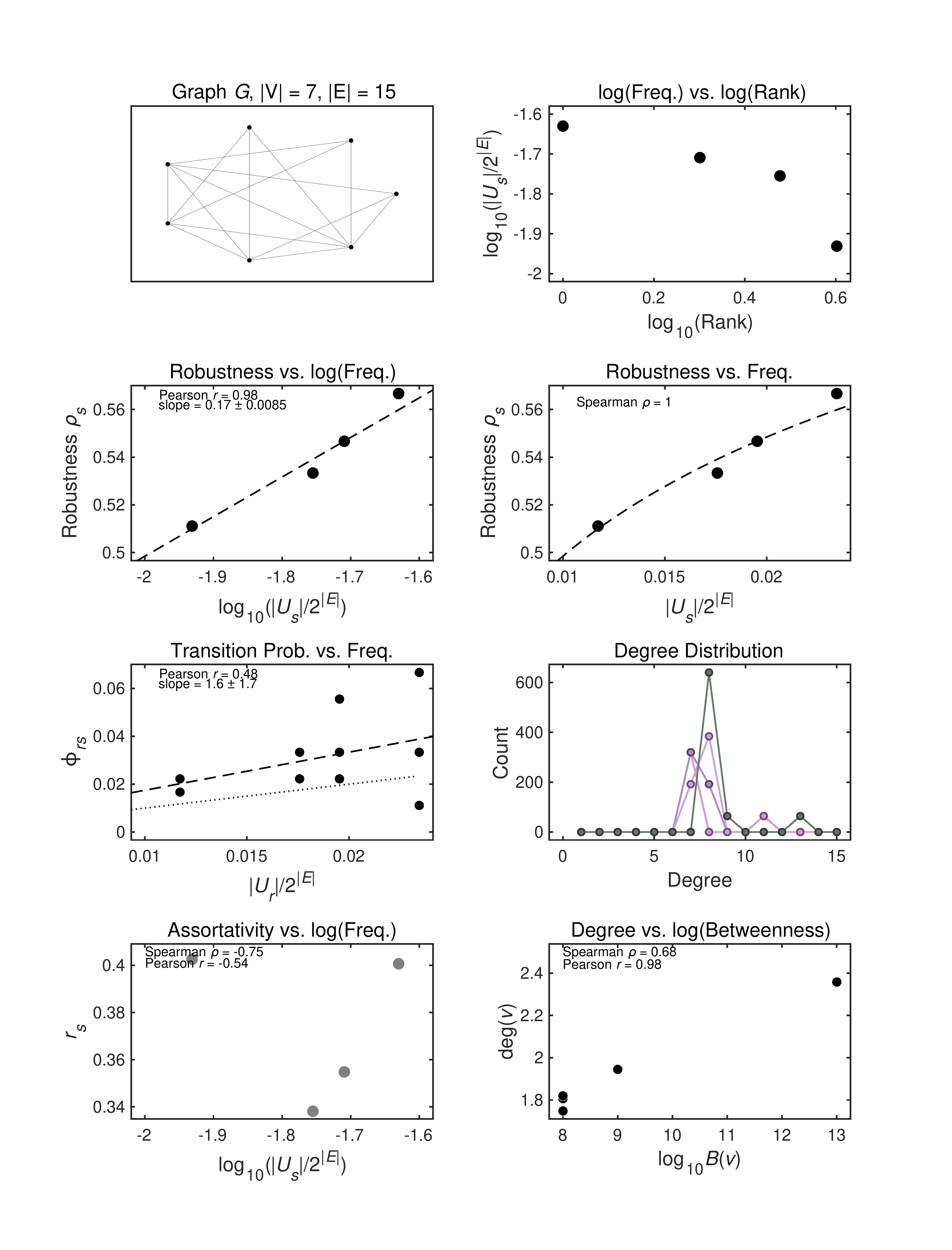}
\pagebreak
\includegraphics[height = \textheight]{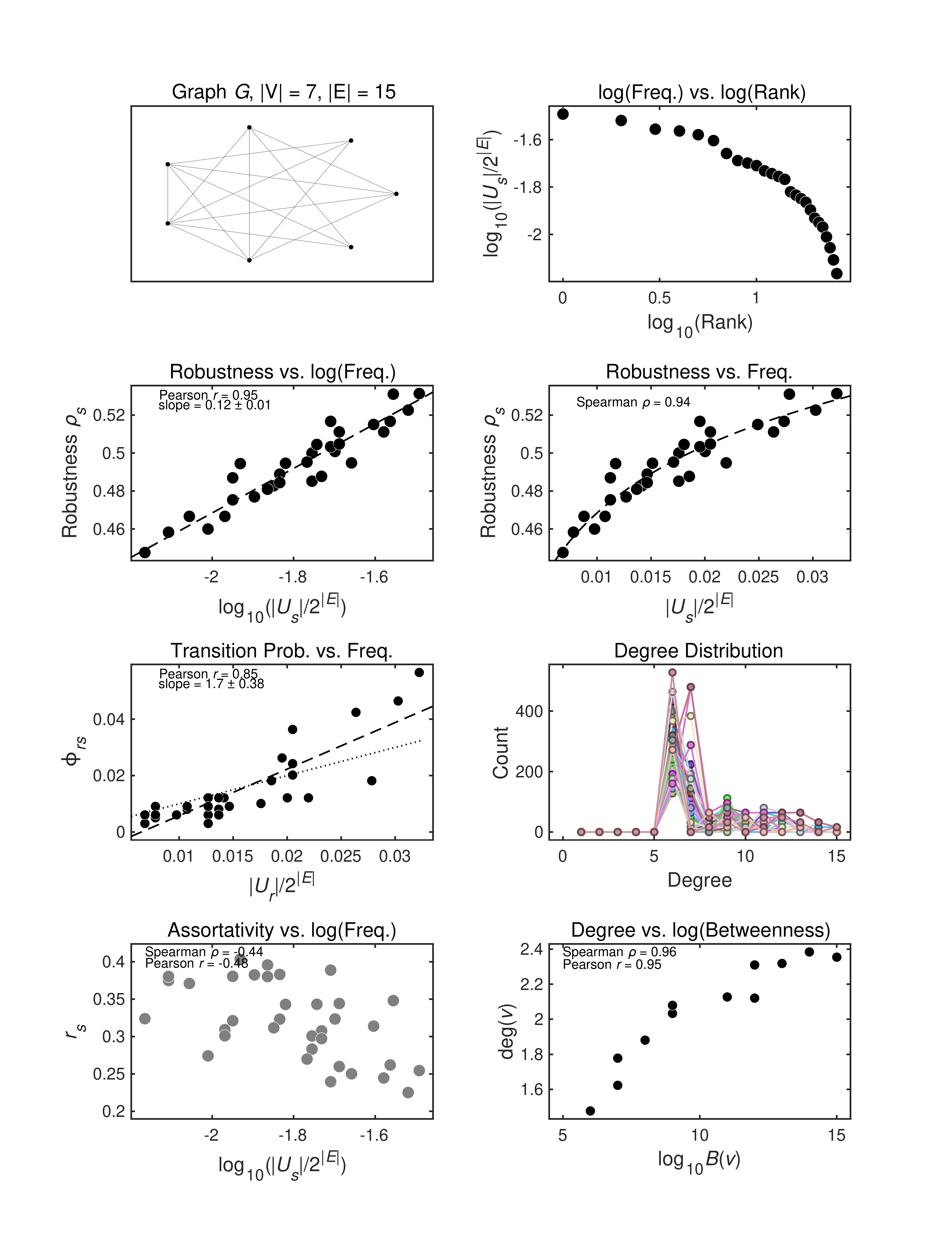}
\pagebreak
\includegraphics[height = \textheight]{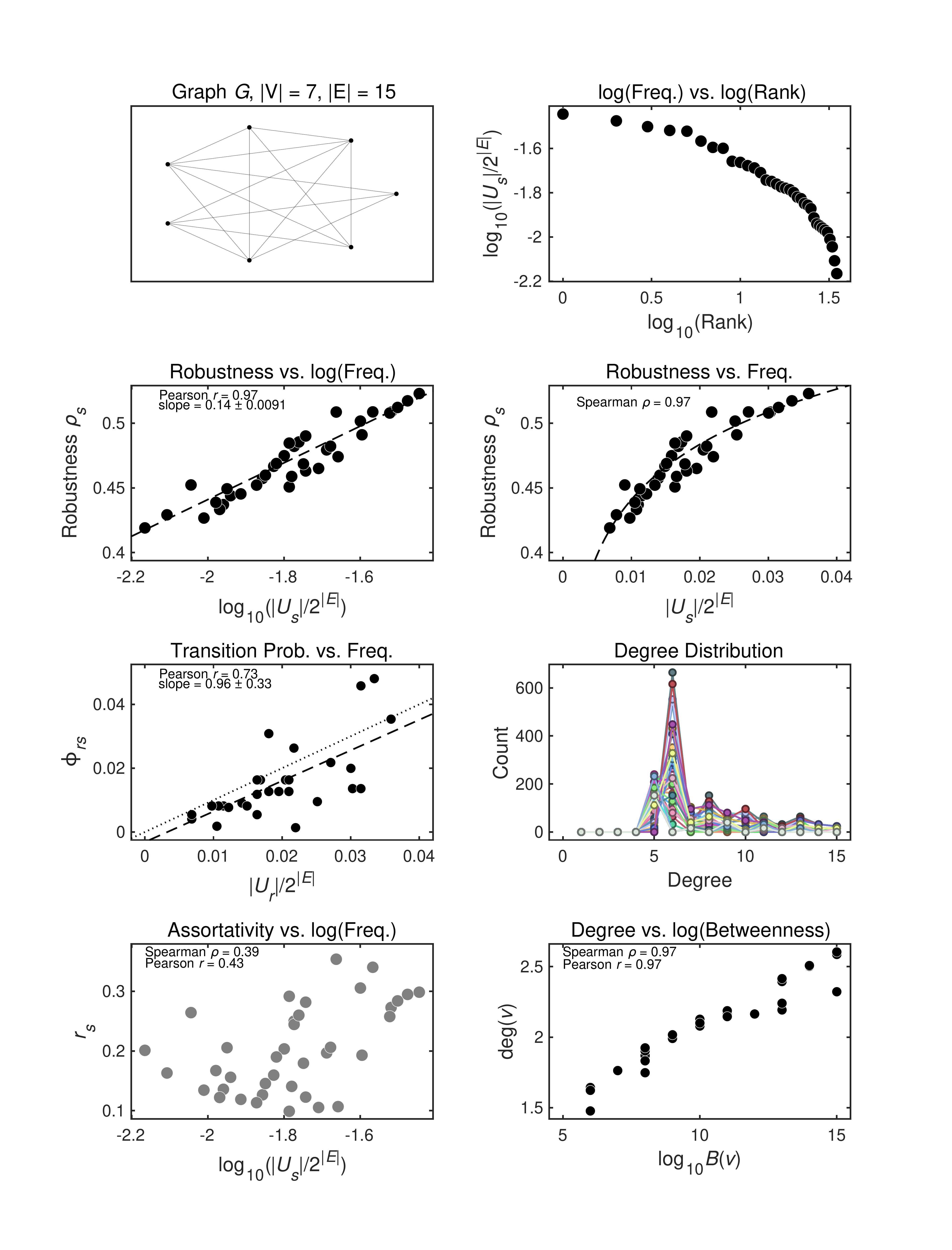}
\pagebreak
\includegraphics[height = \textheight]{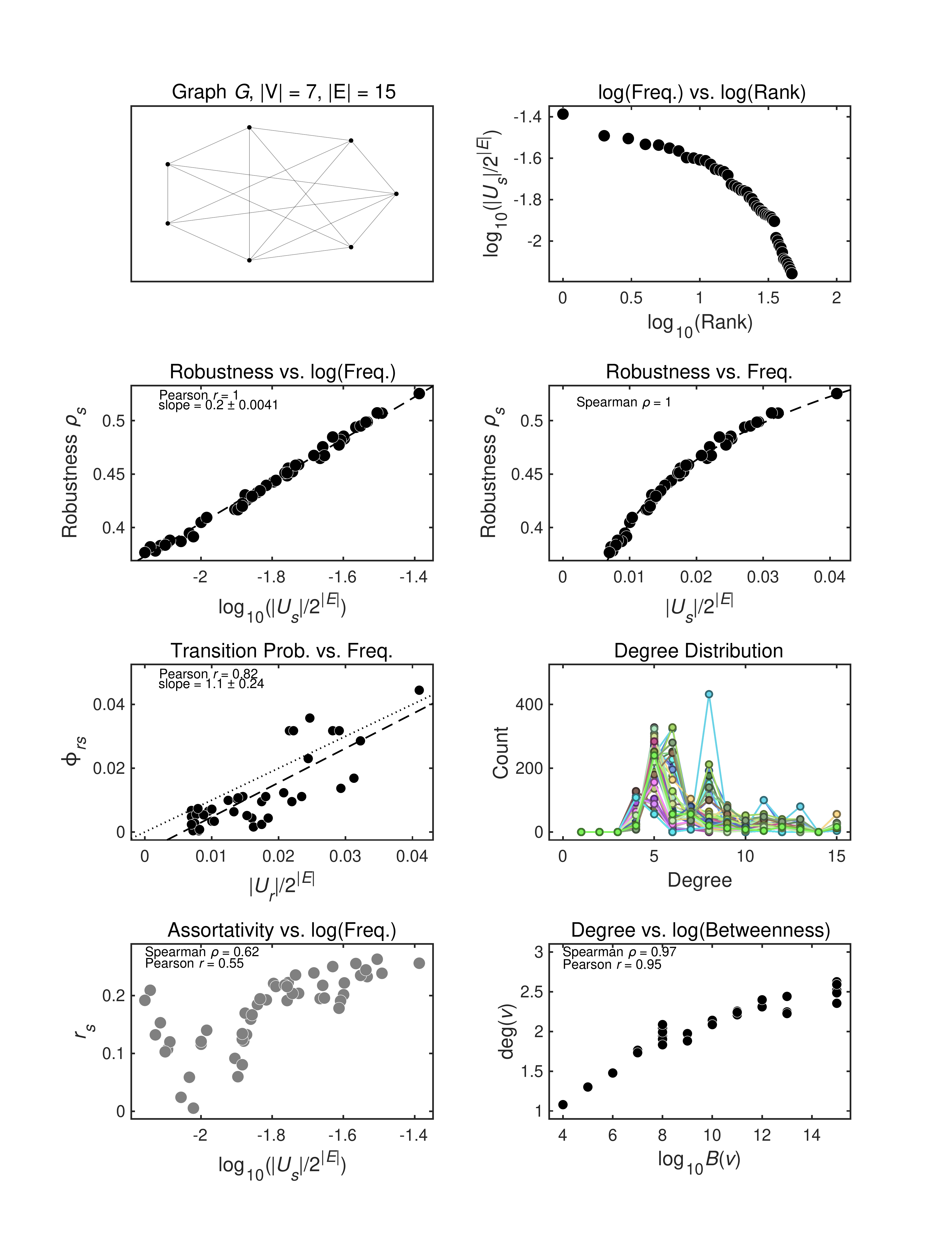}
\pagebreak
\includegraphics[height = \textheight]{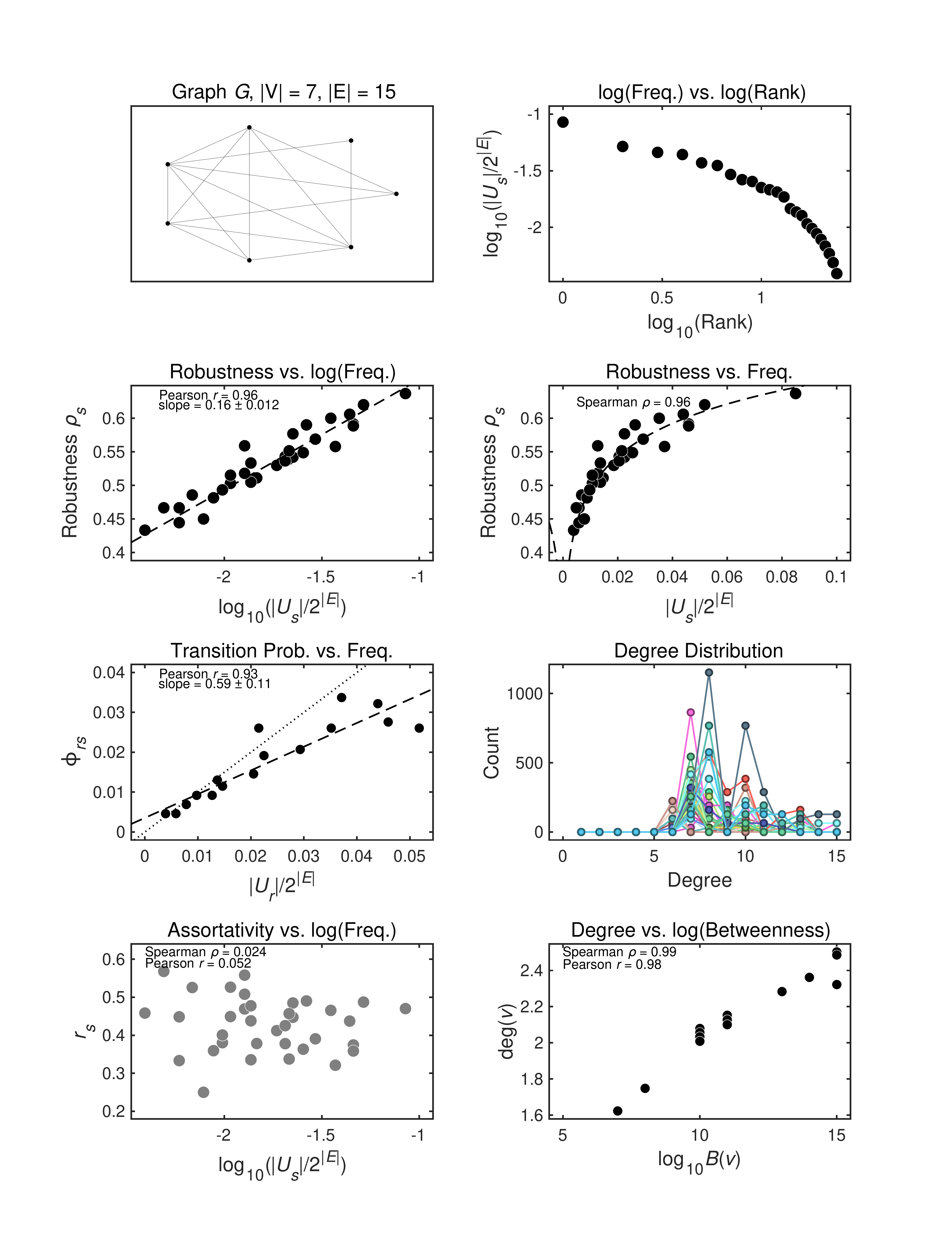}
\pagebreak
\includegraphics[height = \textheight]{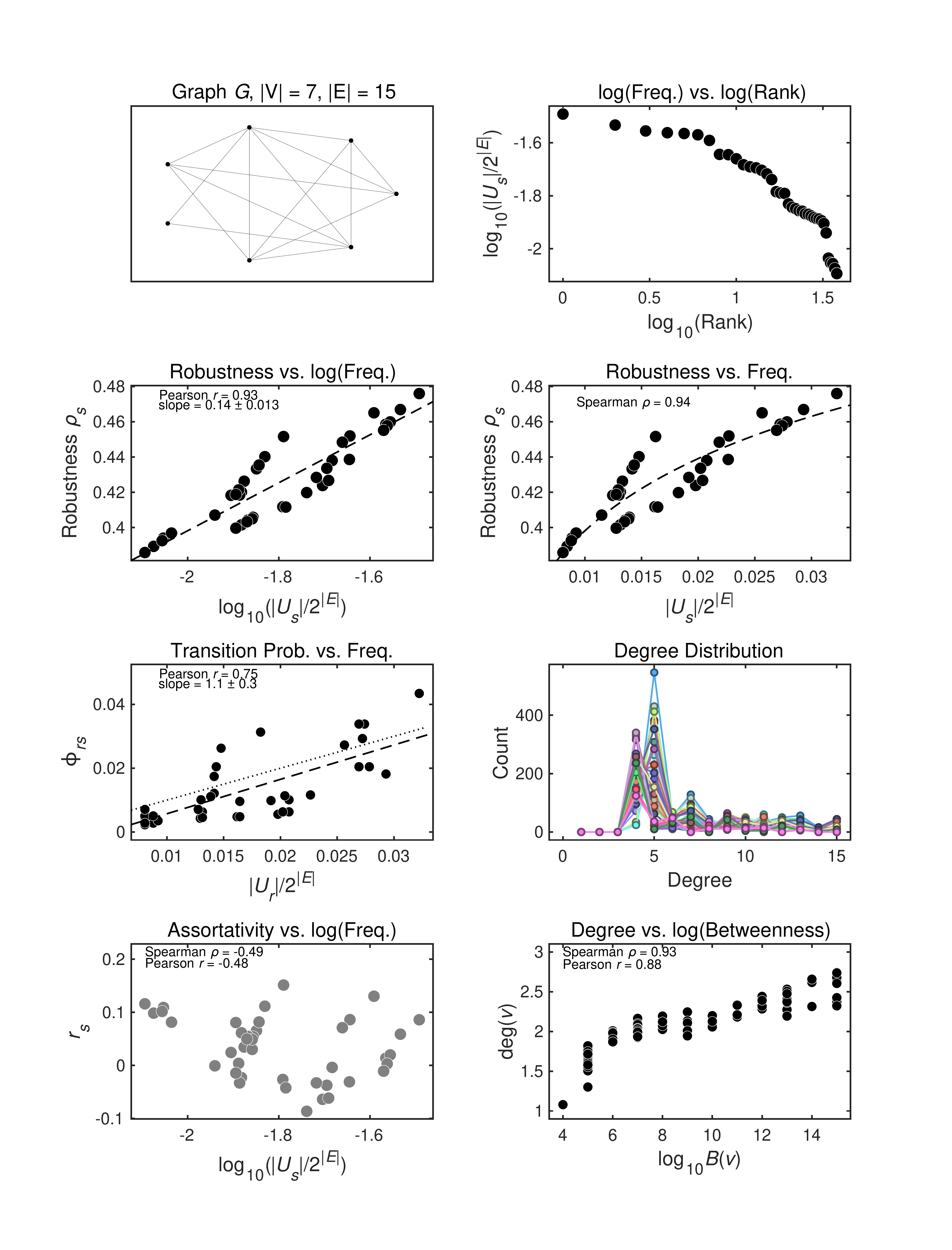}
\pagebreak
\includegraphics[height = \textheight]{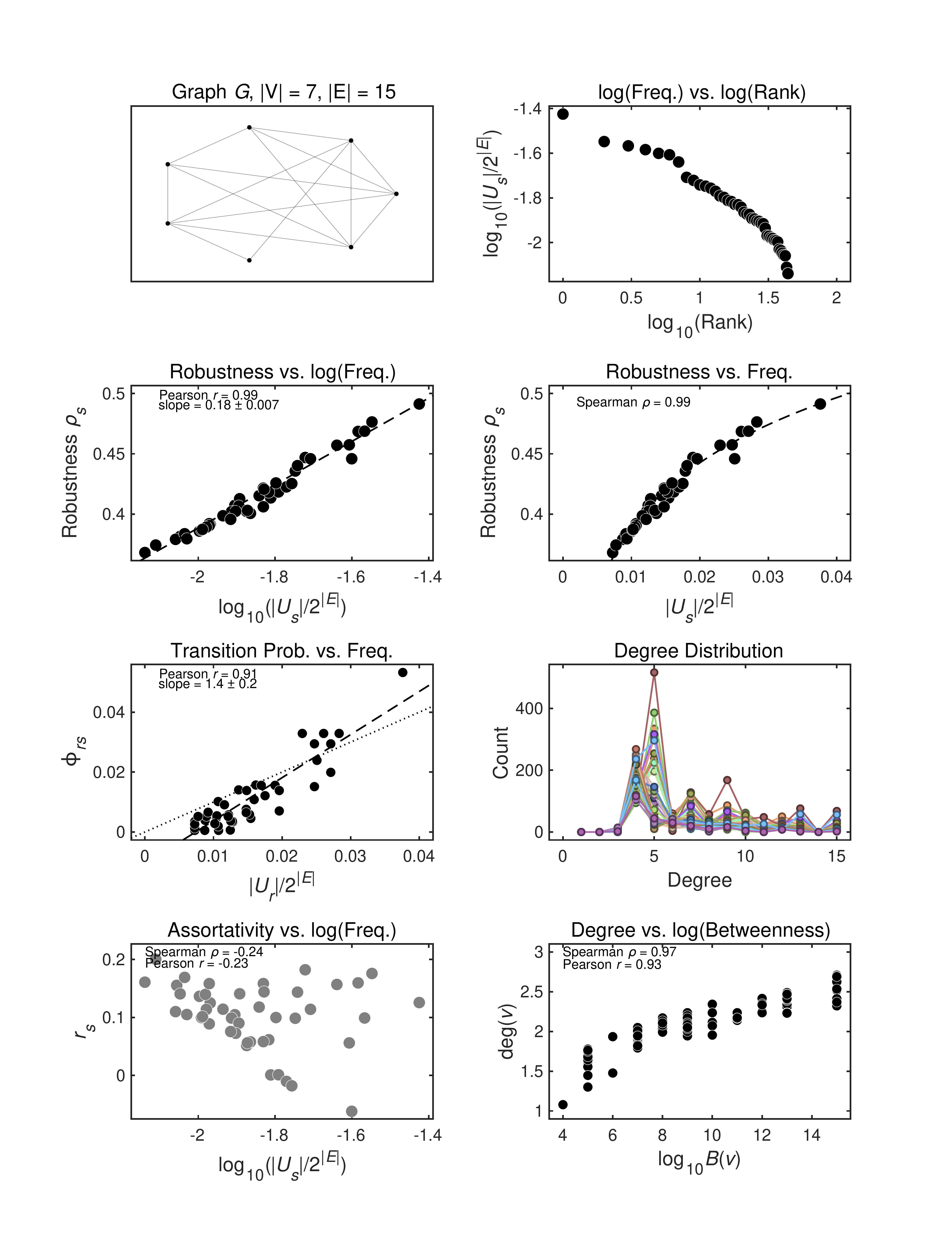}
\pagebreak
\includegraphics[height = \textheight]{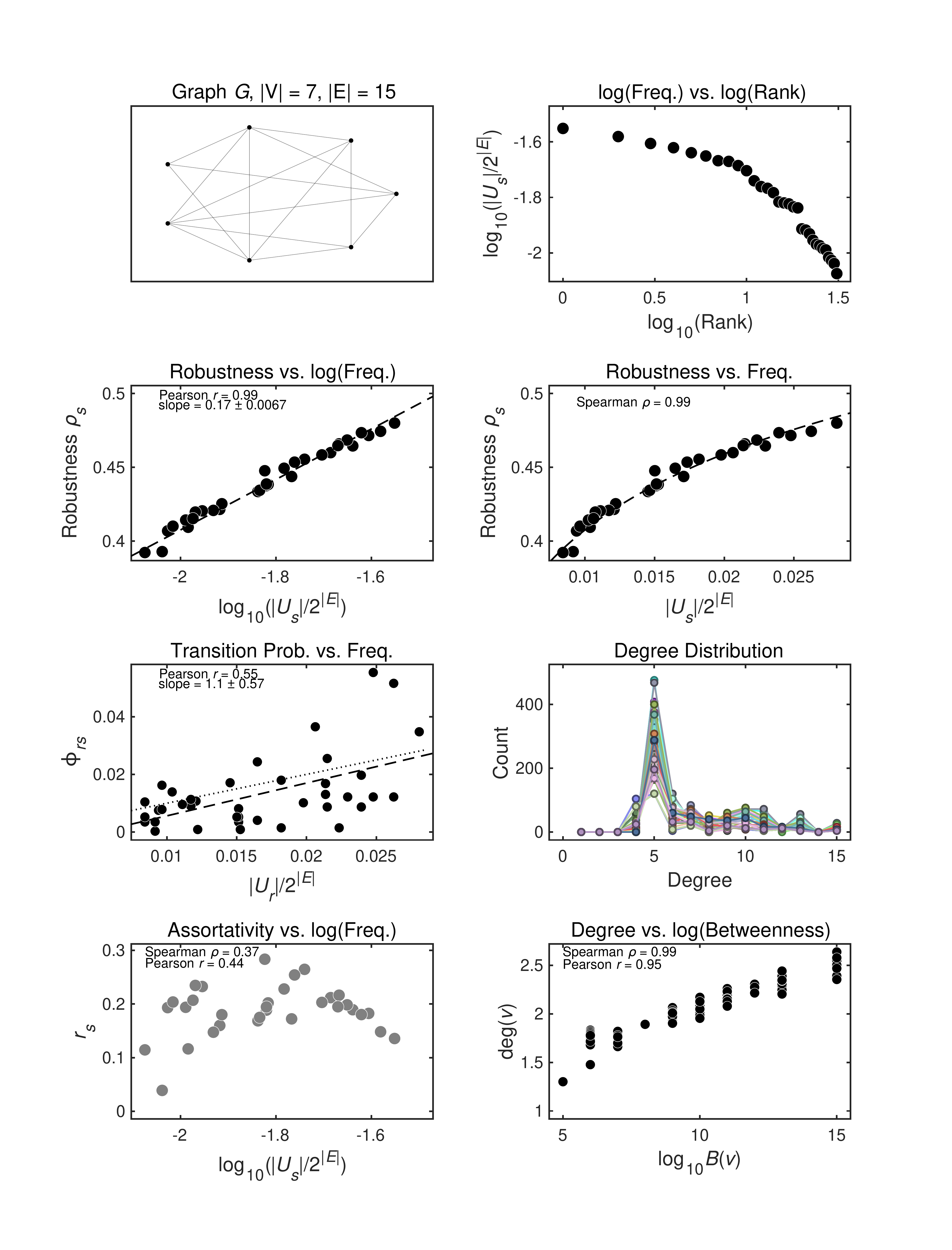}
\pagebreak
\includegraphics[height = \textheight]{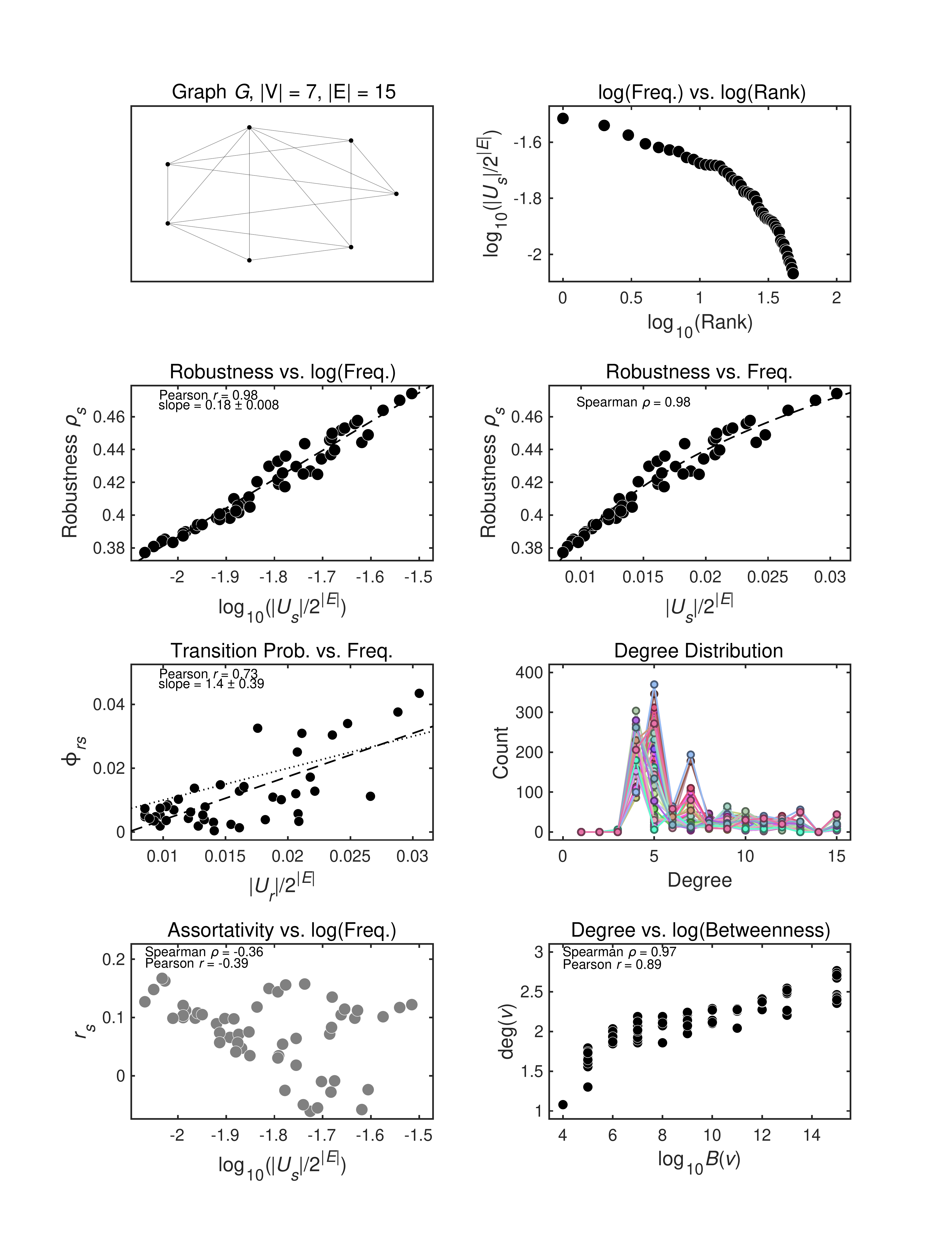}
\pagebreak
\includegraphics[height = \textheight]{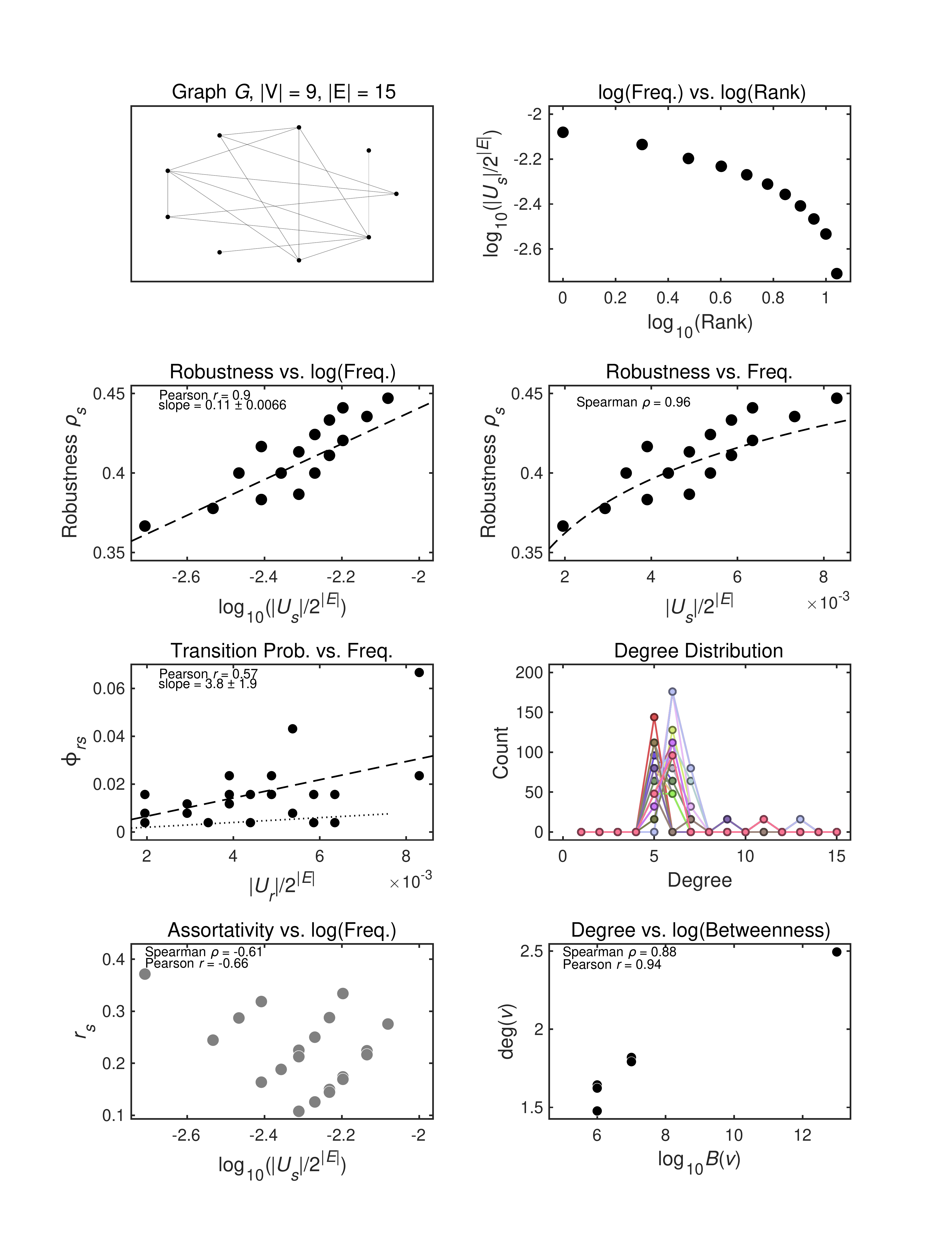}
\pagebreak
\includegraphics[height = \textheight]{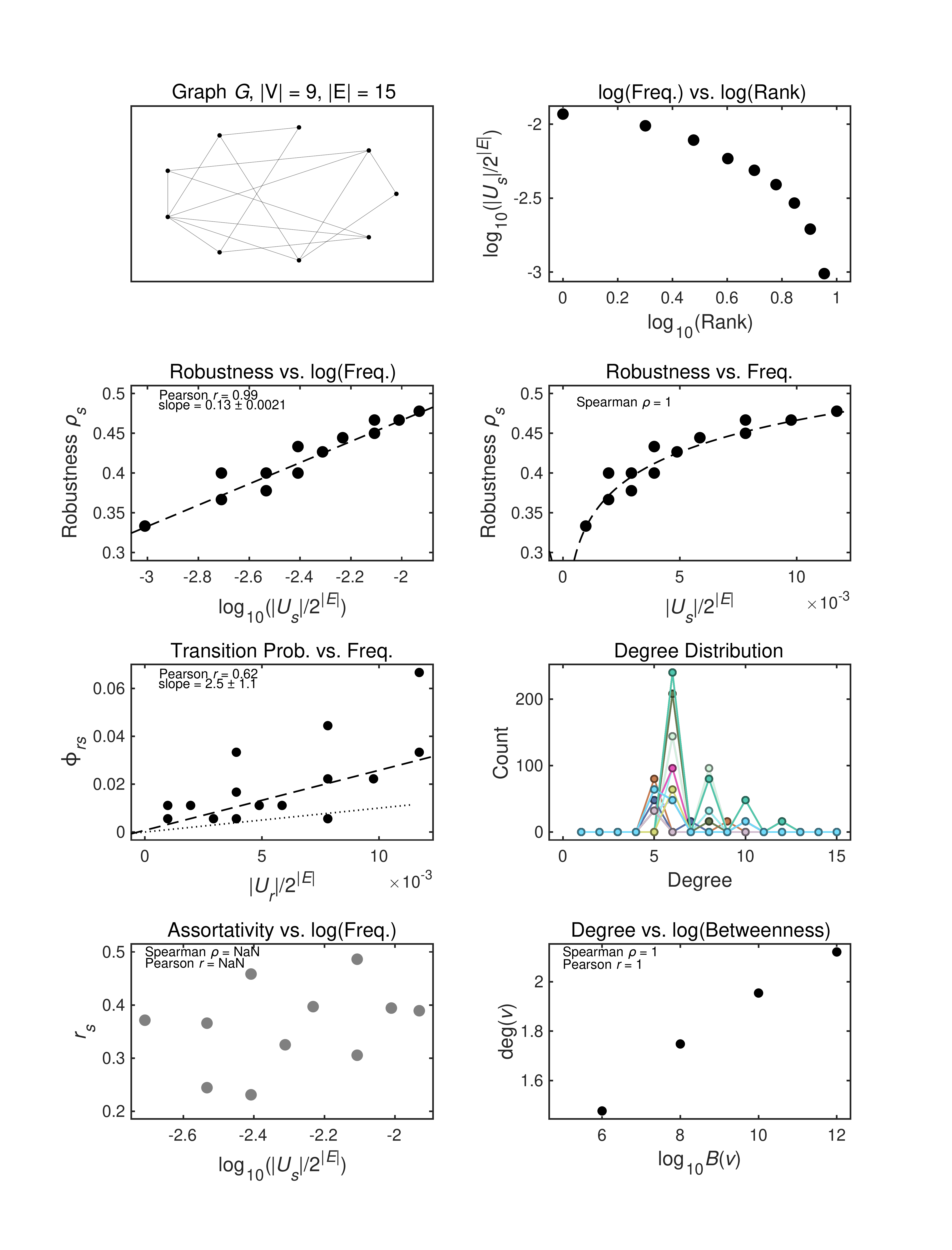}
\pagebreak
\includegraphics[height = \textheight]{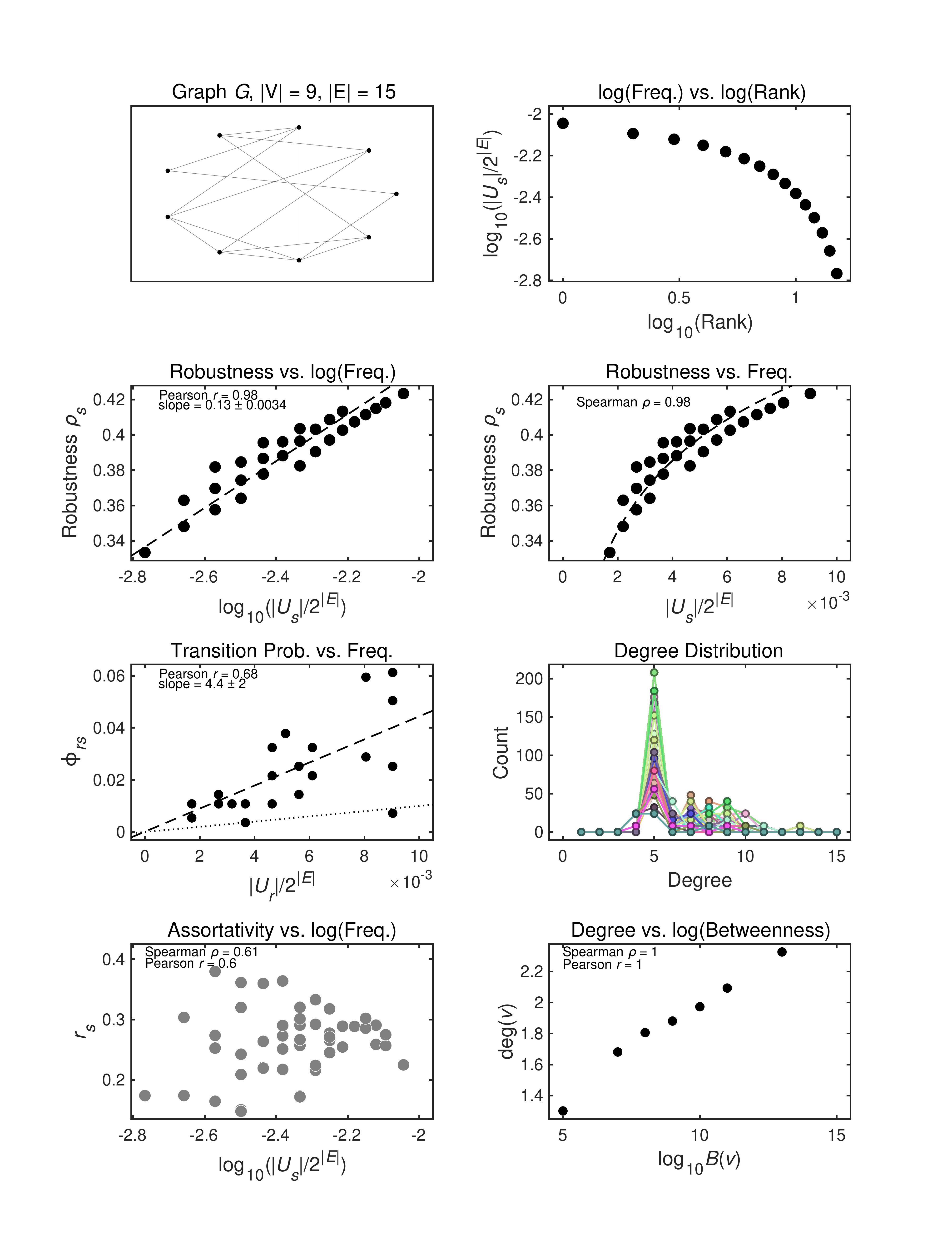}
\pagebreak
\includegraphics[height = \textheight]{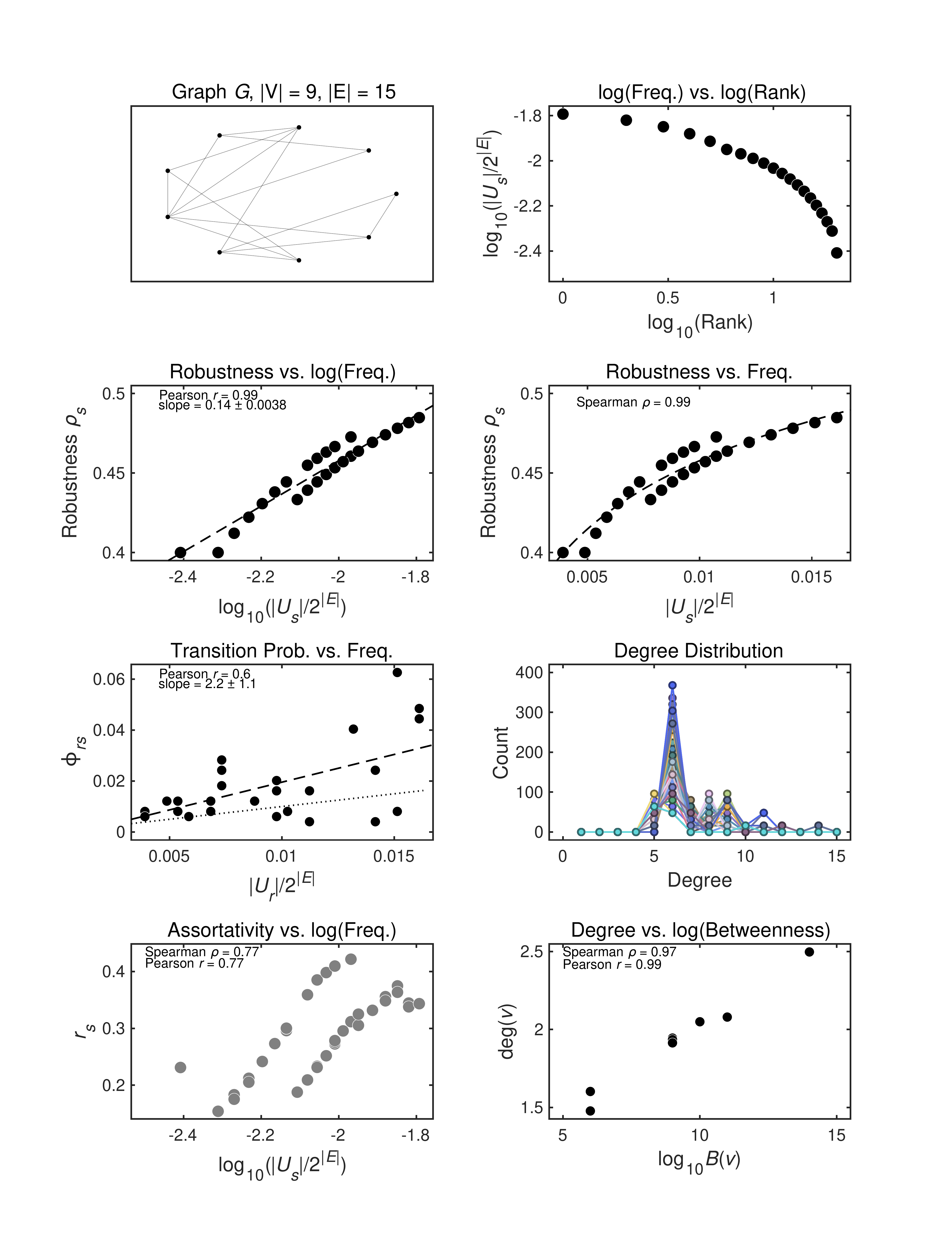}
\pagebreak
\includegraphics[height = \textheight]{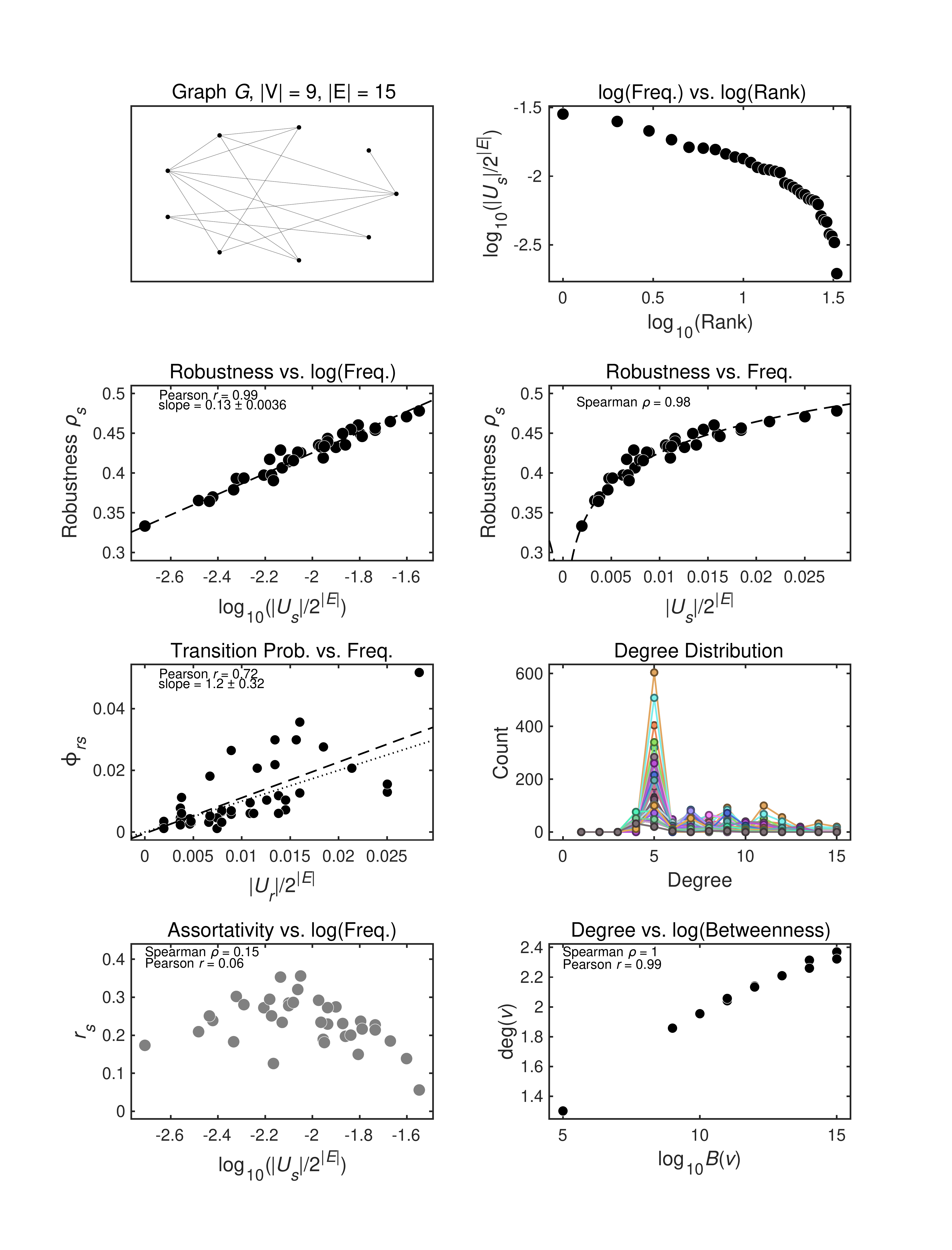}
\pagebreak
\includegraphics[height = \textheight]{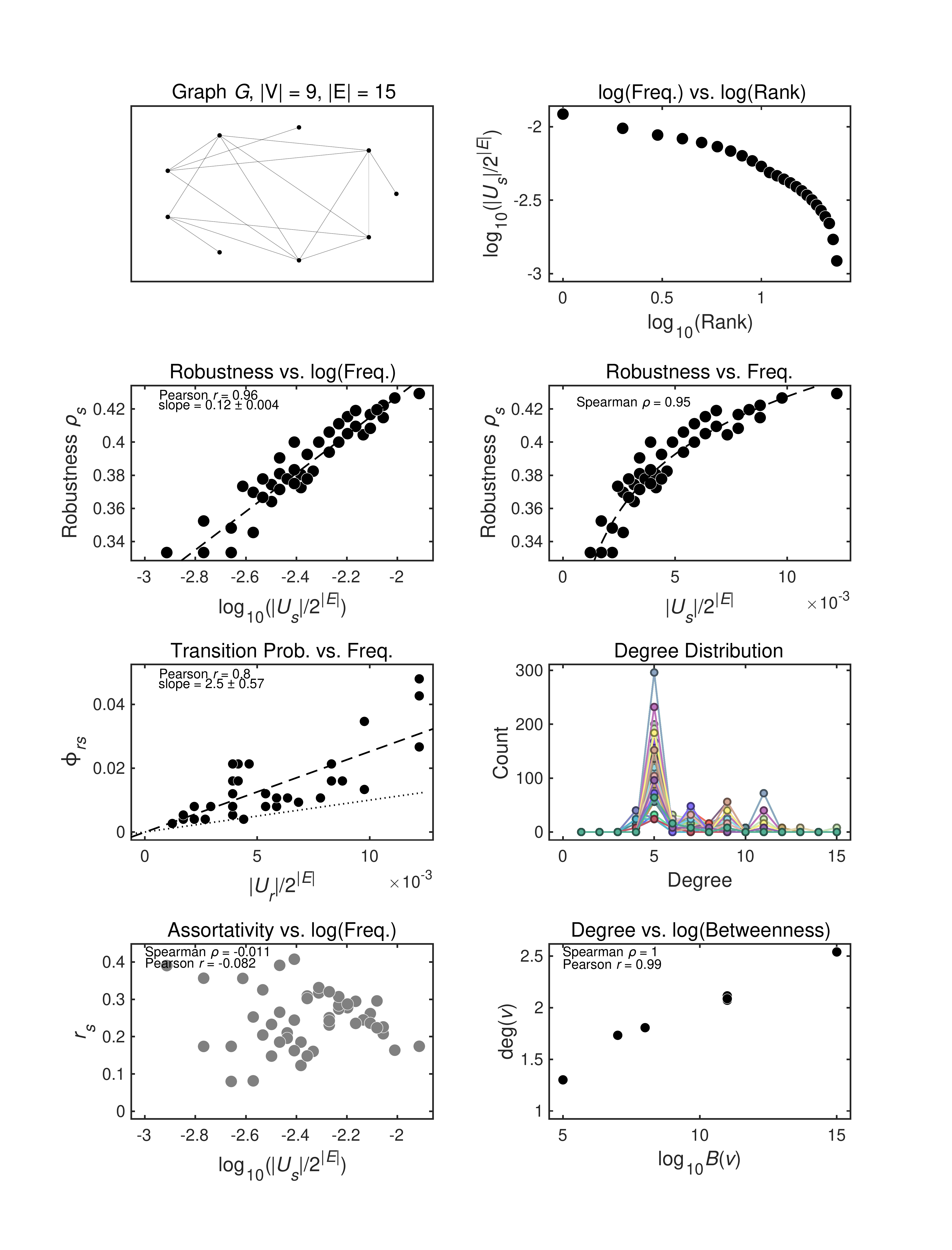}
\pagebreak
\includegraphics[height = \textheight]{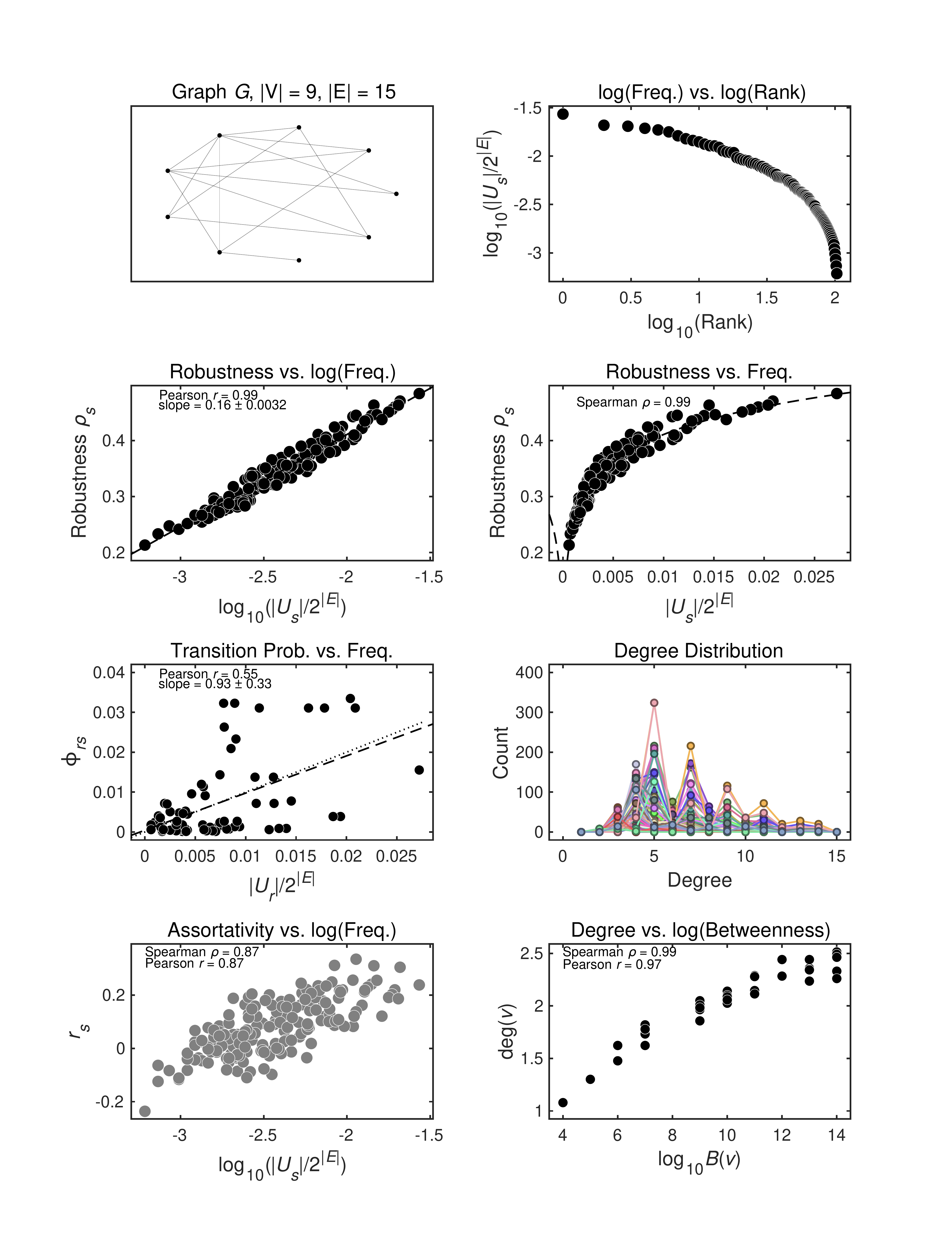}
\pagebreak
\includegraphics[height = \textheight]{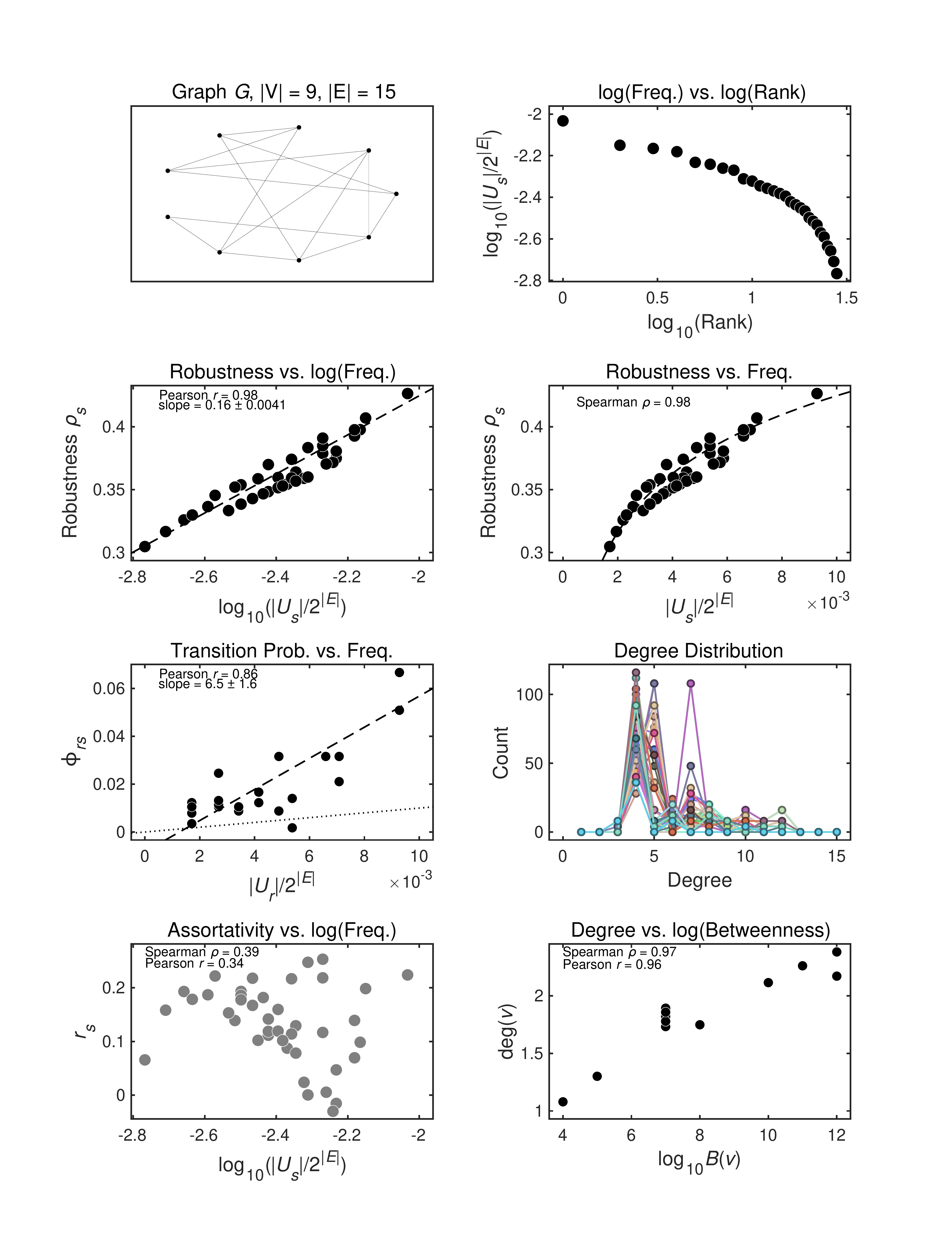}
\pagebreak
\includegraphics[height = \textheight]{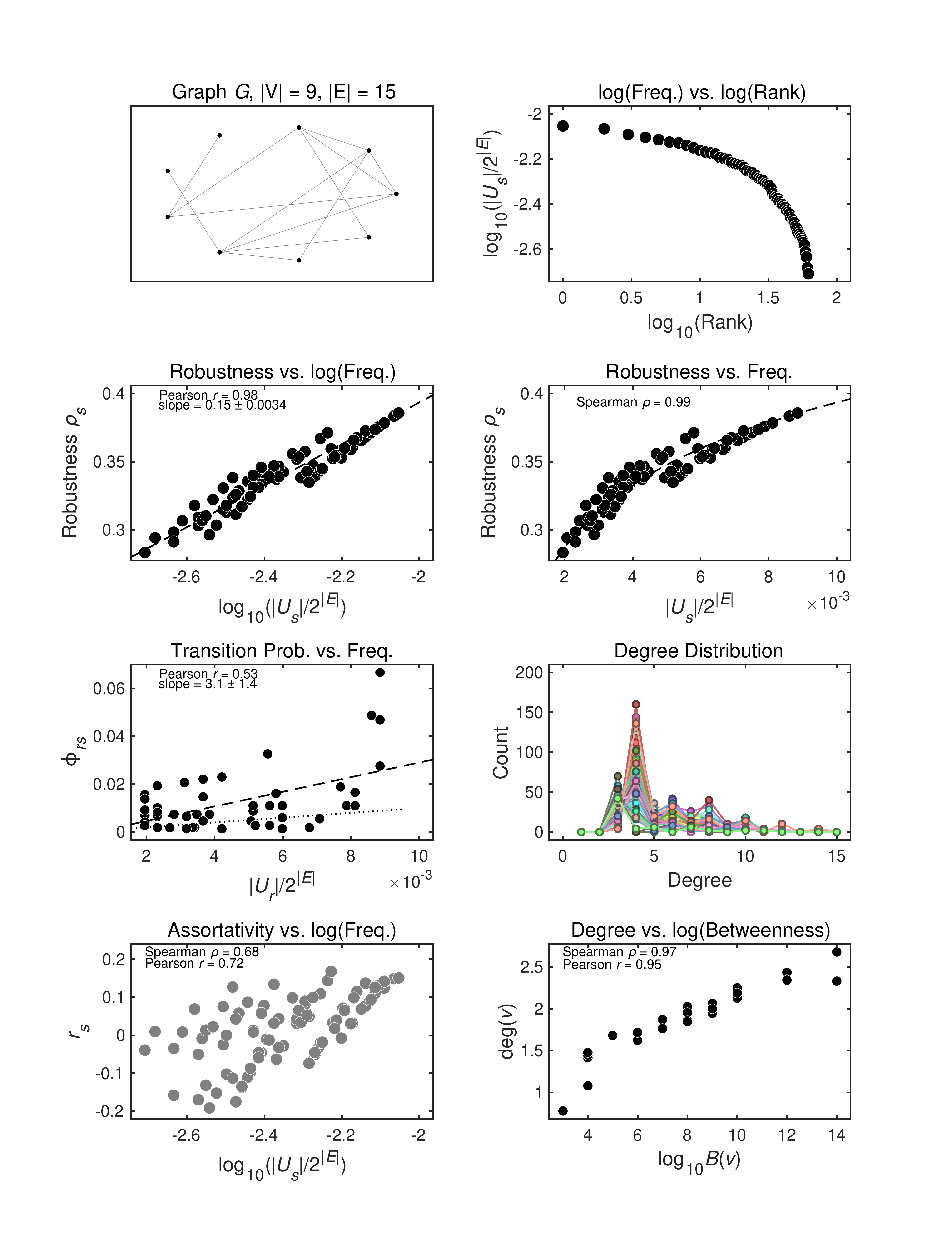}
\pagebreak
\includegraphics[height = \textheight]{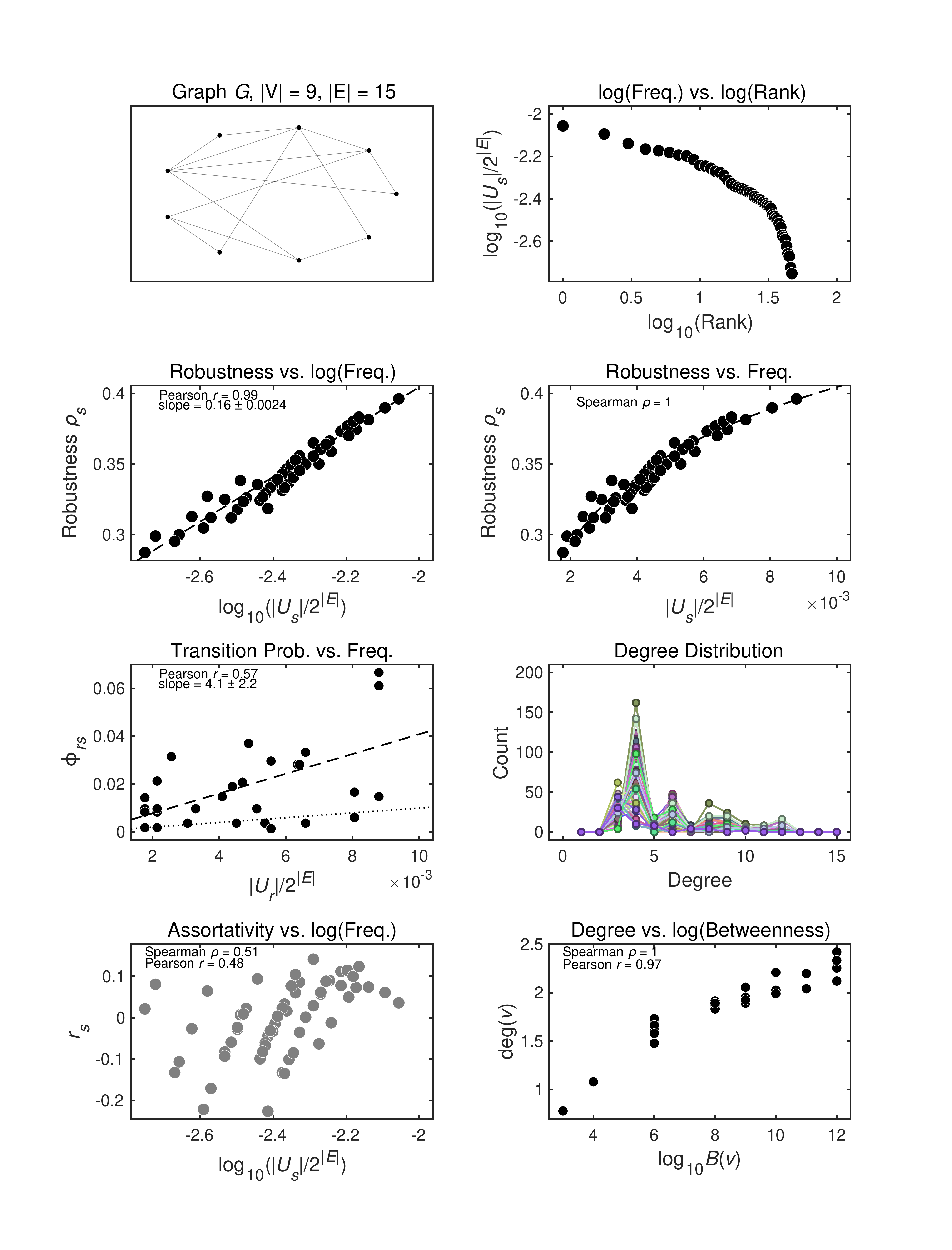}
\pagebreak
\includegraphics[height = \textheight]{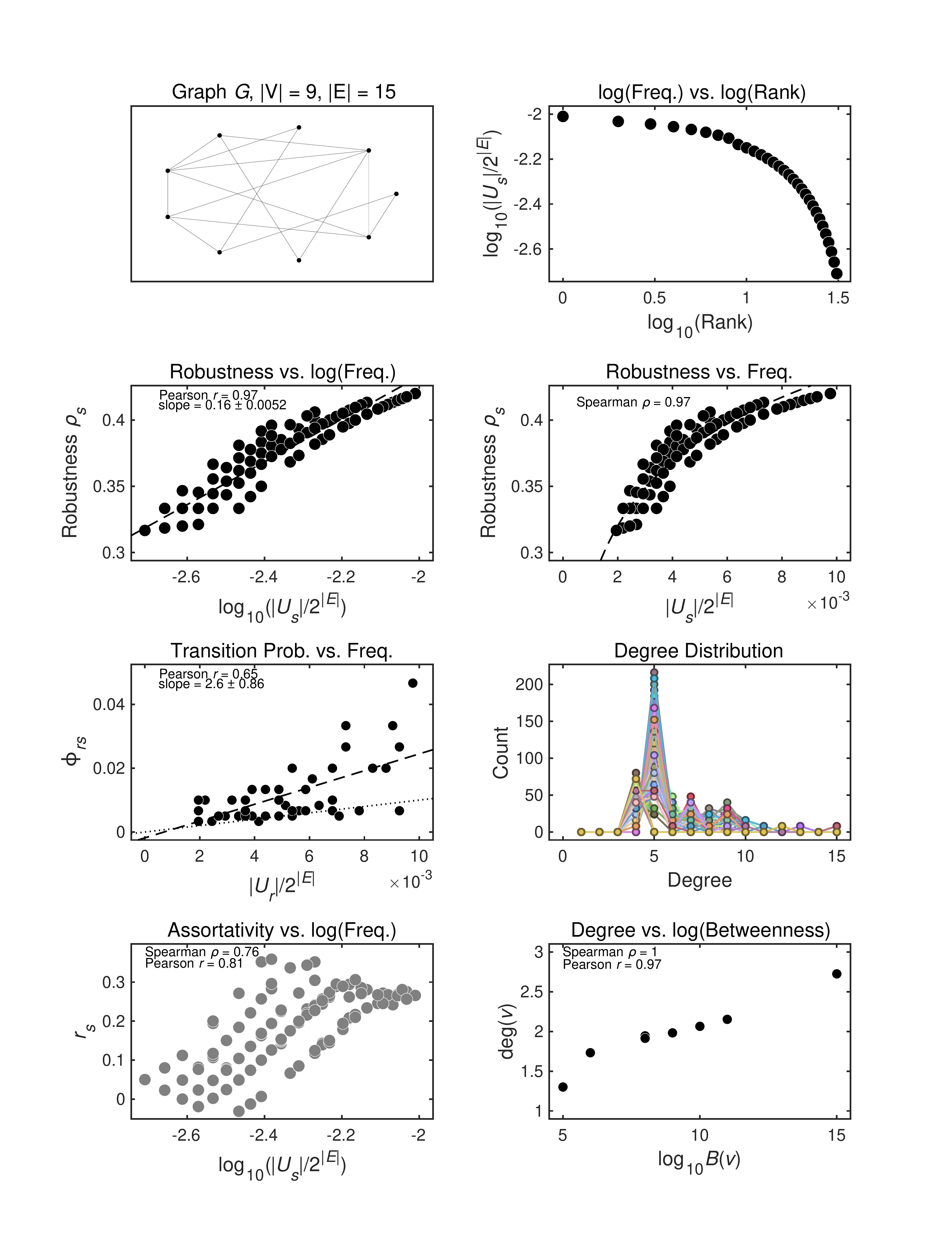}
\pagebreak
\includegraphics[height = \textheight]{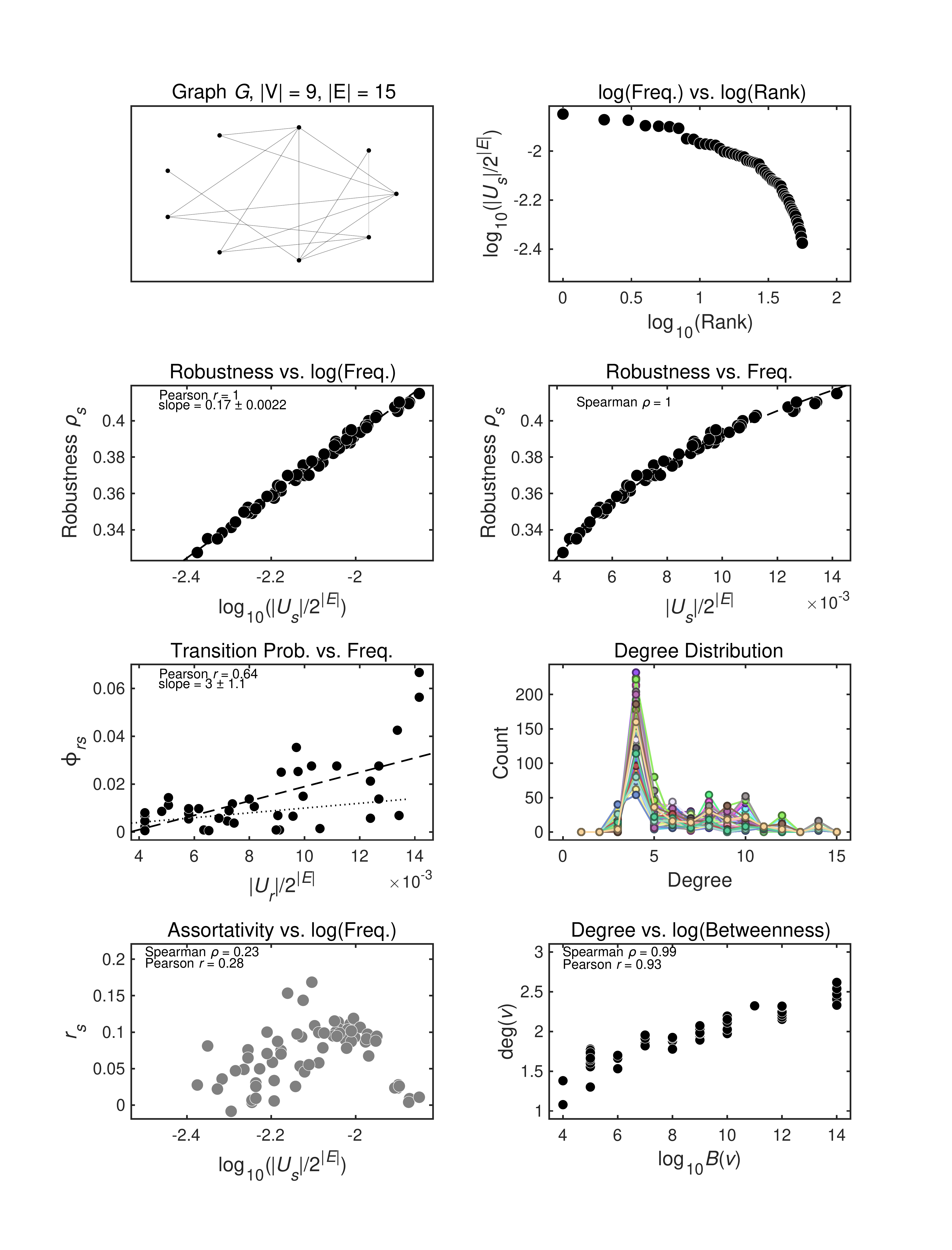}

\end{widetext}

\bibliography{bibs}